\documentclass[times,numbers,sort,compress]{elsarticle}
\usepackage[
    a4paper,
    left=1in,
    right=1in,
    top=1in,
    bottom=1in
]{geometry}
\usepackage[utf8]{inputenc} 
\usepackage[T1]{fontenc}    
\usepackage{url}           
\usepackage{booktabs}       
\usepackage{amsfonts}       
\usepackage{nicefrac}       
\usepackage{microtype}      
\usepackage{float}
\usepackage{caption}
\usepackage{comment}
\usepackage{enumitem}
\usepackage{subcaption}
\usepackage{soul}
\usepackage{amssymb,amsmath,amsthm}
\usepackage{changepage}
\usepackage{color,graphicx,float}
\usepackage{bm}
\usepackage{alltt,listings,tikz}
\usepackage[all]{xy}
\usepackage{hyperref}
\usepackage{fancyhdr}
\usepgflibrary{shadings}
\newcommand{\red}[1]{\ifmmode\mathbf{\textcolor{red}{#1}}\else \textbf{\textcolor{red}{#1}}\fi}

\hypersetup{
    colorlinks,
    linkcolor={blue},
    citecolor={blue},
    urlcolor={black}
}
\begin{document}

\title{\textbf{{\Large Dynamics and stability of volatile thin liquid films on a heated vertical cylinder}}}

\author{Sonam}
\ead{sonamch21@iitk.ac.in}

\author{Souradip Chattopadhyay\corref{cor1}}
\ead{souradipc@iitk.ac.in}

\author{Naveen Tiwari}
\ead{naveent@iitk.ac.in}

\cortext[cor1]{Corresponding author}

\address{Department of Chemical Engineering,  Indian Institute
of Technology Kanpur, Kanpur, UP 208016, India}

\begin{abstract}
Volatile thin liquid films flowing down heated vertical cylinders are central to many heat transfer applications, including evaporative condensers, cooling of electronics, and solar thermal systems. Despite their importance, the coupled dynamics of evaporation, thermocapillarity, and interfacial instabilities on cylindrical geometries remain poorly understood. In this work, a lubrication-based model is developed to capture the nonlinear evolution of a volatile thin film under gravity, surface tension, van der Waals forces, and non-conservative evaporative flux. The model incorporates hydrodynamic pressure contributions to evaporation, enabling a unified description of phase-change effects on curved substrates. A linear stability analysis of the model equation is performed with a non-constant base film thickness. The linear growth rate is first obtained for a spatially uniform film. Then, for a spatially nonuniform film, a coupled system of linearized equations is solved using the finite element method. Further, the full liquid-vapor interface is numerically simulated to demonstrate liquid evaporation. Results reveal that evaporation accelerates film thinning, condensation promotes thickening, and van der Waals forces stabilize ultra-thin films by preventing rupture. The influence of Bond number, wavenumber, and pressure-dependent flux laws on droplet formation and collision dynamics is quantified. These findings provide new mechanistic insights into phase-change driven instabilities in cylindrical thin films and have direct implications for the design of efficient thermal management systems, evaporative cooling devices, and surface-engineered heat exchangers.
\end{abstract}
\maketitle

\section{Introduction}\label{sec:1}
A thin liquid film flowing along a vertical cylinder under thermal influences has been widely investigated because of its relevance to numerous industrial applications \cite{zeng2017experimental,zeng2018thermohydraulic}. The heat transfer process plays a crucial role in technologies involving such flows, for instance, in evaporators \cite{sweetland2000evaporative}. When the liquid film is heated, thermocapillary (Marangoni) stresses arising from temperature differences between the cylinder surface and the surrounding air can alter the flow dynamics \cite{davalos2013stability}.

\par Several studies have been reported in the literature to understand the dynamics of a thin nonvolatile liquid film along a uniformly/non-uniformly heated cylindrical surface. On such surfaces, the Rayleigh-Plateau (RP) instability leads to a wide variety of spatial and temporal dynamics. Studies by Liu \& Liu \cite{liu2014thermocapillary} and Chao et al. \cite{chao2018dynamics} demonstrated that uniform heating of a cylinder induces thermocapillary Marangoni stresses at the liquid–air interface, thereby promoting the Rayleigh–Plateau (RP) instability. The models developed by Chao et al. \cite{chao2020reactive}, Chattopadhyay \& Ji \cite{chattopadhyay2024modeling}, and Chattopadhyay \cite{chattopadhyay2024thermocapillary,chattopadhyay2023thin} examined the behavior of reactive liquid films on a vertical cylinder, where thermocapillary stresses arise from heat generated both by cylinder heating and chemical reaction. While the studies \cite{liu2014thermocapillary,chao2018dynamics,chao2020reactive,chattopadhyay2024modeling,chattopadhyay2024thermocapillary,chattopadhyay2023thin} considered only surface tension to vary linearly with temperature variation, the studies by Khanum \& Tiwari \cite{khanum2020gravity} and Singh \& Tiwari \cite{singh2024stability} additionally assumed the dynamic viscosity to vary exponentially with temperature change. 

\par Although the studies mentioned above investigated how thermal effects influence the dynamics of thin liquid films on a vertical cylindrical surface, they ignored the phase change effects. However, in reality, phase change can significantly affect the stability of a liquid–vapor interface and has wide applications in various physical systems, such as thermal management \cite{miljkovic2013condensation}, food processing \cite{armerding1966evaporation}, and refrigeration \cite{fernandez2014refrigerant}. Evaporative condensers are very useful in engineering applications, as they can meet high cooling demands while using less energy \cite{kim2017cooling}. 

\par For a liquid flowing along a planar substrate, the effect of evaporation/condensation has been reported by Burelbach et al. \cite{burelbach1988nonlinear}, Oron \& Bankoff \cite{oron1999dewetting,oron2001dynamics}, Ajaev \cite{ajaev2005evolution,ajaev2005spreading}, Amini \& Homsy \cite{amini2017evaporation,amini2017evaporation_a}, Joo et al. \cite{joo1991long}, Sultan et al. \cite{sultan2005evaporation}, Ji \& Witelski \cite{ji2018instability} and Shklyaev \& Fried \cite{shklyaev2007stability}. On the other hand, when the liquid flows along an inclined or vertical plane, the effect of interfacial phase change has been reported by Lin \cite{lin1975stability}, Hwang \& Weng \cite{hwang1987finite}, Tiwari et al. \cite{tiwari2014linear}, Tiwari \& Davis \cite{tiwari2011stability,tiwari2009nonmodal,tiwari2009linear} and Jiang et al. \cite{jiang2024three}. In these studies, various forms of evaporative mass flux have been used to study the phase change effects. For instance, $J(h)=E/\left(K_0+h\right)$ ($h$ is the film thickness, $K_0>0$ and $E$ is the scaled evaporation number) was considered in \cite{burelbach1988nonlinear,oron2001dynamics}, while $J(h)=\beta p(h)/\left(K_0+h\right)$ (where $p(h)$ denotes the hydrodynamic pressure and $\beta$ is a scaling constant) was considered by \cite{ji2018instability} and $J(h)=\left[T_0-\delta\left(h_{xx}+\epsilon h^{-3}\right)\right]/\left(K_0+h\right)$ ($h_{xx}$ is the surface tension, $\epsilon h^{-3}$ is the conjoining pressure, $T_0$ denotes a saturation temperature and $\delta$ characterizes the effect due to pressure during phase change) was used in \cite{ajaev2005spreading}. Tiwari \& Davis \cite{tiwari2009linear} derived the one-sided evaporation model assuming that the evaporation is caused by the departure from thermodynamic equilibrium at the liquid-gas interface. They showed that the flux expression remains unchanged when obtained assuming vapor mass transfer as the rate-limiting step.

\par In contrast to planar, inclined, or vertical flat surfaces, theoretical studies of interfacial phase change on vertical cylindrical surfaces are very limited. Hung et al. \cite{hung1996weakly}  studied the linear and weakly nonlinear stability of a thin condensate liquid film flowing down the outer surface of a static vertical cylinder, which was further extended by Chen et al. \cite{chen2004perturbation} to the case of a rotating cylinder. 

\par Motivated by the widespread applications of evaporation and condensation in engineering systems and the limited understanding of volatile thin liquid films on vertical cylinders, the present work develops a theoretical model based on lubrication theory to investigate the dynamics and stability of a volatile thin liquid film flowing over a uniformly heated vertical cylinder (see Sec. \ref{sec:PF}). Compared with a flat substrate, the dynamics on a cylindrical surface are considerably more complex because surface tension plays both stabilizing and destabilizing roles through the axial and azimuthal curvatures of the liquid--vapor interface, respectively. Furthermore, unlike many previous studies on planar/inclined surfaces that employed simplified evaporative flux models \cite{oron2001dynamics,burelbach1988nonlinear}, the present work adopts a more physically realistic non-conservative evaporative flux model following \cite{ajaev2005evolution,ajaev2001steady,ji2018instability}, thereby providing a more accurate description of evaporation effects in conjunction with the complexities introduced by the cylindrical geometry. In this sense, the present formulation extends existing volatile thin-film models by incorporating the effects of cylindrical geometry and phase change within a unified framework. In particular, the coupling between the curvature-induced interfacial dynamics and mass exchange provides insight into how evaporation/condensation can modify the film evolution and stability characteristics on a cylindrical surface, which cannot be captured by models based solely on planar geometry or a prescribed conserved base state. 

\par Moreover, since the mass is not conserved, the linear stability analysis is performed using a non-constant base film thickness in Sec. \ref{sec:stability}. In Sec. \ref{sec:stability_a}, the linear stability of a spatially uniform but time-dependent base film is investigated using a frozen-time framework. Subsequently, in Sec. \ref{spnu}, the analysis is extended to a base state that varies both spatially and temporally, with the base film and perturbation equations solved simultaneously in a finite-element solver. Furthermore, in Sec. \ref{sec:4}, nonlinear simulations of the full reduced evolution equation are performed over a range of parameter values to characterize the interfacial dynamics associated with evaporation-induced film thinning, precursor-film behavior, wave evolution, and droplet coalescence. The conclusions are summarized in Sec. \ref{summary}.

\par To the best of our knowledge, this is the first study to provide a unified theoretical framework for investigating dynamics and stability in volatile thin liquid films flowing over a heated vertical cylinder. The insights obtained from the present analysis are expected to improve the fundamental understanding of interfacial transport in such systems and may contribute to the development of more effective surface-engineering and thermal management strategies for thin volatile liquid films operating under non-isothermal conditions \cite{fu2025low,liu2025enhancing}.

\section{Model}
\label{sec:PF}
\begin{figure}[h!]
\centering
\includegraphics[scale=0.5]{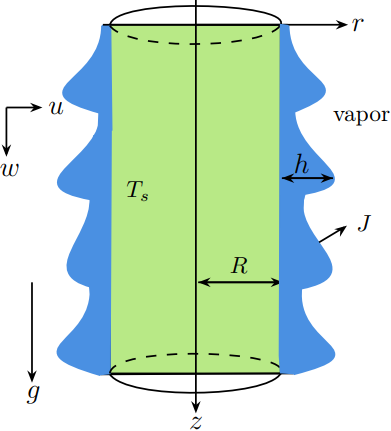} 
\caption{Volatile thin liquid film is flowing down a uniformly heated vertical cylinder.}	
\label{fig1}
\end{figure}

As shown in Fig. \ref{fig1}, an incompressible thin volatile liquid is flowing along the outer surface of a vertical cylinder of radius $R$ under gravity. The liquid is Newtonian, and the flow is assumed to be axisymmetric. The density, dynamic viscosity, surface tension, and thermal conductivity of the liquid are $\rho$, $\mu$, $\sigma$, and $k_{\text{th}}$, respectively. The cylinder is uniformly heated, and a constant temperature $ T_s$ is maintained at its surface, which is higher than the liquid's saturation temperature $T=T_{\text{sat}}$.  Although the density, viscosity, thermal conductivity, specific heat, and surface tension are known to vary with temperature and can influence the flow dynamics, the present study considers only the temperature dependence of surface tension as \cite{chattopadhyay2024thermocapillary,chao2018dynamics,ji2021thermally}: $\sigma(T)=\sigma_0-\sigma_T\left(T-T_{\text{sat}}\right)$, where $\sigma_0$ is the surface tension at $T_{\text{sat}}$ and $\sigma_T=-d\sigma/dT>0$ is a constant. A cylindrical coordinate system $(r,z)$ will be used to describe the flow dynamics, where $r$ denotes the radial direction and $z$ denotes the axial direction (flow direction). The velocity components along the radial and axial directions are $u$ and $w$, respectively. The cylinder wall-liquid interface is denoted by $r=R$. Considering $h=h(z,t)$, the thickness of the liquid at any instant, the equation for the liquid-vapor interface is expressed by $r=R+h$. This interface is surrounded by the vapor, whose density is denoted by $\rho^v\ll\rho$.   

\subsection{Governing Equations}
The flow is governed by the conservation of mass, conservation of momentum, and conservation of energy as given below:
\begin{equation}\label{SC_eqq_1}
\nabla \cdot \textbf{v} = 0,
\end{equation}
\begin{equation}\label{momentum}
\rho\left(\textbf{v}_t+ \textbf{v}\cdot\nabla\textbf{v}\right)= -\nabla p + \mu\nabla^2\textbf{v} +\rho g, 
\end{equation}
\begin{equation}\label{SC_eqq_2}
\rho C_p\left(T_t + \textbf{v}\cdot\nabla T\right) = k_{\text{th}}\nabla^2T,
\end{equation}
where  $\textbf{v} = (u,w)$, $p$ is the liquid pressure, $C_p$ is the specific heat capacity and $g$ is the gravitational acceleration. 

\par At the cylinder wall $r = R$, the usual no-slip and no-penetration conditions are applied for the velocity components, while for the thermal boundary condition, a uniform temperature is used. These boundary conditions are given below: 
\begin{equation}\label{SC_eqq_3}
\textbf{v} =\textbf{0}, \quad T=T_s.
\end{equation}

 \par Considering that $J(z,t)$ denotes the mass flux due to evaporation, the mass balance at the liquid-vapor interface $r=R+h(z,t)$ is given below \cite{delhaye1974jump,burelbach1988nonlinear}: 
 \begin{equation}\label{SC_eqq_3_n}
J=\rho\left(\textbf{v}-\textbf{v}^i\right)\cdot\textbf{n}=\rho^v\left(\textbf{v}^v-\textbf{v}^i\right)\cdot\textbf{n},
\end{equation}
where $\textbf{v}^i$ denotes the velocity of the liquid-vapor interface, $\textbf{v}^v$ is the vapor velocity and $\textbf{n}$ is the unit normal vector to the free surface. The liquid evaporates in the direction that is normal to the liquid-vapor interface. Evaporation (condensation) will occur when $J$ is positive (negative). 

\par At $r=R+h(z,t)$, the tangential stress balance is applied as given below \cite{chao2018dynamics,chao2020reactive,kalliadasis2011falling,khanum2020gravity}:
\begin{equation}\label{SC_eqq_44}
\textbf{t}\cdot\mathbb T\cdot\textbf{n} = \textbf{t}\cdot\nabla_s\sigma(T),
\end{equation}
where $\mathbb T$ is the viscous stress tensor, $\textbf{t}$ is the unit tangential vector to the free surface, $\nabla_s = \left(\textbf{I}-\textbf{nn}\right)\cdot\nabla$ is the surface gradient operator and \textbf{I} is the identity tensor.

\par The normal stress balance at $r=R+h(z,t)$ is imposed as $p^v-p + \textbf{n}\cdot\mathbb T\cdot\textbf{n} +J\left(\textbf{v}^v-\textbf{v}\right)\cdot\textbf{n}  =
-\sigma(T)\nabla_s\cdot\textbf{n}$. Using (\ref{SC_eqq_3_n}), the normal stress balance at the liquid-vapor interface becomes 
\begin{equation}\label{SC_eqq_4}
\phi(h)+p^v-p + \textbf{n}\cdot\mathbb T\cdot\textbf{n} -\frac{J^2}{\rho^v}(\chi-1)  =
-\sigma(T)\nabla_s\cdot\textbf{n},
\end{equation}
where $J^2/\rho^v$ represents the vapor recoil and $\chi=\rho^v/\rho$. Following \cite{ji2019dynamics,ji2021thermally}, the functional form of $\phi(h)$ in (\ref{SC_eqq_4}) is considered as $\phi(h)=-\widetilde\Pi/\left(6\pi h^3\right)$, where $\widetilde\Pi>0$ denotes the dimensional Hamaker number. 

\par The kinematic boundary condition at $r = R+h(z,t)$ is imposed as \cite{burelbach1988nonlinear}
\begin{equation}\label{kinematic}
\left(u-h_t-wh_z\right)\left(1+h_z^2\right)^{-1/2}=\frac{J}{\rho}.
\end{equation}

\par Finally, the energy balance at $r = R+h(z,t)$ is imposed as \cite{oron1997long}: 
\begin{equation}\label{SC_eqq_3_n2}
JH_{\text{vap}}=-k_{\text{th}}\nabla T\cdot\textbf{n}.
\end{equation}

\par The rate of evaporation (or condensation) depends on several factors, such as the pressure difference at the liquid–vapor interface, the temperature of the liquid, and the relative humidity of the surrounding air. Considering $p_{\text{ve}}$ and $p^v$ as the equilibrium vapor pressure and vapor pressure, respectively, the evaporative mass flux adopted in this study is expressed as \cite{ajaev2001steady,karapetsas2016evaporation,craster2009pinning,ji2023mathematical}
\begin{subequations}\label{SC_eqq_3_n1}
\begin{equation}\label{SC_eqq_3_n1a}
J=\widehat\alpha\rho^v\left(\frac{R_gT_{\text{sat}}}{2\pi}\right)^{1/2}\left(\frac{p_{\text{ve}}}{p^v}-1\right),
\end{equation}
\begin{equation}\label{SC_eqq_3_n1b}
\frac{p_{\text{ve}}}{p^v}-1=\frac{p-p^v}{\rho R_gT_{\text{sat}}}+\frac{H_{\text{vap}}}{R_gT_{\text{sat}}}\left(\frac{T^i}{T_{\text{sat}}}-1\right),
\end{equation}
where $\widehat\alpha$ denotes an accommodation coefficient for evaporation, $H_{\text{vap}}$ is the internal latent heat of vaporization, $T^i$ is the temperature at the liquid-vapor interface and $R_g$ is the universal gas constant.
\end{subequations}

\subsection{Scalings and Construction of Evolution Equation}
\par To discuss the phase change effect in the presence of both streamwise and azimuthal curvatures, a thin film model will be constructed by assuming the mean film thickness $\mathcal H \ll R$ \cite{kalliadasis1994drop,chattopadhyay2023thin,chattopadhyay2024modeling}. Consequently, a thin film parameter $\epsilon\ll1$ is introduced as $\epsilon=\mathcal H/R$. The following scalings are introduced:
\begin{equation}\label{scale}
(r,z)=R\left(\widetilde r,\widetilde z\right),~h=\mathcal H\widetilde h, ~(u,w)=\mathcal U\left(\epsilon \widetilde u,\widetilde w\right),~t=t_{\text{con}}\widetilde t,~p=p^v+\rho gR\widetilde p,~T=T_{\text{sat}}+\widetilde T\Delta T,~J=\mathcal J_0\widetilde J,
\end{equation}
where $\mathcal U=\rho g\mathcal H^2/\mu$ denotes the characteristic velocity scale, $t_{\text{con}}=R/\mathcal U$ is the convective timescale, $\mathcal J_0=k_{\text{th}}\Delta T/\left(\mathcal HH_{\text{vap}}\right)$ denotes the characteristic scale for interfacial mass flux and $\Delta T=\left(T_s-T_{\text{sat}}\right)>0$. The tilde-decorated variables are dimensionless. 

\par Further, following \cite{khanum2020gravity,mayo2013gravity} a new radial coordinate  $\widetilde r=1+\epsilon\widetilde y$ is introduced. Using this radial transformation together with (\ref{scale}) in the  governing equations (\ref{SC_eqq_1})--(\ref{SC_eqq_2}) and boundary conditions (\ref{SC_eqq_3}), (\ref{SC_eqq_44})--(\ref{SC_eqq_3_n2}),  the following dimensionless leading order system of equations is obtained after dropping the tilde sign:
\begin{subequations}\label{leading0}
\begin{equation}\label{eerd}
u_y+w_z= 0, \quad p_y=0, \quad p_z=1+w_{yy}, \quad T_{yy}=0,
\end{equation}
\begin{equation}\label{eerd1}
u=0, \quad w=0, \quad T=1 \quad\text{at}~y=0,
\end{equation}
\begin{equation}\label{gghc}
w_y=-MaT_z^i, \quad p=\frac{1}{Bo}\left(\frac{1}{\epsilon}-h-h_{zz}\right) -\frac{\Pi}{h^3}\quad\text{at}~y=h(z,t),
\end{equation}
\begin{equation}\label{gghc1}
h_t+EJ+wh_z=u, ~T_y=-J \quad\text{at}~y=h(z,t),
\end{equation}
\end{subequations}
where $\widehat{Bo}=\rho g R^2/\sigma_0$ represents the Bond number, $Bo=\epsilon^{-1}\widehat{Bo}$ represents the scaled Bond number, $\widehat{Ma}=\sigma_T\Delta T/\left(\rho g \mathcal H^2\right)$ represents the Marangoni number, $Ma=\epsilon\widehat{Ma}$ represents the scaled Marangoni number, $\widehat E = \mathcal J_0/\left(\rho\mathcal U\right)$ is the evaporation number, $E = \epsilon^{-1}\widehat E$ is the scaled evaporation number and $\Pi=\widetilde\Pi/\left(6\pi\rho g R\mathcal H^3\right)$ denotes the scaled Hamaker constant. Note that this scaled evaporation number can be expressed as $E=t_{\text{con}}/t_{\text{eva}}$, where $t_{\text{eva}}=\rho\mathcal H/\mathcal J_0$ is the evaporative timescale. While deriving the system of equations (\ref{leading0}), the inertial contributions are neglected by assuming $\epsilon^2\ll1$ and the Reynolds number $Re=\mathcal UR/\nu$ is of $O(1)$.

\par Using (\ref{scale}) in the normal stress balance (\ref{SC_eqq_4}), the scaled normal stress balance was obtained as $p+\frac{\mathcal J_0^2(\chi-1)}{\rho gR\rho^v}J^2=\frac{1}{Bo}\left(\frac{1}{\epsilon}-h-h_{zz}\right)-\frac{\Pi}{h^3}$. The second term on the left-hand side of this equation can be expressed as
\begin{equation}\label{gghc2}
\frac{\mathcal J_0^2(\chi-1)}{\rho gR\rho^v}=\frac{(\chi-1)Nd}{Re},
\end{equation}
 where $Nd=\beta_0^2/\left(\chi Pr^2\right)$, $\beta_0=C_p\Delta T/H_{\text{vap}}$ and $Pr=\nu/\kappa$ is the Prandtl number. In this study, it is assumed that $Nd\sim O\left(\epsilon^2\right)$ and this assumption is consistent with previous thin film studies with phase change effect \cite{uma2004interfacial,usha2003weakly,u1978linearized}. Consequently, the term in (\ref{gghc2}) is neglected in (\ref{gghc}). In (\ref{gghc1}), $J$ is the scaled interfacial evaporative mass flux, which is determined by interfacial thermodynamics and heat transfer. 

\par Using (\ref{scale}), the interfacial evaporation law [Eq. (\ref{SC_eqq_3_n1})] reduces to the following form:
\begin{subequations}\label{SC_eqq_3_n1a_scaled}
\begin{equation}
K_0J=bp+T^i,
\end{equation}
where 
\begin{equation}\label{b_def}
K_0=\frac{k_{\text{th}}\left(2\pi R_g\right)^{1/2}T_{\text{sat}}^{3/2}}{\widehat\alpha\rho^v\mathcal H H_{\text{vap}}^2}, ~b=\frac{gRT_{\text{sat}}}{H_{\text{vap}}\Delta T}.
\end{equation}
\end{subequations}

\par Solving (\ref{leading0}), the temperature solutions are obtained as 
\begin{equation}
T=1-Jy,
\end{equation}
where the scaled evaporative mass flux $J$ is given below:
\begin{equation}\label{mass_flux}
J=\frac{1}{K_0+h}+\frac{b}{K_0+h}\underbrace{\left[\frac{1}{Bo}\left(\frac{1}{\epsilon}-h-h_{zz}\right)-\frac{\Pi}{h^3}\right]}_{p(h)}.
\end{equation}
It can be observed from (\ref{mass_flux}) that when $b=0$, the evaporation flux does not account the contribution from the hydrodynamic pressure $p(h)$.  However, when $b\neq0$, the curvatures and intermolecular forces influence the evaporative mass flux through $p(h)$. 

\par Using (\ref{mass_flux}), the temperature of the liquid-vapor interface $y=h$ is obtained as follows:
\begin{equation}\label{interface_temp}
T^i=\frac{K_0}{K_0+h}\left[1-\frac{bh}{BoK_0}\left(\frac{1}{\epsilon}-h-h_{zz}\right)+\frac{b\Pi}{K_0h^2}\right].
\end{equation}

\par Solving the velocity field, the axial velocity $w$ is obtained as
\begin{equation}\label{eq:ndvel}
w = \left[1+\left(\frac{1}{Bo}-\frac{3\Pi}{h^4}\right)h_z+\frac{h_{zzz}}{Bo}\right]\left(hy-\frac{y^2}{2}\right)-MaT_z^iy.
\end{equation}

\par Inserting (\ref{eq:ndvel}) into the mass conservation form of the kinematic condition [first equation in (\ref{gghc1})], which is $h_t+\left(\int_0^hw\text{d}y\right)_z=-EJ$,  the liquid-vapor interface is obtained as given below: 
 \begin{equation}\label{evolution_equation}
 h_t+\left[\frac{h^3}{3}\left(1+\frac{h_z+h_{zzz}}{Bo}\right)-\frac{\Pi}{h}h_z-\frac{Mah^2}{2}T_z^i\right]_z=-EJ.
  \end{equation}
Throughout the rest of paper, the results will be based on the nonlinear evolution equation (\ref{evolution_equation}), which is obtained from the leading-order approximation with respect to $\epsilon$. In the experiments, a very small Reynolds number, typically of the order of $10^{-2}$, is observed \cite{kliakhandler2001viscous}. In Sec. \ref{parameter}, the Reynolds number is also estimated for weak and strong volatile liquids and is found to be extremely small. Therefore, Eq. (\ref{evolution_equation}) is considered to be valid in the regime $\epsilon^2\ll1$.

\par It can be observed from (\ref{evolution_equation}) that the liquid-vapor interface is influenced by the curvatures along the streamwise as well as the azimuthal direction, van der Waals forces (via the scaled Hamaker constant $\Pi$), and thermocapillary force (via Marangoni number $Ma$). The right-hand side of Eq. (\ref{evolution_equation}) captures the phase change effects at the interface, and the expression of $J$ is given in (\ref{mass_flux}). In the rest of the discussion, the strength of evaporation or condensation will be characterized by the parameter $E$. A positive value of $E$ corresponds to liquid evaporation, while a negative value simply indicates vapor condensation. It is worth noting that, unless stated otherwise, the present study assumes $\epsilon=0.1$ throughout. However, when $b=0$, the parameter $\epsilon$ has no effect. Therefore, $\epsilon$ will not be specified in the figure captions corresponding to cases where $b=0$.

 \par When $b=\Pi=0$ in (\ref{mass_flux}), then Eq. (\ref{evolution_equation}) reduces to the following form:
 \begin{equation}\label{evolution_equation_valid}
 h_t+\frac{E}{K_0+h}+\left[\frac{h^3}{3}\left(1+\frac{h_z+h_{zzz}}{Bo}\right)+\frac{MaK_0h^2}{2\left(K_0+h\right)^2}h_z\right]_z=0.
 \end{equation}
\par It can be noted that when $E=0$ with $Bo=1$ and $K_0=1/Bi$, where $Bi$ denotes the Biot number \cite{kalliadasis2011falling}, Eq. (\ref{evolution_equation_valid}) is consistent with previous studies for thin liquid films on heated cylindrical surfaces \cite{chao2018dynamics,chattopadhyay2023thin,chattopadhyay2024modeling}. Furthermore, for $E=0$ and for a non-heated surface $(Ma=0)$ with a transformation $(t,z)=Bo^{-1/3}(t,z)$, Eq. (\ref{evolution_equation_valid}) takes the following form 
\begin{equation}\label{evolution_equation_valid0}
 h_t+\left[\frac{h^3}{3}\left(1+Bo^{-2/3}h_z+h_{zzz}\right)\right]_z=0.
 \end{equation} 
 Eq. (\ref{evolution_equation_valid0}) is similar to Eq. (7) in Kalliadasis \& Chang \cite{kalliadasis1994drop}, where the parameter $\beta$ in \cite{kalliadasis1994drop} corresponds to $Bo^{-2/3}$ in the azimuthal curvature term used here. Neglecting the contribution from the azimuthal curvature, the resulting equation reduces to that for thin-film flow over a planar substrate \cite{oron1997long,pereira2007dynamics,tiwari2009nonmodal}.

\subsection{Parameter Estimation}\label{parameter}
At this point, it is useful to estimate the ranges of the three key dimensionless parameters $\left(Bo, Ma, E\right)$, so that the results can be analyzed. For both liquids, the mean film thickness is taken as $\mathcal{H}=2\times10^{-5}$ m, while the radius of the vertical cylinder is fixed at $R=2\times10^{-4}$ m. The corresponding aspect ratio is therefore $\epsilon=\mathcal{H}/R=0.1$. The thermophysical properties of hexane and water at $25^\circ$C, used to estimate these dimensionless parameters, are listed in Table \ref{liquid_properties}.

\begin{table}[htbp]
\centering
\caption{Thermophysical properties of hexane and water at $25^\circ$C 
\cite{sultan2005evaporation,lide1995crc}.}
\label{liquid_properties}
\begin{tabular}{lccccc}
\toprule
\textbf{Property} & \textbf{Symbol} & \textbf{Unit} & 
\textbf{Hexane} & \textbf{Water} \\
\midrule

Density 
& $\rho$ 
& $\mathrm{kg\,m^{-3}}$ 
& $656$ 
& $10^{3}$ \\

Dynamic viscosity 
& $\mu$ 
& $\mathrm{kg\,m^{-1}\,s^{-1}}$ 
& $3\times10^{-4}$ 
& $8.9\times10^{-4}$ \\

Surface tension 
& $\sigma_0$ 
& $\mathrm{kg\,s^{-2}}$ 
& $18.42\times10^{-3}$ 
& $7.2\times10^{-2}$ \\

Temperature coefficient of surface tension 
& $\sigma_T$ 
& $\mathrm{kg\,s^{-2}\,K^{-1}}$ 
& $1.02\times10^{-4}$ 
& $1.5\times10^{-4}$ \\

Thermal conductivity 
& $k_{\mathrm{th}}$ 
& $\mathrm{kg\,m\,s^{-3}\,K^{-1}}$ 
& $0.12$ 
& $0.6$ \\

Latent heat of vaporization 
& $H_{\mathrm{vap}}$ 
& $\mathrm{J\,kg^{-1}}$ 
& $3.22\times10^{5}$ 
& $2.4\times10^{6}$ \\

\bottomrule
\end{tabular}
\end{table}

\par Based on the values reported in Table \ref{liquid_properties}, the dimensionless parameters are first estimated for hexane. The Bond number $\widehat{Bo}=\rho g R^2/\sigma_0$ is estimated to be $1.4\times10^{-2}$. For a temperature difference in the range $\Delta T=0.1-1$ K, the Marangoni number $\widehat{Ma}=\sigma_T\Delta T/\left(\rho g \mathcal H^2\right)$ lies in the range 3.96--39.62 and the evaporation number $\widehat E = \mathcal J_0/\left(\rho\mathcal U\right)$ is estimated in the range $3.31\times\left[10^{-4}-10^{-3}\right]$. Similarly, the values of $\widehat{Bo}$, $\widehat{Ma}$ and $\widehat{E}$ are estimated for water. The resulting scaled dimensionless parameters for hexane and water are summarized in Table \ref{scaled_parameters}. 

\begin{table}[htbp]
\centering
\caption{Estimated ranges of the scaled dimensionless parameters for hexane and water for $\Delta T=0.1$-$1$ K.}
\label{scaled_parameters}
\begin{tabular}{c c c c}
\toprule
\textbf{Parameter} & \textbf{Definition} & \textbf{Hexane} & \textbf{Water} \\
\midrule
$Bo$ & $\epsilon^{-1}\widehat{Bo}$ 
     & $1.4\times10^{-1}$ & $5.45\times10^{-2}$ \\
$Ma$ & $\epsilon\widehat{Ma}$ 
     & $0.39-3.96$ & $0.382-3.82$ \\
$E$  & $\epsilon^{-1}\widehat{E}$ 
     & $3.31\times\left[10^{-3}-10^{-2}\right]$ 
     & $2.84\times\left[10^{-3}-10^{-2}\right]$ \\
\bottomrule
\end{tabular}
\end{table}

\par It can be observed that, for $\epsilon=0.1$, the Bond number ($Bo$) remains relatively small, with $Bo=O\left(10^{-1}\right)$ for hexane and $Bo=O\left(10^{-2}\right)$ for water. However, for a smaller value of $\epsilon$, say $\epsilon=0.04$, corresponding to a cylinder of larger radius $R=5\times10^{-4}$ m, values of $Bo\approx2.18$ for hexane and $Bo\approx0.85$ for water are obtained, thereby bringing the Bond number into the $O(1)$ range. Importantly, this reduction in $\epsilon$ does not significantly affect the orders of magnitude of the other key dimensionless parameters, $Ma$ and $E$.

\par Using the representative parameters, the velocity scale $\mathcal U=\rho g\mathcal H^2/\mu$ is estimated to be $8.58\times10^{-3}$ m s$^{-1}$ for hexane and  $4.41\times10^{-3}$ m s$^{-1}$ for water. Consequently, the value of the Reynolds number, $Re=\rho\mathcal U R/\mu$, is found to be approximately $3.75$ for hexane and $1$ for water. Moreover, when $\epsilon=0.04$, the value of $Re\approx 9.38$ is obtained for hexane and $Re\approx2.48$ is obtained for water. Thus, $Re$ remains of $O(1)$ across all the considered datasets. During the derivation of the scaled system of equations (\ref{leading0}), the radial and axial momentum equations are obtained, respectively, in the forms $\epsilon^4Re\left(u_{t}+uu_{y}+wu_{z}\right) =-p_y+O\left(\epsilon^2\right)$ and $\epsilon^2Re\left(w_{t}+uw_{y}+ww_{z}\right)=1-p_z+w_{yy}+O(\epsilon)$, respectively. Since $\epsilon^2\ll1$ and $Re=O(1)$, these inertial terms are negligible under the lubrication approximation.

\par The parameter sets for the two liquids, hexane and water, in the aforementioned dataset indicate that the neglect of the inertial terms is consistent with the lubrication approximation used in the present model. For the present study, $Bo=1$ is set for simplicity, as the primary focus is placed on the isolated effect of $E$ on the cylindrical geometry. The effect of $Bo$ has been discussed in previous studies \cite{liu2014thermocapillary,frenkel1992nonlinear}. However, the effect of a lower $Bo$ will also be demonstrated in Figs. \ref{fig13} and \ref{fig14}. For the theoretical discussion presented in the remainder of this study, $E=O\left(10^{-4}\right)-O\left(10^{-2}\right)$ will be considered.

\section{Linear stability analysis}\label{sec:stability}
 In this section, the linear stability analysis of Eq. (\ref{evolution_equation}) will be performed. The liquid-vapor evolution equation (\ref{evolution_equation}) for $Bo=1$ takes the following form:
  \begin{equation}\label{evolution_equation_stability}
 h_t+\left[\frac{h^3}{3}\left(1+h_z+h_{zzz}\right)-\frac{\Pi}{h}h_z\right]_z=-EJ,
 \end{equation}
  where $J$ is given in (\ref{mass_flux}). 

\par The primary focus of this study is placed on the isolation and examination of the fundamental effects of evaporation and condensation on the dynamics and stability of thin-film flow over a cylindrical substrate. Since the influence of thermocapillary (Marangoni) stresses on film dynamics in cylindrical geometries has already been extensively investigated in previous studies \cite{chattopadhyay2026stability,chao2018dynamics,khanum2020gravity}, the Marangoni number ($Ma$) is not treated as a primary focus in the present analysis. As discussed in Sec. \ref{parameter}, for representative temperature differences of $\Delta T = 0.1-1$ K, Marangoni numbers ranging from $Ma = 0.39$ to $3.96$ for hexane and from $Ma = 0.382$ to $3.82$ for water are obtained. Although thermocapillarity is known to influence instabilities, including the amplification of the RP instability, these values are considered relatively modest under small-temperature-difference conditions. 
Accordingly, in Eq. (\ref{evolution_equation_stability}), the thermocapillary contribution is deliberately neglected as a simplifying assumption. However, in the presence of $Ma$, the linear stability analysis is provided in the \nameref{appendix}.
  
\par As mass is not conserved, the base state can evolve in both space and time. Consequently, the film thickness $h$ is perturbed as follows:
\begin{equation}\label{perturb0}
h(z,t)= H(z,t)+\zeta\eta(z,t),
 \end{equation}
 where $h=H$ is the base state, $\eta\sim O(1)$ is the perturbation at any instant and $\zeta<E\ll1$ is the initial disturbance amplitude. 

\subsection{Spatially Uniform Films}\label{sec:stability_a}
In this section it is assumed that $H(z,t)$ in Eq. (\ref{perturb0}) is spatially uniform \cite{ji2018instability,shklyaev2007stability}, i.e.,
\begin{equation}\label{weak_eva2}
H(z,t) = H_0(t)
 \end{equation}
 and $H_0(0)=H_b$ is the initial film thickness. Unless stated otherwise, $H_b = 0.2$ will be used throughout this section.
\par Using (\ref{perturb0}) with (\ref{weak_eva2}) in (\ref{evolution_equation_stability}) and following \cite{ji2017finite,ji2018instability}, $O(1)$ and $O(\zeta)$ equations are obtained as follows: 
\begin{subequations}\label{sc11}
 \begin{equation}\label{sc0}
O(1)~: \quad H_{0,t}=-EJ_0\left(H_0\right),
 \end{equation}
\begin{equation}\label{sc}
O(\zeta)~: \quad\eta_t+H_0^2\eta_z+\left(\frac{H_0^3}{3}-\frac{\Pi}{H_0}\right)\eta_{zz}+\frac{H_0^3}{3}\eta_{zzzz}=EJ_1\left(H_0\right),
 \end{equation}
where $H_0$ in (\ref{sc}) will be determined by solving the Eq. (\ref{sc0}) and
\begin{equation}\label{sscc}
J_0\left(H_0\right)=\frac{1}{K_0+H_0}\left[1+b\left(\frac{1}{\epsilon}-H_0\right)-\frac{b\Pi}{H_0^3}\right],
\end{equation}
\begin{equation}
J_1\left(H_0\right)=\frac{1}{K_0+H_0}\left[b\left(\eta+\eta_{zz}\right)-\frac{3b\Pi}{H_0^4}\eta\right]+\frac{J_0\left(H_0\right)}{K_0+H_0}\eta.
\end{equation}
\end{subequations}
Eq. (\ref{sc11}) describes the case in which a thin liquid film remains uniform on the cylinder initially. At $O(1)$, the mass transfer across the interface is assumed to occur uniformly over the entire surface. The axial flux would enter at $O(\zeta)$ through interfacial perturbations. In addition, it is noted that when $b=0$, the $O(1)$ equation is not influenced by the van der Waals force.  

\begin{figure}[h!]
\centering
\includegraphics[scale=0.42]{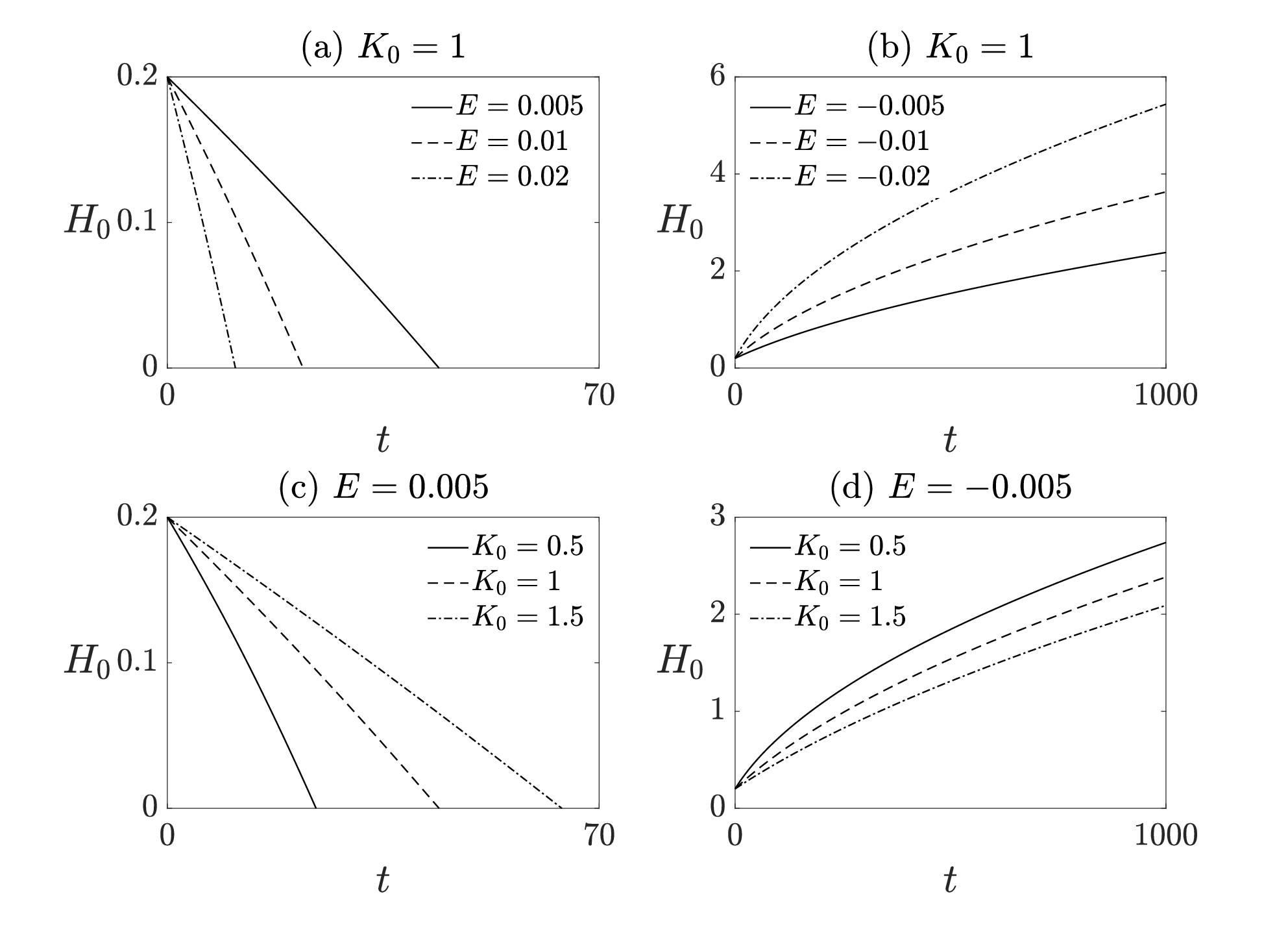}
\caption{Influence of phase change on the temporal evolution of the base film thickness when $b=0$.}
\label{fig2}
\end{figure}

In Fig. \ref{fig2}, the effect of phase change on the base film thickness $H_0$ is shown as time varies. The figure is plotted when the mass flux at the interface is entirely governed by the thermal effects, i.e., for $b=0$ [see Eq. (\ref{SC_eqq_3_n1a_scaled})]. When $b=0$, the leading order film thickness can be obtained as: 
\begin{equation}\label{crit}
H_0(t)=-K_0+\sqrt{\left(K_0+H_b\right)^2-2Et}.
\end{equation}
Eq. (\ref{crit}) shows that increasing $E~ (>0)$ accelerates evaporation and thus mass loss is happening from the liquid-vapor interface. Consequently, the thinning of the liquid occurs earlier.  However, for a given $E~(>0)$, a higher $K_0$ slows down the thinning due to evaporation. Eq. (\ref{crit}) is consistent with the results reported by Burelbach et al. \cite{burelbach1988nonlinear} for a thin evaporating/condensing liquid film on a flat substrate. For $E>0$, Eq. (\ref{crit}) yields a critical time 
\begin{equation}\label{zxcv}
t_c=\frac{\left(2K_0+H_b\right)H_b}{2E}
\end{equation}
at which the liquid will be fully evaporated, consistent with the previous studies on thin evaporating liquid film on flat/inclined surfaces \cite{burelbach1988nonlinear,mohamed2020linear}. Additionally, when condensation occurs ($E<0$), the term under the square root in Eq. (\ref{crit}) increases with time. This shows that the rate of liquid thickening increases as mass accumulates at the interface. In this case, a larger $K_0$ slows down the thickening. This behavior of $E$ and $K_0$ on $H_0$ is displayed in Fig. \ref{fig2}.

\par
\begin{figure}[h!]
\centering
\includegraphics[scale=0.42]{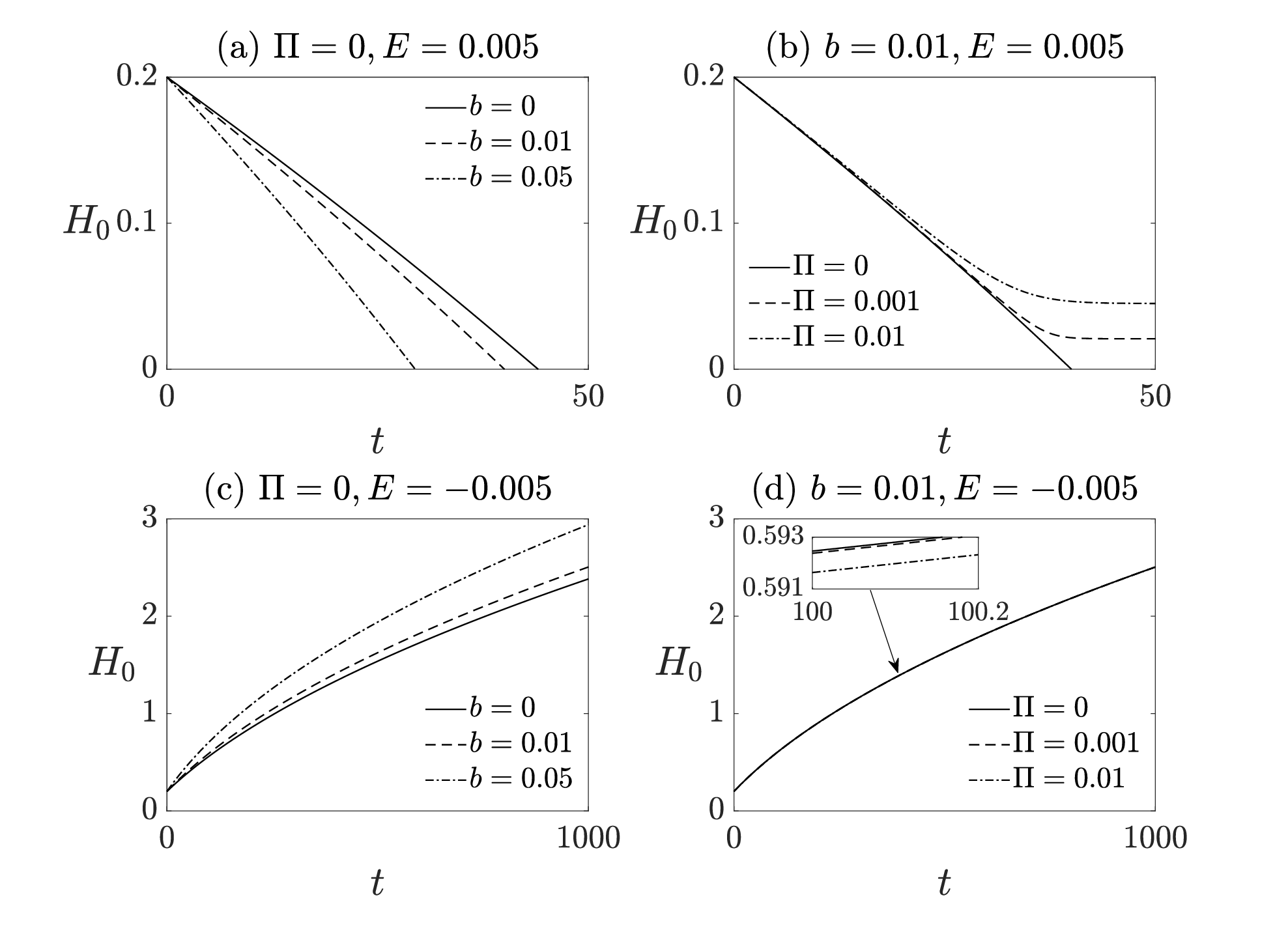}
\caption{Temporal evolution of the base film thickness for different $b$ and $\Pi$. The other fixed parameters are  $\epsilon=0.1$ and $K_0=1$.}
\label{fig3}
\end{figure}

 While Fig. \ref{fig2} was plotted for the case when $b=0$ (consequently $\Pi=0$), the effect due to $b$ and $\Pi$ on the temporal evolution of the base film thickness $H_0$ is shown in  Fig. \ref{fig3}. For hexane, the parameter $b=gRT_{\text{sat}}/\left(H_{\text{vap}}\Delta T\right)$ [see Eq. (\ref{b_def})] is obtained to lie in the range  $2.08\times\left[10^{-5},10^{-6}\right]$ for $\Delta T\in[0.1,1]$ K with $T_{\text{sat}}\approx341.9$ K, while for water, the value of $b$ is estimated as $3.05\times\left[10^{-6},10^{-7}\right]$ over the same range of $\Delta T$ with $T_{\text{sat}}\approx373.15$ K. Here, $25^\circ$C ($298.15 \mathrm{K}$) and $1$ atm are taken as the reference conditions for the tabulated liquid thermophysical properties used in the dimensional estimates in Sec. \ref{parameter}. The saturation temperature, $T_{\mathrm{sat}}$, is treated separately as the thermodynamic equilibrium temperature corresponding to the operating pressure of $1$ atm and is used explicitly in the parameter $b$. Thus, the reference temperature at which the liquid thermophysical properties are specified and the saturation temperature associated with the phase-change condition represent distinct physical quantities and are not required to be identical. These estimates indicate that, for both hexane and water datasets considered, $b\in\left[O\left(10^{-7}\right),O\left(10^{-5}\right)\right]$. However, when such small values of $b$ are used, negligible variations are observed, whereas slightly noticeable effects are obtained when $b=O\left(10^{-3}\right)$. Previous studies \cite{ajaev2005spreading,ajaev2001steady} on volatile liquid films have also considered values of $b$ in the range $O\left(10^{-4}\right)$ to $O\left(10^{-3}\right)$ for different setups. In the present study, the effects are appreciable for relatively large values of $b\geq O\left(10^{-2}\right)$.  Therefore, in order to clearly demonstrate the effect of $b$, values of $O\left(10^{-2}\right)$ and larger will be adopted in Fig. \ref{fig3} and throughout the subsequent analysis.

\par The top panel of Fig. \ref{fig3} is shown for the evaporation of the liquid. It is observed that, when $b\neq0$, i.e., when the interfacial evaporation law is pressure-dependent [see Eq. (\ref{SC_eqq_3_n1a_scaled})], an acceleration in the thinning of the liquid is observed with increasing $b$ [see Fig. \ref{fig3}a]. In contrast, for the condensation case shown in Fig. \ref{fig3}c, an acceleration in the thickening of the liquid is observed as $b$ is increased. While Fig. \ref{fig3}a corresponds to the case $\Pi=0$, the effect of $\Pi$ for a fixed $b$ is illustrated in Fig. \ref{fig3}b. Eq. (\ref{sc0}) shows that the behavior of the leading-order system is strongly influenced by the parameter $\Pi$ when $H_0\to0$. It is found that the dynamics near $H_0\to0$ is dominated by the $\Pi$ term. The steady-state condition $H_{0,t}=0$ yields a non-zero equilibrium thickness $H_0=H^*$, where
\begin{equation}\label{ntrx}
H^*\sim\left(\frac{b\Pi}{1+(b/\epsilon)}\right)^{1/3}.
\end{equation}
Eq. (\ref{ntrx}) shows that when $\Pi=0$, the base film thickness $H_0$ continues to thin monotonically and ultimately reaches zero thickness for an evaporating liquid $(E>0)$, as is observed in Fig. \ref{fig2}a. The behavior of $\Pi$ is illustrated in Fig. \ref{fig3}b for an evaporating liquid $(E=0.005)$. It is observed from Fig. \ref{fig3}b that the presence of $\Pi$ suppresses the thinning due to evaporation. For $\Pi = 0$, the film reaches zero thickness, whereas for $\Pi \neq 0$, the base film retains a finite thickness, $H^*$, which increases with higher values of $\Pi$. For the chosen parameter set, $H^* \approx 0.02$ for $\Pi = 0.001$, and $H^* \approx 0.044$ for $\Pi = 0.01$. Further, for the condensing liquid, it is observed from Fig. \ref{fig3}d that the presence of $\Pi$ reduces the thickness of the base liquid.  From Figs. \ref{fig3}b and \ref{fig3}d, it is observed that the effect of $\Pi$ is significant for an evaporating liquid as $H_0\to0$. In this limit, the term $b\Pi/H_0^3$ becomes significant. In contrast, during condensation, the effect of $\Pi$ on the evolution of $H_0$ with time is found to be negligible. This is because $H_0$ increases with time during condensation, causing the term $b\Pi/H_0^3$ to become progressively smaller. Consequently, the influence of $\Pi$ on the evolution of $H_0$ becomes negligible.

\begin{figure}[h!]
\centering
\includegraphics[scale=0.41]{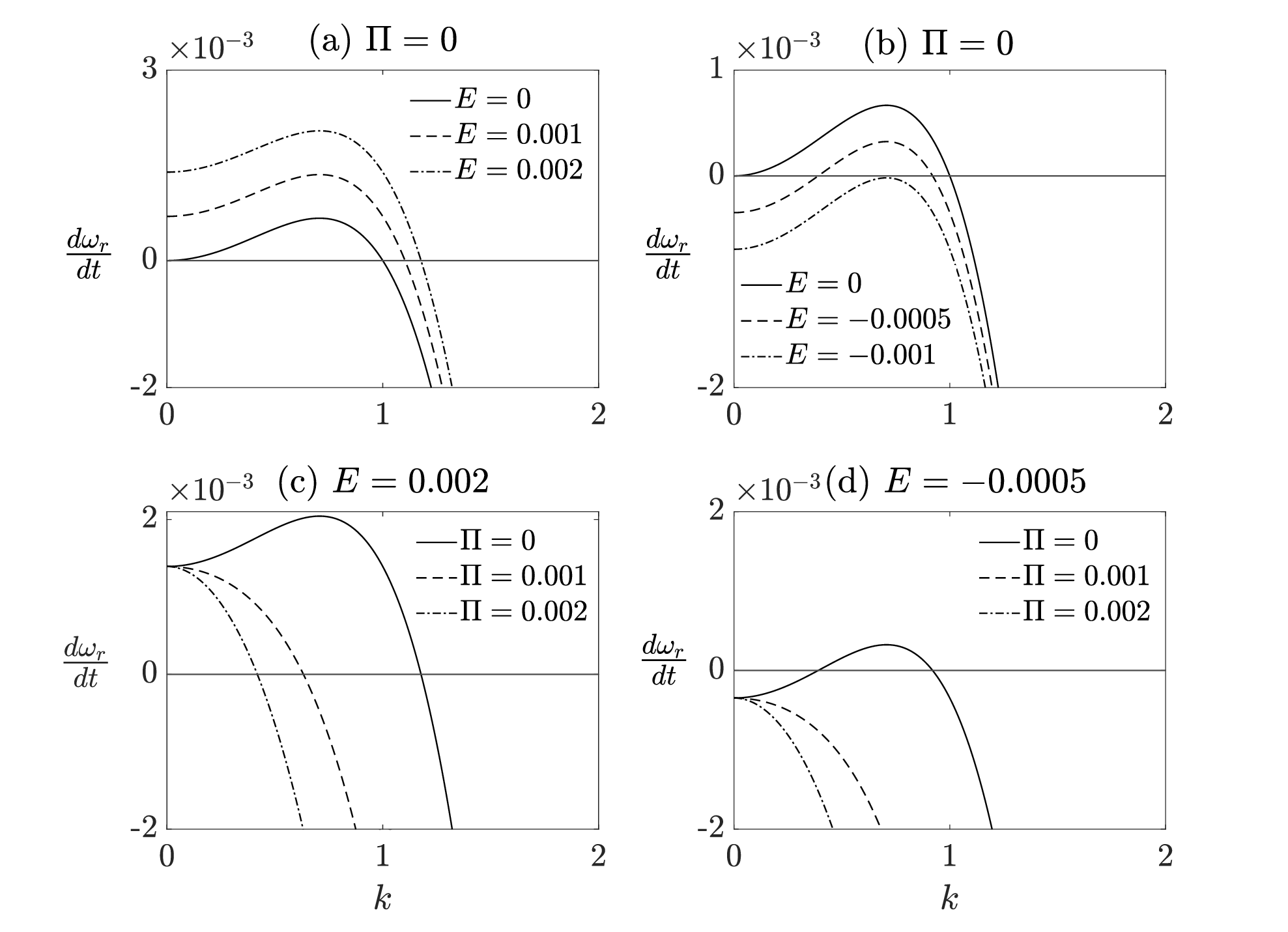}
\caption{Effect of $E$ and $\Pi$ on the linear growth rate as a function of wavenumber $k$. The $H_0$ is evaluated from Eq. (\ref{sc0}) at $t=1$. The other fixed parameters are $b=0$ and $K_0=1$.}
\label{fig4}
\end{figure}
\begin{figure}[h!]
\centering
\centering
\includegraphics[scale=0.41]{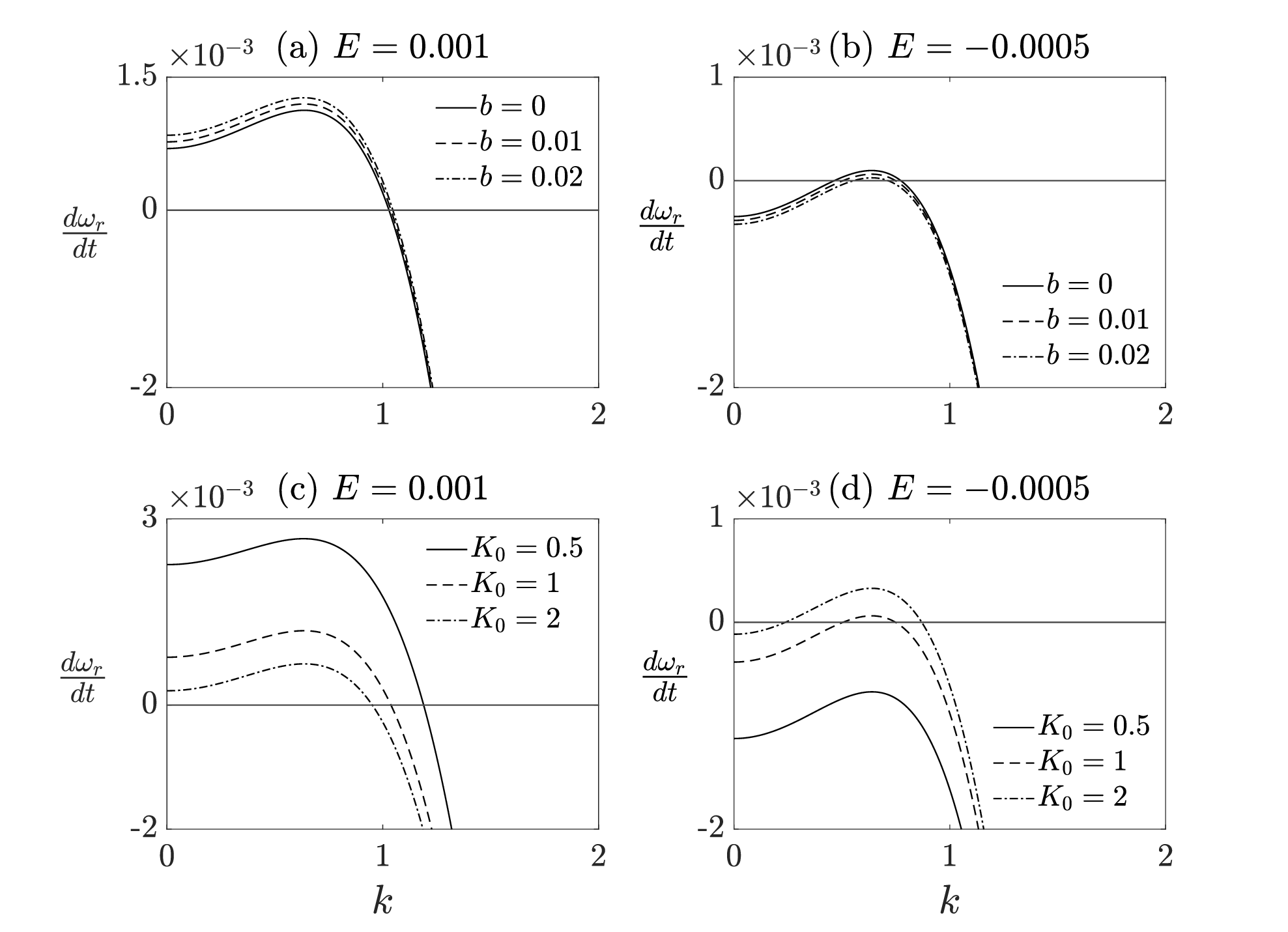}
\caption{Effect of $b$ and $K_0$ on the linear growth rate as a function of wavenumber $k$. The top panel is plotted for $K_0=1$ and the bottom panel is plotted for $b=0.01.$ The $H_0$ is evaluated from Eq. (\ref{sc0}) at $t=1$. The other fixed parameters are $\Pi=0.0001$ and $\epsilon=0.1$.}
\label{fig5}
\end{figure}

\par Now, consider 
\begin{equation}
\eta\propto\exp[ikz+\omega(t)]
\end{equation}
in (\ref{sc}), where  $k\in\mathbb R$ is the wavenumber and $\omega(t)\in\mathbb C$ is the growth rate of perturbation. The real part of $d\omega(t)/dt$ is denoted by $d\omega_r/dt$. Inserting $\eta$ in (\ref{sc}) yields the effective growth rate as follows:
\begin{subequations}\label{growth1}
\begin{equation}
\frac{d\omega_r}{dt}=Ak^2-Bk^4+\frac{bE}{K_0+H_0}\left(1-\frac{3\Pi}{H_0^4}\right)+\frac{EJ_0\left(H_0\right)}{K_0+H_0},
\end{equation}
\begin{equation}\label{bytr}
A=\frac{H_0^3}{3}-\frac{\Pi}{H_0}-\frac{Eb}{K_0+H_0},~B=\frac{H_0^3}{3}.
\end{equation}
 \end{subequations}
The flow instability will be amplified or reduced, depending on the sign of $d\omega_r/dt$. If $d\omega_r/dt>0$, the flow instability increases, whereas for $d\omega_r/dt<0$, it is reduced. The growth rate given in Eq. (\ref{growth1}) represents the instantaneous rate at which perturbations grow or decay for a film of thickness $H_0$. Therefore, the flow's stability characteristics depend on the instantaneous base film thickness. It is important to note that the growth-rate expression in Eq. (\ref{growth1}) is obtained using the frozen-time stability analysis \cite{ji2017finite, ji2018instability} and thus, the base state is assumed to evolve slowly due to weak evaporation $(E\ll1)$, while the instability develops much faster than evaporation alters the film thickness.

\par It is further obtained from the solution of Eq. (\ref{sc0}) that, for $0\leq E\leq 0.002$, $0\leq\Pi\leq0.01$, $0\leq b\leq0.01$ and $0\leq K_0\leq 1$, the variation in the base-state film thickness $H_0$ remains negligible (up to two decimal places) from its initial value $0<H_b\leq 1$ over the interval $0< t\leq2$. Therefore, to examine the effect of $E$ on the linear growth rate, $H_0$ will be evaluated at $t=1$ with parameter ranges chosen as $0\leq E\leq 0.002$, $0\leq\Pi\leq0.01$, $0\leq b\leq0.01$, and $0\leq K_0\leq 1$. For illustration, the corresponding values of $H_0$ at $t=1$ for Fig. \ref{fig4} are given here. In Fig. \ref{fig4}a, the values of $H_0$ for $E=0,0.001$ and 0.002 are 0.200, 0.199, and 0.198, respectively. In Fig. \ref{fig4}b, the values of $H_0$ for $E=0,-0.0005$ and $-0.001$ are 0.200, 0.200, and 0.201, respectively. In Fig. \ref{fig4}c, the values of $H_0$ for $\Pi=0,0.001$ and $0.002$ are 0.198, 0.198 and 0.198, respectively. In Fig. \ref{fig4}d, the values of $H_0$ for $\Pi=0,0.001$ and $0.002$ are 0.200, 0.200 and 0.200, respectively.  

\par The influence of the phase change (via the parameter $E$) and $\Pi$ on the linear growth rate $d\omega_r/dt$ is displayed in Fig. \ref{fig4} when the wavenumber $k$ varies. The results are shown for $b=0$. Although the $O(1)$ equation (\ref{sc0}) is not affected by $\Pi$ when $b=0$, it is indicated by Eq. (\ref{growth1}) that the growth rate is affected by $\Pi$ through the coefficient $A$ even when $b=0$ as follows:
\begin{equation}\label{xcvb}
\frac{d\omega_r}{dt}=\frac{k^2}{3}\left[H_0^3\left(1-k^2\right)-\frac{3\Pi}{H_0}\right]+\frac{E}{\left(K_0+H_0\right)^2}.
\end{equation}
When $E=0$, Eq. (\ref{xcvb}) agrees well with Eq. (4.4) obtained by Ji et al. \cite{ji2019dynamics}. Eq. (\ref{xcvb}) shows that the term related to $E$ is independent of the wavenumber $k$ $\left(\text{i.e.,}~d\omega_r/dt=E/\left(K_0+H_0\right)^2~\text{as}~k\to0\right)$ and thus will act as an additive term whether or not the $\Pi$ is present. For evaporation $(E>0)$,  the growth rate curves will be shifted upward, and for condensation $(E<0)$,  the curves will be shifted downward. This behavior is illustrated in the top panel of Fig. \ref{fig4}. Further, the effect of $\Pi$ on the linear growth rate curves is shown in the bottom panel of Fig. \ref{fig4} for $E\neq0$. It is observed that the presence of $\Pi$ reduces the instability for both evaporating and condensing liquids.

\par In Fig. \ref{fig5}, the linear growth rate will be shown as a function of wavenumber $k$ for different values of $b$ and $K_0$. While it was observed from Fig. \ref{fig4} that the flow instability amplifies due to evaporation, the instability is reinforced when $b\neq0$ [see Fig. \ref{fig5}a]. However, it is found from Fig. \ref{fig5}c that the instability due to evaporation can be suppressed with an increasing $K_0$.  On the other hand, when the condensation occurs, it is found that the reduced instability due to condensation can further be suppressed when $b\neq0$. However, it is noted that, although the flow instability is suppressed in the condensation case, a higher $K_0$ can amplify the instability [see Fig. \ref{fig5}d].  

\par Setting Eq. (\ref{growth1}) to zero, the cutoff wavenumber $k=k_c$ is obtained as given below:
 \begin{equation}\label{cutoff}
k_{c}=\sqrt{\frac{A+\sqrt{A^2+4B\left[\frac{bE}{K_0+H_0}\left(1-\frac{3\Pi}{H_0^4}\right)+\frac{EJ_0\left(H_0\right)}{K_0+H_0}\right]}}{2B}},
\end{equation}
where the flow is linearly unstable for $0<k<k_{c}$. Eq. (\ref{cutoff}) shows that the cutoff wavenumber depends on the parameter $E$. For $E=0$ with $H_0=1$ and in the absence of $\Pi$, $k_c=1$, and this cutoff wavenumber corresponds to the classical RP mode for the capillary instability. 

\par Further, setting $\frac{d}{dk}\left(\frac{d\omega_r}{dt}\right)=0$ yields the most unstable mode $k=k_m=\sqrt{A/(2B)}$. For a nonvolatile $(E=0)$ thin liquid film flowing along the vertical cylinder with $H_0=1$ and $\Pi=0$, the most unstable mode $k_m=1/\sqrt{2}$, consistent with the previous thin film studies on similar geometries \cite{chao2018dynamics,ding2019breakup}. It can be noted that for a volatile thin liquid film, $k_m$ is influenced by the parameter $E$ via the coefficient $A$ in (\ref{bytr}). Consequently,  the maximal effective growth rate $\frac{d\omega_{r,m}}{dt}= \max_{k \in \mathbb{R}^+} \left[\frac{d\omega_r}{dt}(k) \right]$ is obtained as follows: 
\begin{equation}\label{max-wave}
\frac{d\omega_{r,m}}{dt}=\frac{A^2}{4B}+\frac{bE}{K_0+H_0}\left(1-\frac{3\Pi}{H_0^4}\right)+\frac{EJ_0\left(H_0\right)}{K_0+H_0}.
\end{equation}

\begin{figure}[h!]
\centering
\subfloat[]{
\includegraphics[scale=0.22]{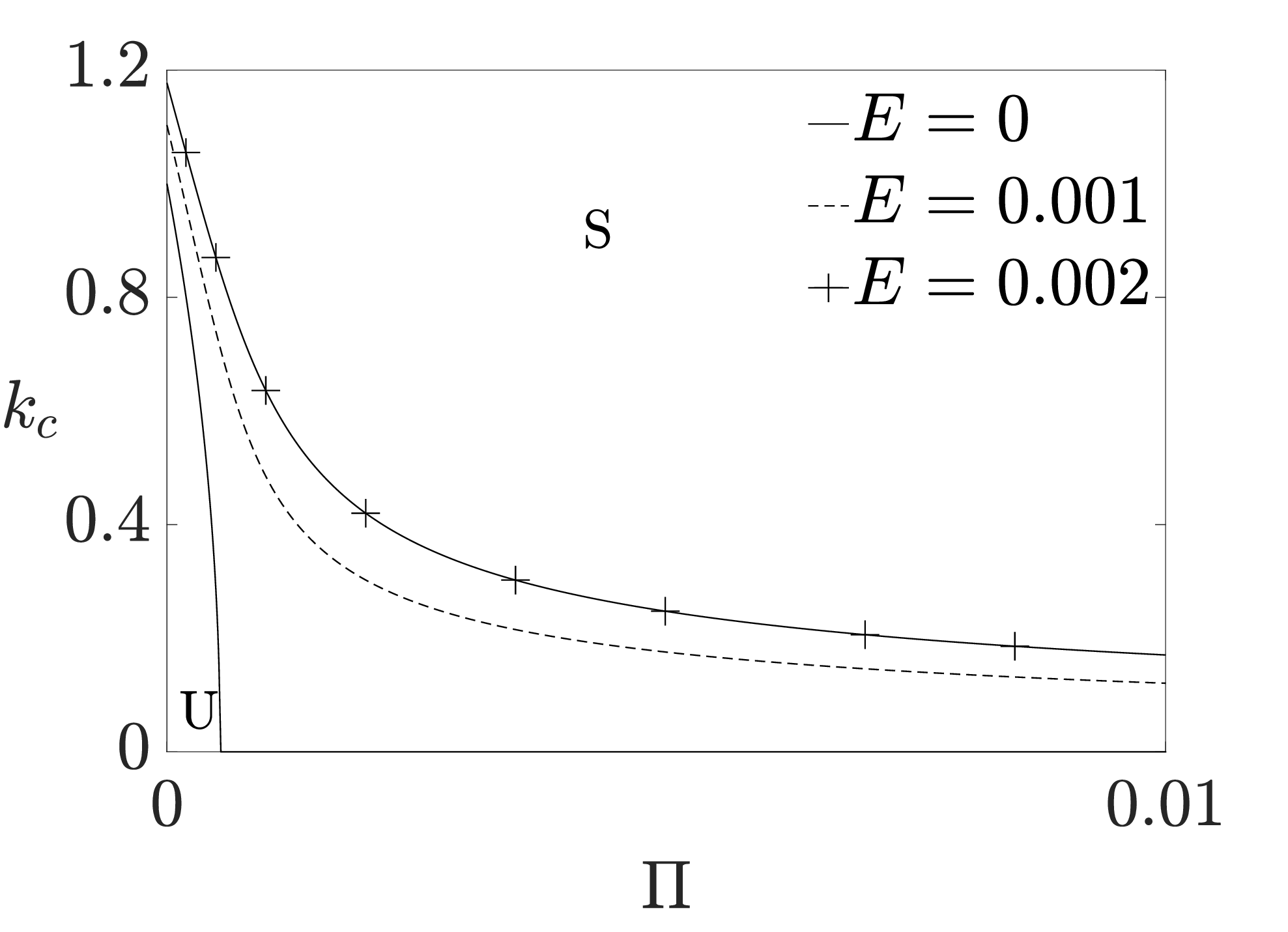}}
\subfloat[]{
\includegraphics[scale=0.22]{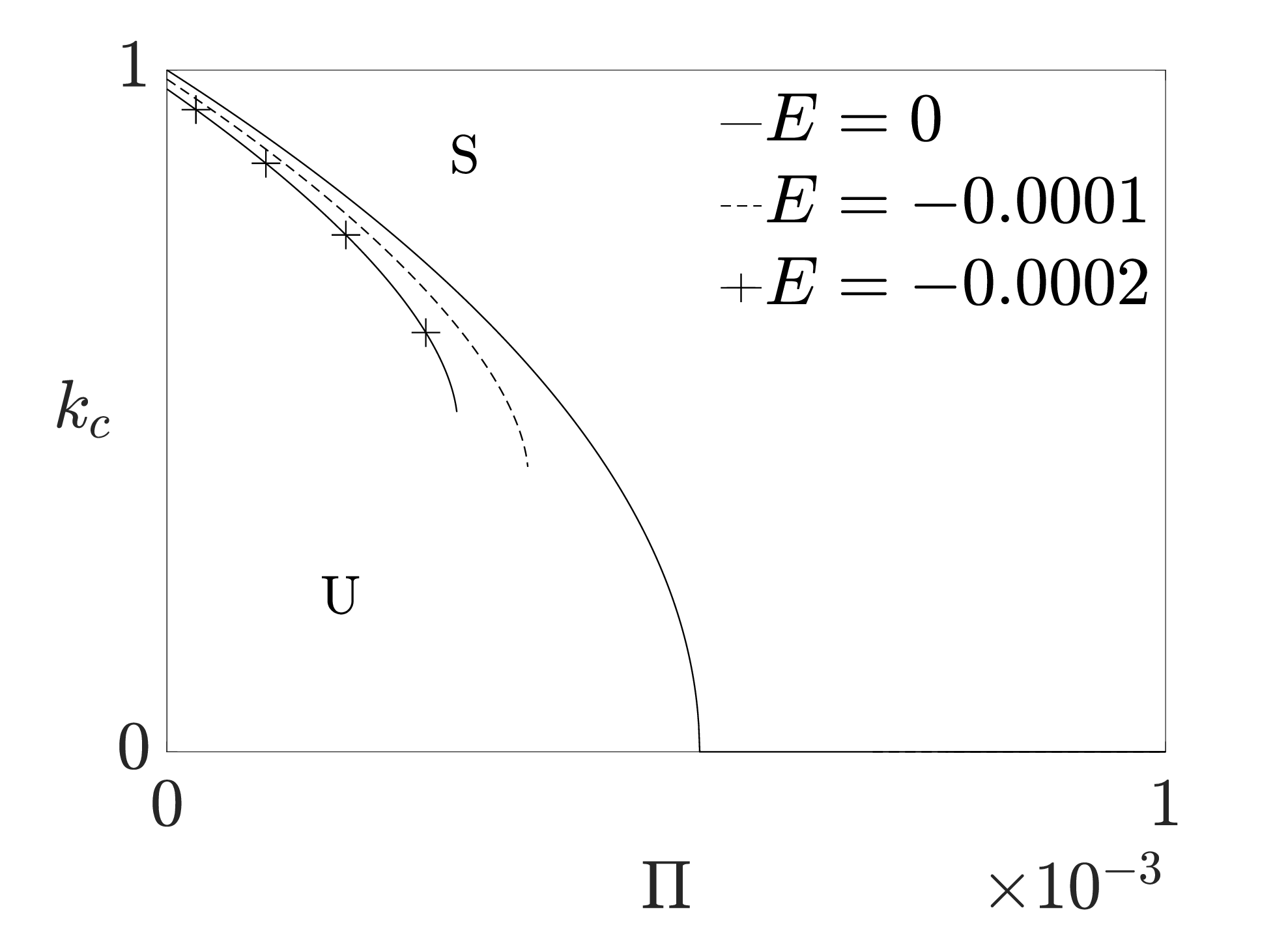}}
\caption{Neutral stability curve as a function of $\Pi$. The $H_0$ is evaluated from Eq. (\ref{sc0}) at $t=1$. The other fixed parameters are $b=0$ and $K_0=1$. Here, `U' and `S' represent unstable and stable regimes, respectively.}
\label{fig6}
\end{figure}

\par In Fig. \ref{fig6}, the variation of cutoff wavenumber $k_c$ is displayed as a $\Pi$ when $b=0$. Fig. \ref{fig6}a is plotted when the liquid is evaporating. Three typical $E>0$ values are chosen: $E=0, 0.001$ and 0.002. It is observed that when $E=0$, that is, in the absence of evaporation, the $k_c$ starts from 1 when $\Pi=0$, corresponding to RP instability. Moreover, when $E=0$, the Eq. (\ref{cutoff}) gives $k_c=\left[1-\left(3\Pi/H_0^4\right)\right]^{1/2}$ and this yield a critical value of $\Pi=\Pi_c=H_0^4/3$ such that $0<k<k_c$ holds when $\Pi<\Pi_c$ and $k_c=0$ when $\Pi=\Pi_c$. Consequently, the $E = 0$ curve decreases with increasing $\Pi$ for $\Pi < \Pi_c \approx 0.00053$ (for $E=0$, $H_0=H_b=0.2$) and drops to zero at $\Pi = \Pi_c$. This indicates that, in the absence of evaporation, the film is unstable only within a very narrow range of $\Pi$ values and is stable beyond this range. When evaporation occurs $(E > 0)$, the $k_c$ initially follows the same trend as for $E = 0$, decaying with increasing $\Pi$. However, while the $E = 0$ curve exhibits a concave decay with $\Pi$, the $k_c$ curve for $E > 0$ shows a convex decay. Further, unlike the non-evaporating case, there always exists a regime where the flow remains unstable for all $\Pi$. Furthermore, higher values of $E$ enhance this unstable regime. While for a given $E>0$, the $k_c$ decays rapidly for small $\Pi$ values, the rate of decay slows as $\Pi$ increases. In Fig. \ref{fig6}b, the variation of the $k_c$ is shown when condensation occurs $(E<0)$. It is observed that the $E<0$ curves follow the same trend as that of $E=0$, and a higher $E<0$ enhances the stable regime. 
\begin{figure}[h!]
\centering
\subfloat[$H_b=0.5$]{
\includegraphics[scale=0.22]{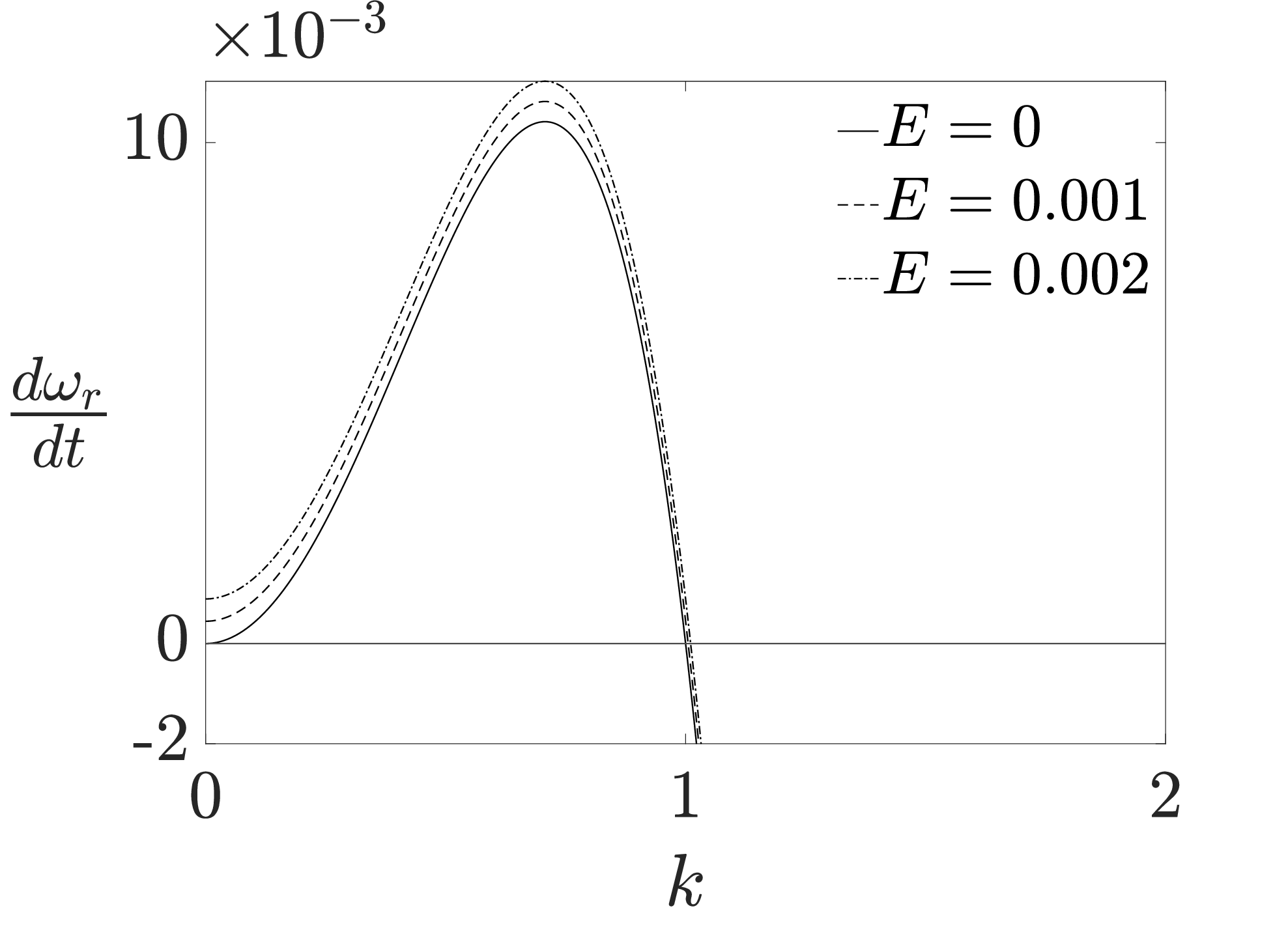}}
\subfloat[$H_b=1$]{
\includegraphics[scale=0.22]{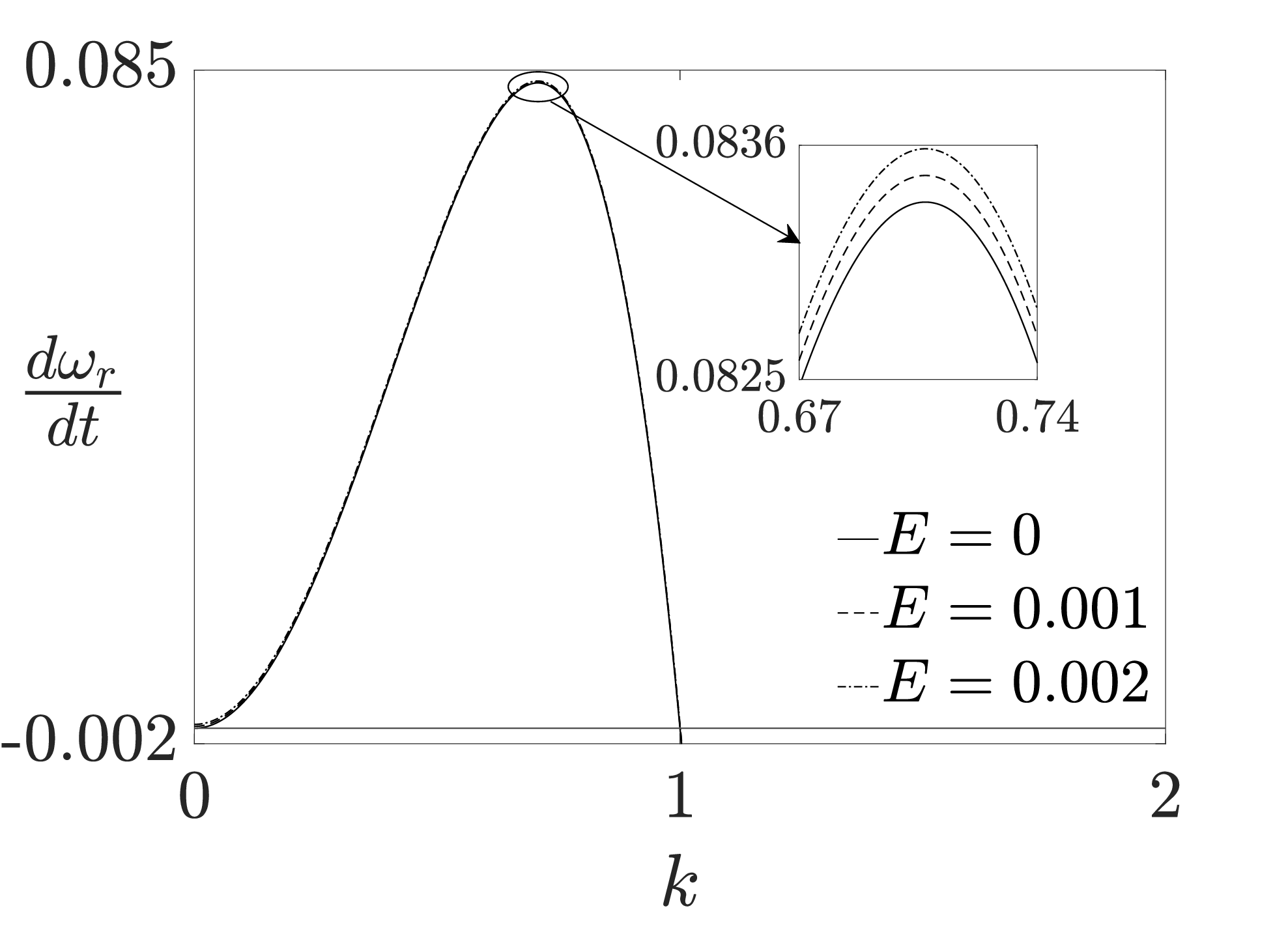}}
\caption{Linear growth rate as a function of wavenumber $k$ for different $E$ when $H_0$ is evaluated from Eq. (\ref{sc0}) at $t=1$. The other fixed parameters are $b=0$, $\Pi=0$ and $K_0=1$.}
\label{fig5nn}
\end{figure}
 
\par The effects of various parameters $(E, \Pi, b, K_0)$ on the linear growth rate are illustrated in Fig. \ref{fig4} and Fig. \ref{fig5}, with $H_0$ being evaluated in each case by solving Eq. (\ref{sc0}). These figures show the effect of $E$ on the linear growth rate for a uniform layer $H_0$ with $H_0(0)=H_b=0.2$. In Fig. \ref{fig5nn}, the growth rate is plotted for different $E>0$, where now $H_0$ is calculated with an initial film thickness which is higher than $H_b=0.2$. Comparing Fig. \ref{fig4}a with Fig. \ref{fig5nn}, it is observed that the growth rate is amplified with a thicker $H_0$. It is also observed that when $H_0$ is thicker, the film becomes unstable over a narrower range of wavenumbers $k$, with or without $E$.

\subsection{Spatially Non-uniform Films}\label{spnu}
In Sec. \ref{sec:stability_a}, the stability analysis was carried out for spatially uniform films, where the base state was assumed to depend only on time [see Eq. (\ref{weak_eva2})]. Furthermore, the effects of the parameter $E$ and hydrostatic pressure (through the parameter $b$) were examined using a frozen-time analysis. Under this approximation, evaporation (condensation) was found to have a destabilizing (stabilizing) effect on the flow, while the presence of $b$ further amplified (reduced) the instability.

In this section, the stability analysis is extended by allowing the base state to vary in both space and time. This extension enables the spatial variation of the base-state thickness and its associated effects on the stability of the film to be accounted for. A coupled set of evolution equations governing the base state $H$ and the perturbation $\eta$ is derived and solved numerically. This formulation provides a more complete description of the effects of $E$ and $b$ on the evolution and stability of the film. Throughout the remainder of this study, attention is restricted to the evaporating case $(E>0)$, both in the absence and presence of hydrostatic pressure effects $(b=0$ and $b\neq0)$.

\par Using (\ref{perturb0}), the PDE (\ref{evolution_equation_stability})  is expanded about $h=H$. The $O(1)$ and $O(\zeta)$ equations are obtained as given below:
 \begin{subequations}\label{hjsc}
 \begin{equation}\label{leading}
O(1)~: \quad H_t+\left[\frac{H^3}{3}\left(1+H_z+H_{zzz}\right)-\frac{\Pi}{H}H_z\right]_z=-EJ_0,
 \end{equation}
\begin{equation}\label{zeta}
O(\zeta)~: \quad\eta_t+\left[H^2\left(1+H_z+H_{zzz}\right)\eta+\frac{H^3}{3}\left(\eta_z+\eta_{zzz}\right)-\Pi\left(\frac{\eta_z}{H}-\frac{H_z}{H^2}\eta\right)\right]_z=EJ_1,
 \end{equation}
  where 
 \begin{equation}
J_0=\frac{1}{K_0+H}\left[1+b\left(\frac{1}{\epsilon}-H-H_{zz}\right)-\frac{b\Pi}{H^3}\right],
\end{equation}
\begin{equation}
J_1=\frac{b}{K_0+H}\left(\eta+\eta_{zz}-\frac{3\Pi}{H^4}\eta\right)+\frac{J_0}{K_0+H}\eta.
\end{equation}
\end{subequations}
When $H$ is spatially independent, Eq. (\ref{hjsc}) recovers Eq. (\ref{sc11}).  

\par To quantify the growth of the perturbation, the amplification ratio is defined as 
\begin{equation}\label{nn}
N(t)=\ln\left[\frac{\left||\eta(z,t)|\right|_2}{\left||\eta(z,t=0)|\right|_2}\right],
\end{equation}
where $\left||\cdot|\right|$ represents the $L^2$ norm. When $N(t)>0$, the amplification of disturbances over time is represented, whereas $N(t)<0$ corresponds to decay. The case $N(t)=0$ signifies that the perturbation remains unchanged. 

\par Moreover, considering $\omega_r(t)=\lambda t$, $H_0=1$ and in the absence of $E$, the growth rate Eq. (\ref{growth1}) reduces to the form 
\begin{equation}\label{linear_growth11}
\lambda=\frac{k^2}{3}\left(1-3\Pi-k^2\right).
\end{equation}
In the linear regime, the temporal growth rate $\lambda_1$ is related to the amplification factor as follows: 
\begin{equation}
\lambda_1\sim\frac{dN}{dt}.
\end{equation}

\par To obtain the profiles of the base film thickness $(H)$ and disturbance $(\eta)$, Eq. (\ref{hjsc}) is solved using COMSOL Multiphysics 6.3, with the initial conditions prescribed as $H(t=0)=1$ and $\eta(t=0)=\sin(kz)$. A quadratic finite-element discretization is employed for the spatial domain. For all computations, the domain length is taken to be $50\times(2\pi/k)$. Spatial convergence is verified by comparing simulations performed using 4000 and 10,000 spatial grid points. The corresponding $N$ curves are found to be indistinguishable. Therefore, 4000 spatial grid points are used for all simulations presented in this study. Furthermore, an adaptive time-stepping scheme is employed. The effect of the absolute error tolerance is also examined for a fixed number of spatial grid points. It is found that an absolute error tolerance of $10^{-9}$ is sufficient to ensure the required accuracy. Accordingly, an absolute error tolerance of $10^{-9}$ is used for all simulations. Moreover, the results are first verified by solving the coupled system of equations (\ref{hjsc}) for $E=0$, from which the corresponding growth rate is obtained. This growth rate is then compared with the growth rate given in Eq. (\ref{linear_growth11}).

\begin{figure}[h!]
\centering
\subfloat[$E=0.001$]{
\includegraphics[scale=0.36]{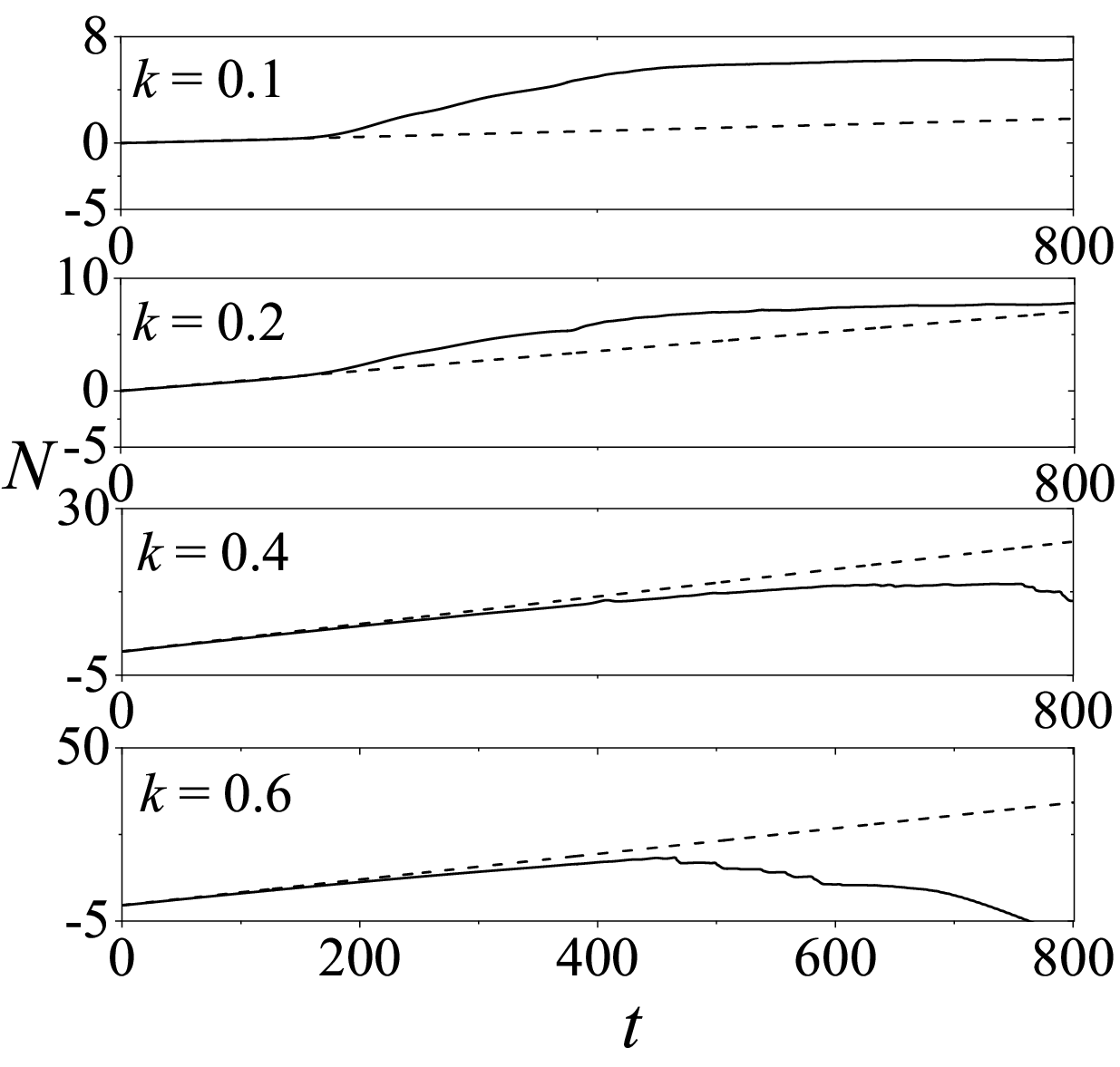}}
\quad
\subfloat[$E=0.01$]{\includegraphics[scale=0.36]{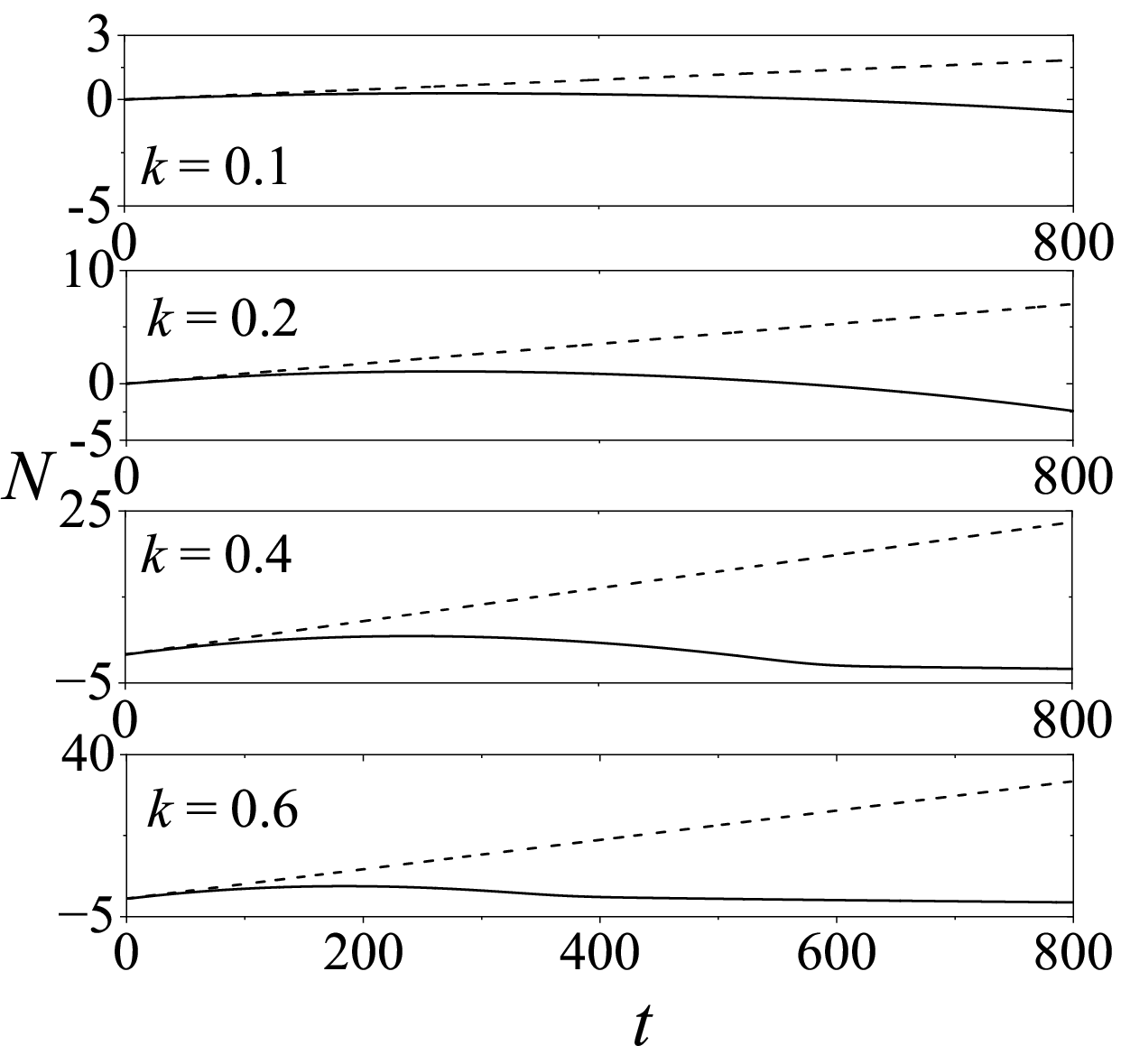}}
\caption{Temporal evolution of $N$ for evaporating liquid (solid lines) compared with the classical normal mode growth (dashed lines). The other fixed parameters are  $b=0$, $\Pi=0.1$, $K_0=10$. }
\label{fig7_n}
\end{figure}

\begin{figure}[h!]
\centering
\subfloat[$E=0.001$]{
\includegraphics[scale=0.34]{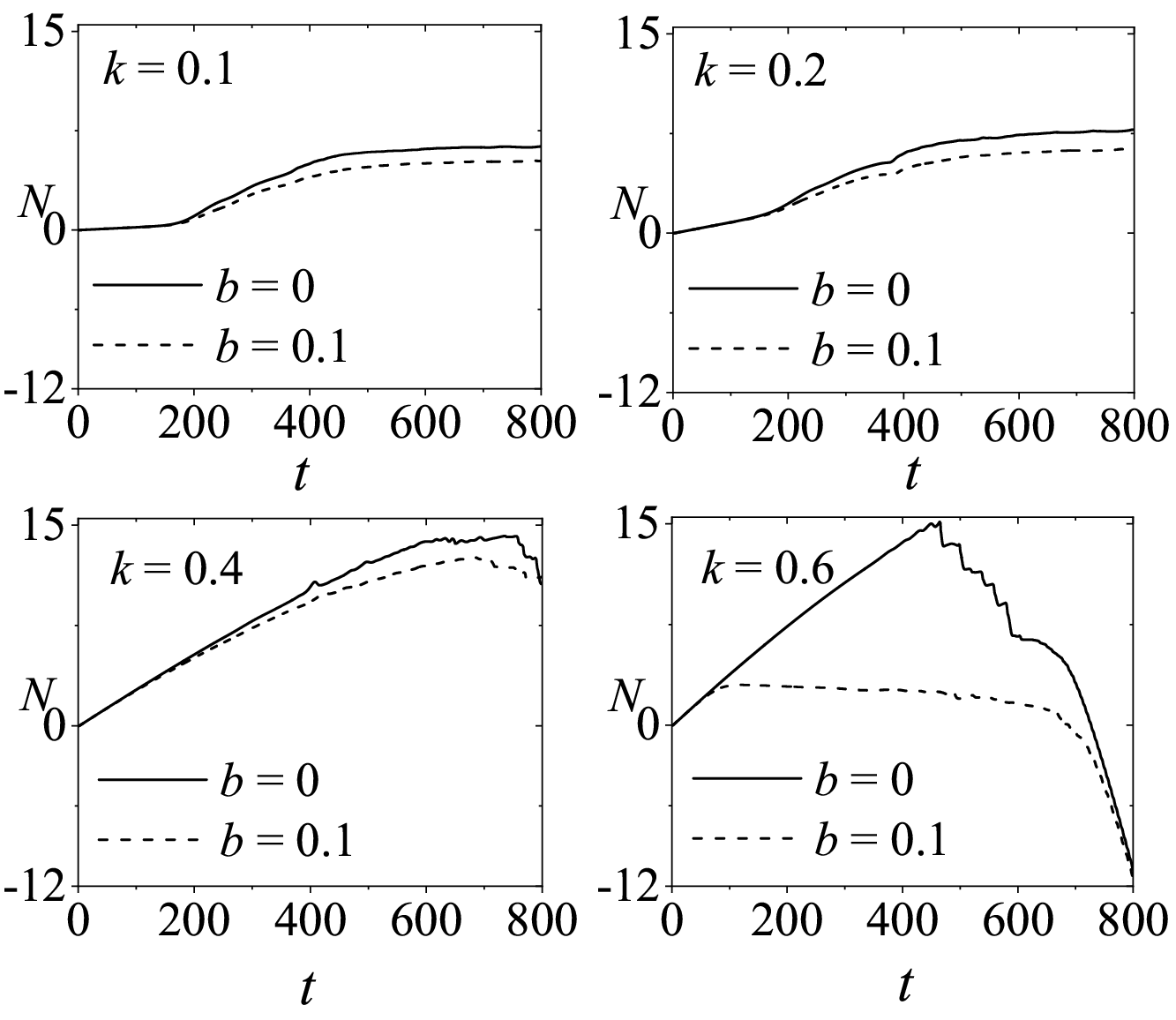}}
\quad
\subfloat[$E=0.01$]{\includegraphics[scale=0.34]{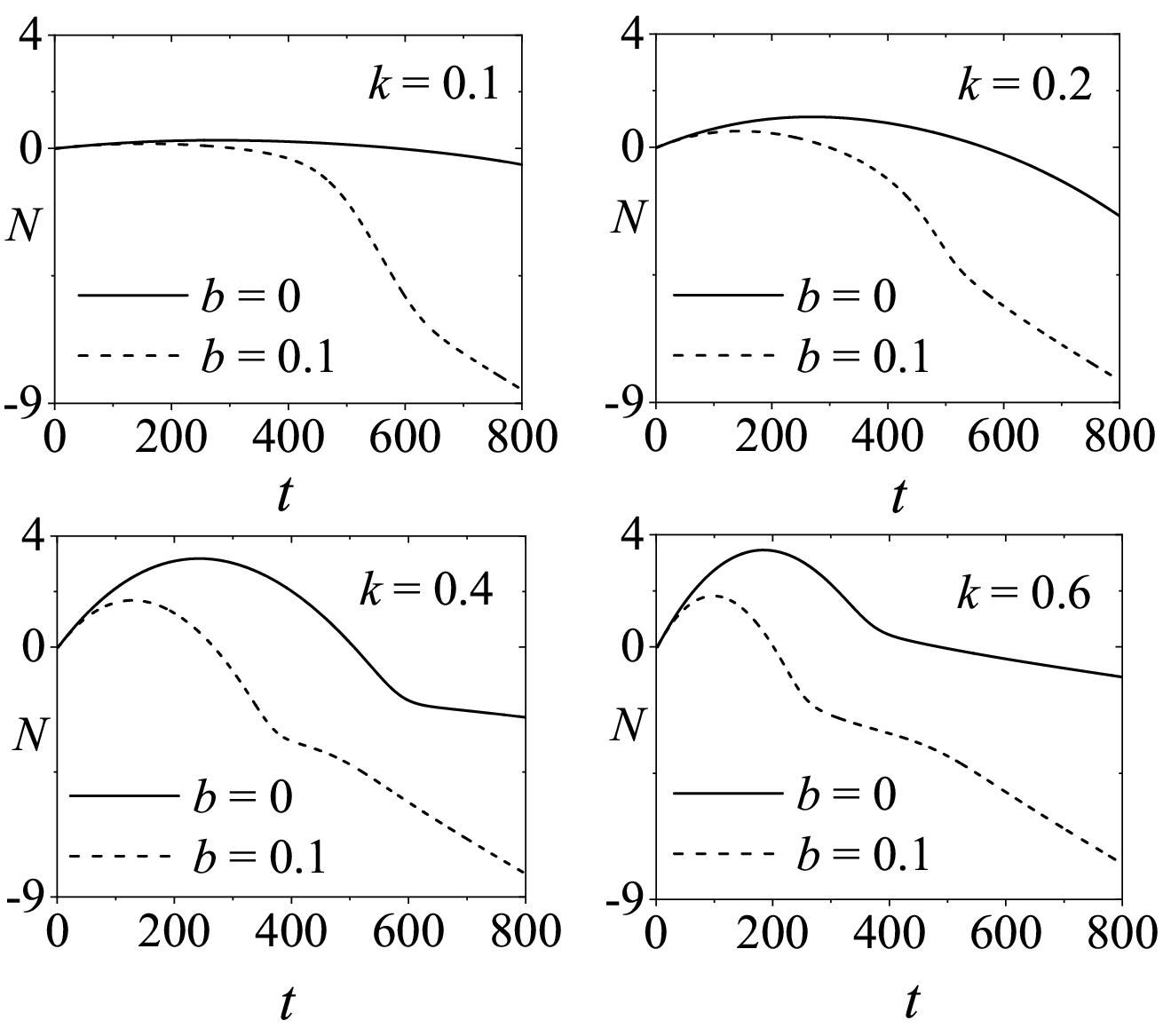}}
\caption{Effect of $b$ on the temporal evolution of $N$ for evaporating liquid. The other fixed parameters are  $\epsilon=0.1$, $\Pi=0.1$, $K_0=10$. }
\label{fig7_nnn_scc}
\end{figure}

\begin{figure}[h!]
\centering
\includegraphics[scale=0.42]{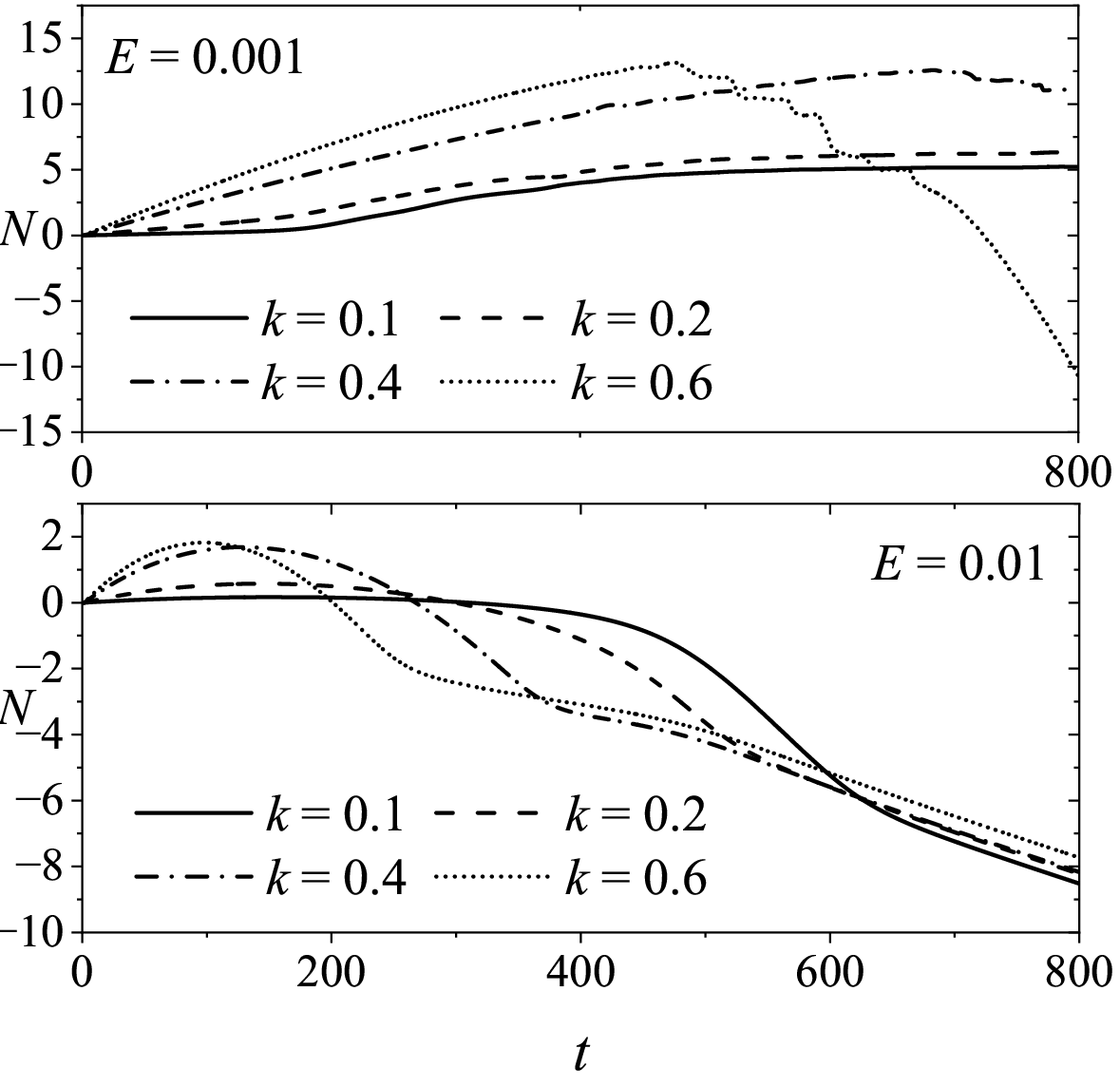}
\caption{Temporal evolution of $N$ for an evaporating liquid at different wavenumbers $k$. The other fixed parameters are $b=0.1$, $\epsilon=0.1$, $\Pi=0.1$, and $K_0=10$.}
\label{fig7_n_sc}
\end{figure}

\par The effect of the parameter $E>0$ is presented in Fig. \ref{fig7_n} to demonstrate the temporal evolution of the growth rate of perturbation $N$. Two typical $E$ values are chosen, $E=0.001$ and $E=0.01$, corresponding to moderate and strong evaporation. In both cases, the growth predicted by the linearized system (\ref{hjsc}) initially agrees well with the normal-mode growth for a non-evaporating film [Eq. (\ref{linear_growth11})], but deviates after a certain time. It is further noted from Fig. \ref{fig7_n}a that for moderate evaporation $(E=0.001)$ and small $k$, the profile of $N$ remains higher than the normal-mode prediction. However, for larger $k$, the profile of $N$ consistently lies below the corresponding normal-mode growth rate. A similar trend is observed for a weak evaporation case $(E=0.0001)$ (figure is not shown here). These observations conclude that evaporation plays a negligible role at early times, but as it becomes dominant, the behavior deviates from classical behavior. However, it is found from Fig. \ref{fig7_n}b that when evaporation is moderate, the $N$ given by Eq. (\ref{nn}) consistently lies below the corresponding normal-mode growth rate for any $k$.  

\par In Fig. \ref{fig7_nnn_scc}, the effect of $b$ on the profile $N$ is illustrated. It is observed that, for $b\neq0$, the growth-rate curve remains below that corresponding to $b=0$ under both moderate and strong evaporation conditions across all the wavenumbers considered. Although this difference becomes more pronounced at later times, the profiles are nearly overlapping at early times. In Fig. \ref{fig7_n_sc}, the temporal evolution of the growth rate $N$ is presented for $b\neq0$ under two different evaporation conditions (moderate when $E=0.001$ and strong when $E=0.01$). It is observed that the growth rate is higher at higher wavenumbers. When the evaporation is moderate $(E=0.001)$, an increase in the growth rate is observed for smaller wavenumbers. The growth rate curves are observed to reach a maximum for $k=0.4$ around $t=700$, whereas for higher wavenumbers, the onset of maximum is observed at earlier times. However, the initial increasing trends are found to be similar for all wavenumbers. Moreover, when the evaporation is strong $(E=0.01)$, an increase in the growth rate with increasing $k$ is observed at the initial time, similar to the case of $E=0.001$. However, in this case, the peak of the amplification ratio $N$ is found at an earlier time for any $k$, compared with the $E=0.001$ case.

\par
\begin{figure}[h!]
\centering
\subfloat[$k=0.1, b=0$]{\includegraphics[scale=0.28]{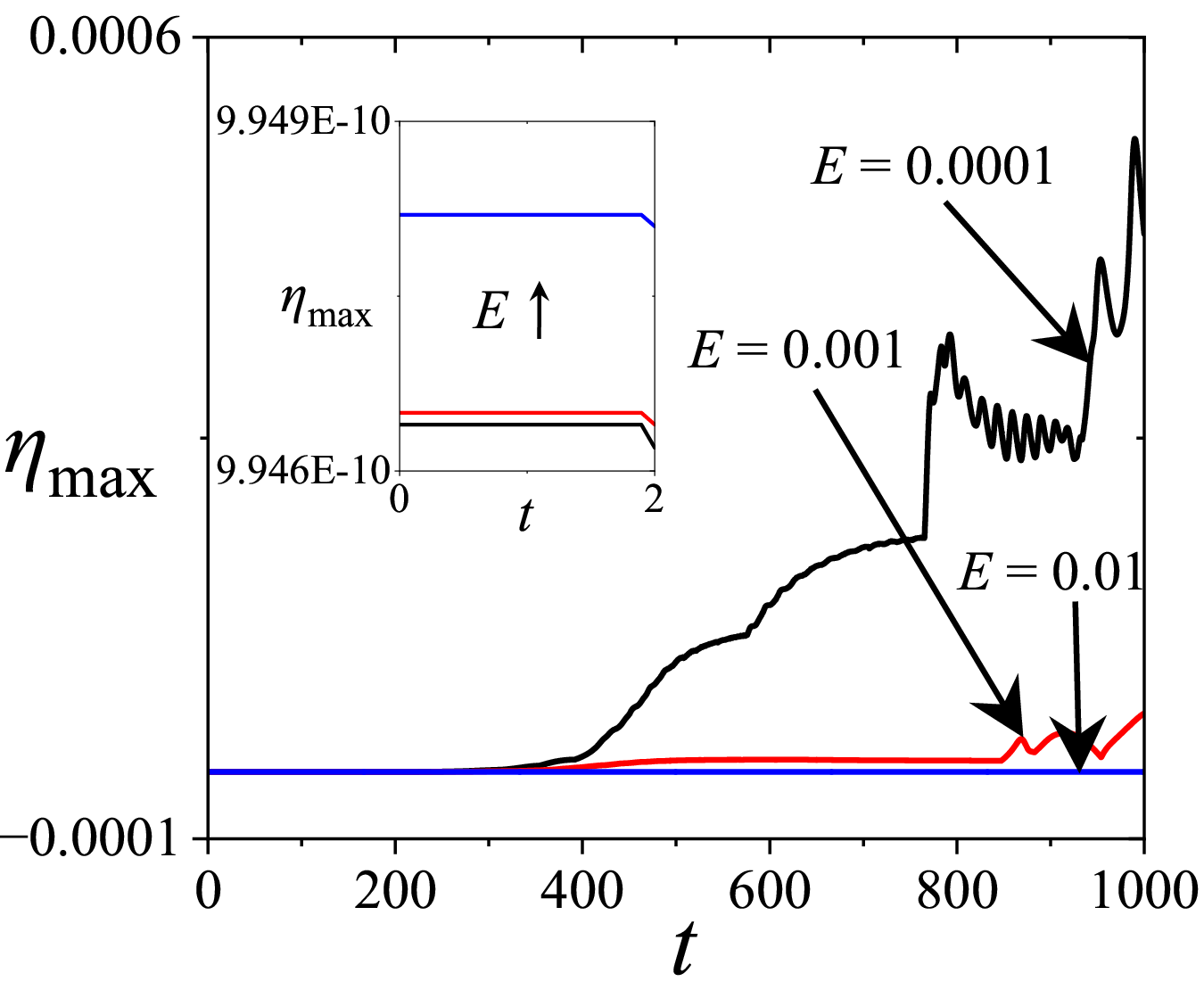}}
\subfloat[$k=0.6, b=0$]{\includegraphics[scale=0.28]{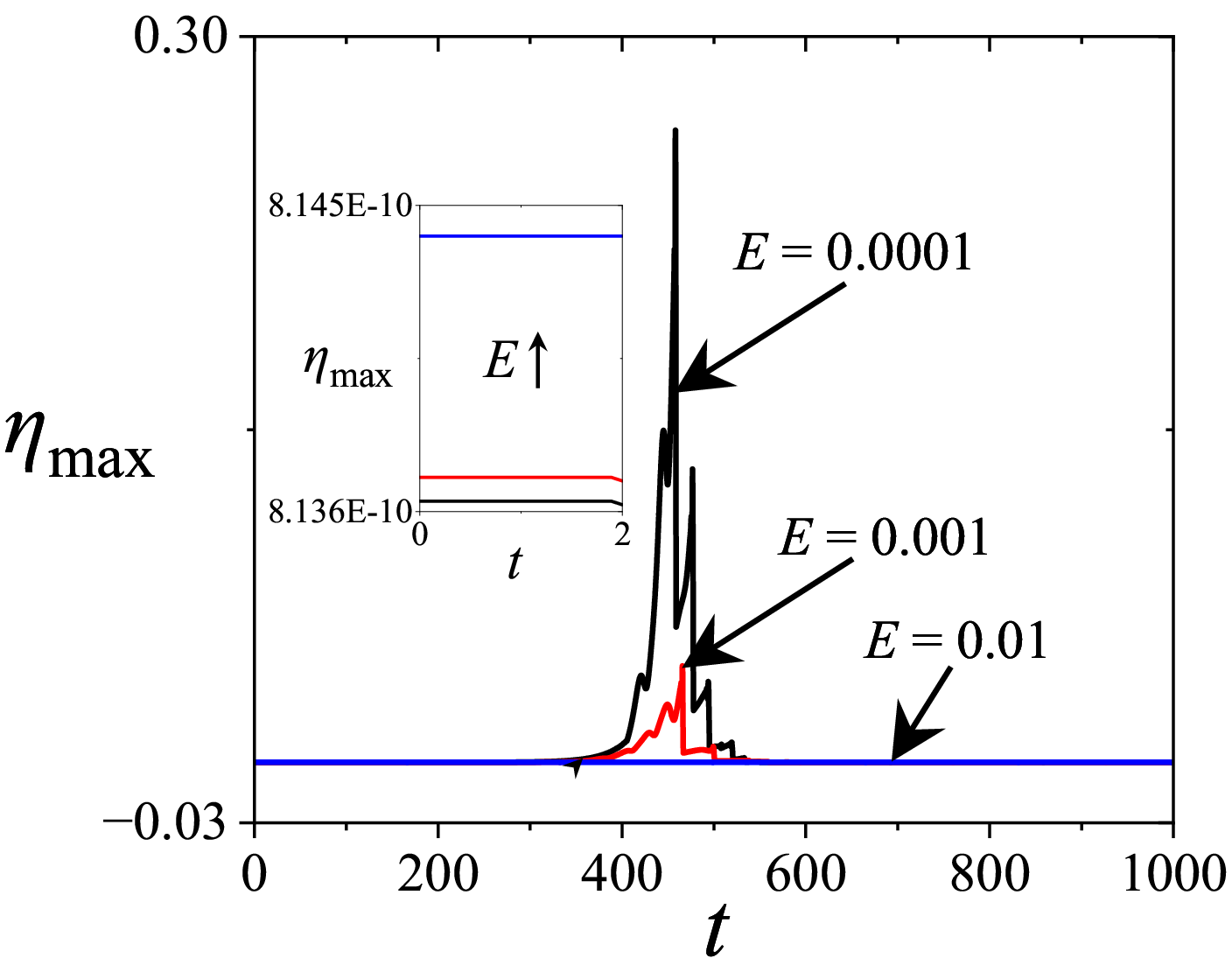}}
\qquad
\subfloat[$k=0.1, E=0.001$]{\includegraphics[scale=0.28]{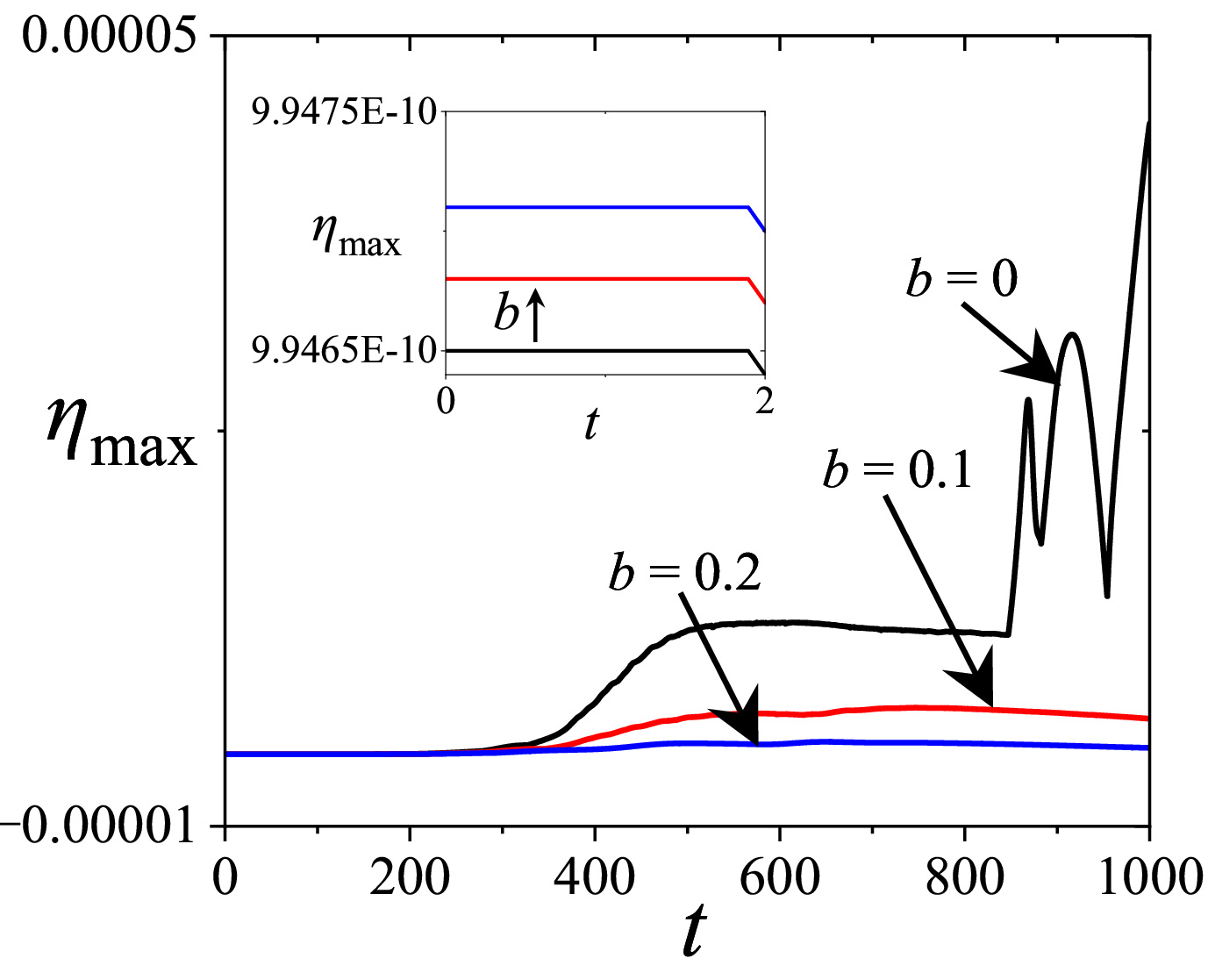}}
\subfloat[$k=0.6, E=0.001$]{\includegraphics[scale=0.28]{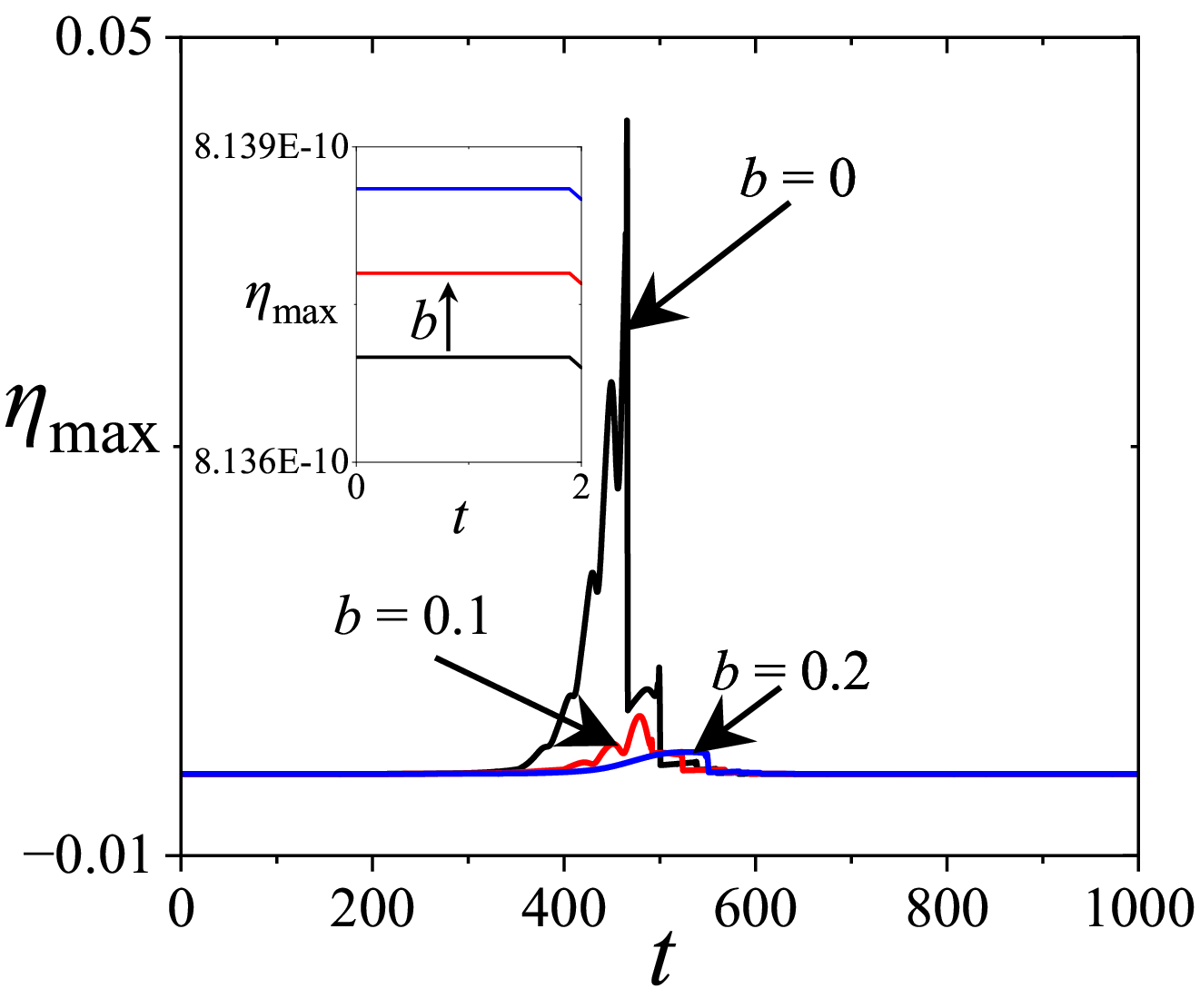}}
\caption{Evolution of $\eta_{\max}/10^{9}$. Top panel: For different values of $E$ (weak to strong). Middle panel: For different values of $b$.  The other fixed parameters are $\Pi=0.1$, $\epsilon=0.1$ and $K_0=10$.}
\label{eta}
\end{figure}

\begin{figure}[h!]
\centering
\includegraphics[scale=0.5]{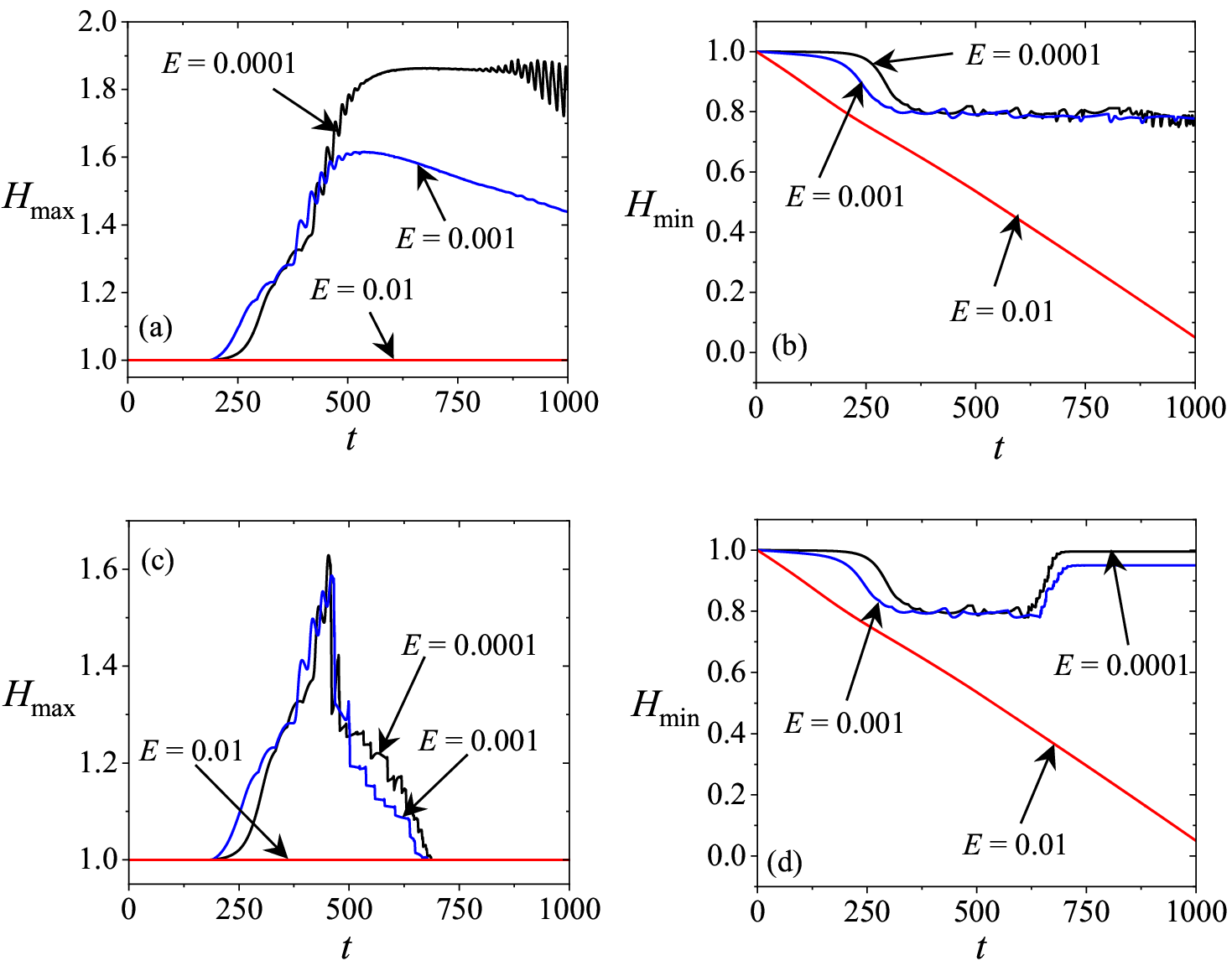}
\caption{Evolution of $H_{\max}$ and $H_{\min}$ for evaporating liquid. (a,b) $k=0.1$ and (c,d) $k=0.6$. The other fixed parameters are $b=0$, $\Pi=0.1$ and $K_0=10$.}
\label{Hmax}
\end{figure}

 In Fig. \ref{eta}, the evolution of the maximum amplitude of the perturbation $\eta_{\max}(t)=\max_z\eta(z,t)$ is presented from the linearized system of equations  (\ref{hjsc}) for weak to strong evaporation. Fig. \ref{eta}a is plotted for $k=0.1$, whereas Fig. \ref{eta}b is plotted for $k=0.6$. From both figures, it can be seen that as evaporation increases, $\eta_{\max}$ decreases. For both wavenumbers, the perturbation amplitude exhibits a weak transient enhancement with increasing $E$ during the initial stage $(0\leq t\leq 2)$, consistent with the destabilizing effect of evaporation identified in the frozen-time stability analysis in Sec. \ref{sec:stability_a} [see the inset figures]. However, this trend is reversed at later times, with larger values of 
$E$ reduces $\eta_{\max}$. The suppression of the perturbation amplitude can be attributed to the evolution of the base state $H(z,t)$. With increasing evaporation $(E>0)$, liquid is removed more rapidly, and the base state is consequently driven toward smaller film thicknesses. This thinning is shown in Fig. \ref{Hmax} as the temporal evolution of $H_{\max}(t)$ and $H_{\min}(t)$, which are found to decrease with increasing $E$. Although the base state considered in this section is both spatially and temporally evolving, the frozen-time stability analysis in Sec. \ref{sec:stability_a} provides an instantaneous measure of the instability tendency of the evolving film. In the frozen-time analysis, larger growth rates are obtained for thicker base states [see Fig. \ref{fig5nn}]. Hence, as the local film thickness is progressively reduced by evaporation, the evolving system is driven through a sequence of thinner base states for which lower instantaneous growth rates are obtained from the frozen-time analysis. This provides a physical interpretation of the observed suppression of $\eta_{\max}$ at later times with increasing evaporation rate. The initial enhancement of $\eta_{\max}$ with increasing $E$ therefore reflects the transient destabilizing effect of evaporation, whereas the subsequent suppression is consistent with the influence of the evaporation-induced thinning of the base state.

\par While the top panel of Fig. \ref{eta} is presented for $b=0$, the effect of $b\neq0$ is shown in Figs. \ref{eta}c and \ref{eta}d for $E=0.001$. It is found that $\eta_{\max}$ decreases with higher $b\neq0$. However, the inset figures show that at the very initial time $(0\leq t\leq 2)$, $\eta_{\max}$ increases with higher $b\neq0$. This observation is consistent with the one in Sec. \ref{sec:stability_a}, where it was observed that a higher $b\neq0$ enhances the flow instability [see Fig. \ref{fig5}a].

\begin{figure}[h!]
\centering
\subfloat[$E=0.0001$]{
\includegraphics[scale=0.35]{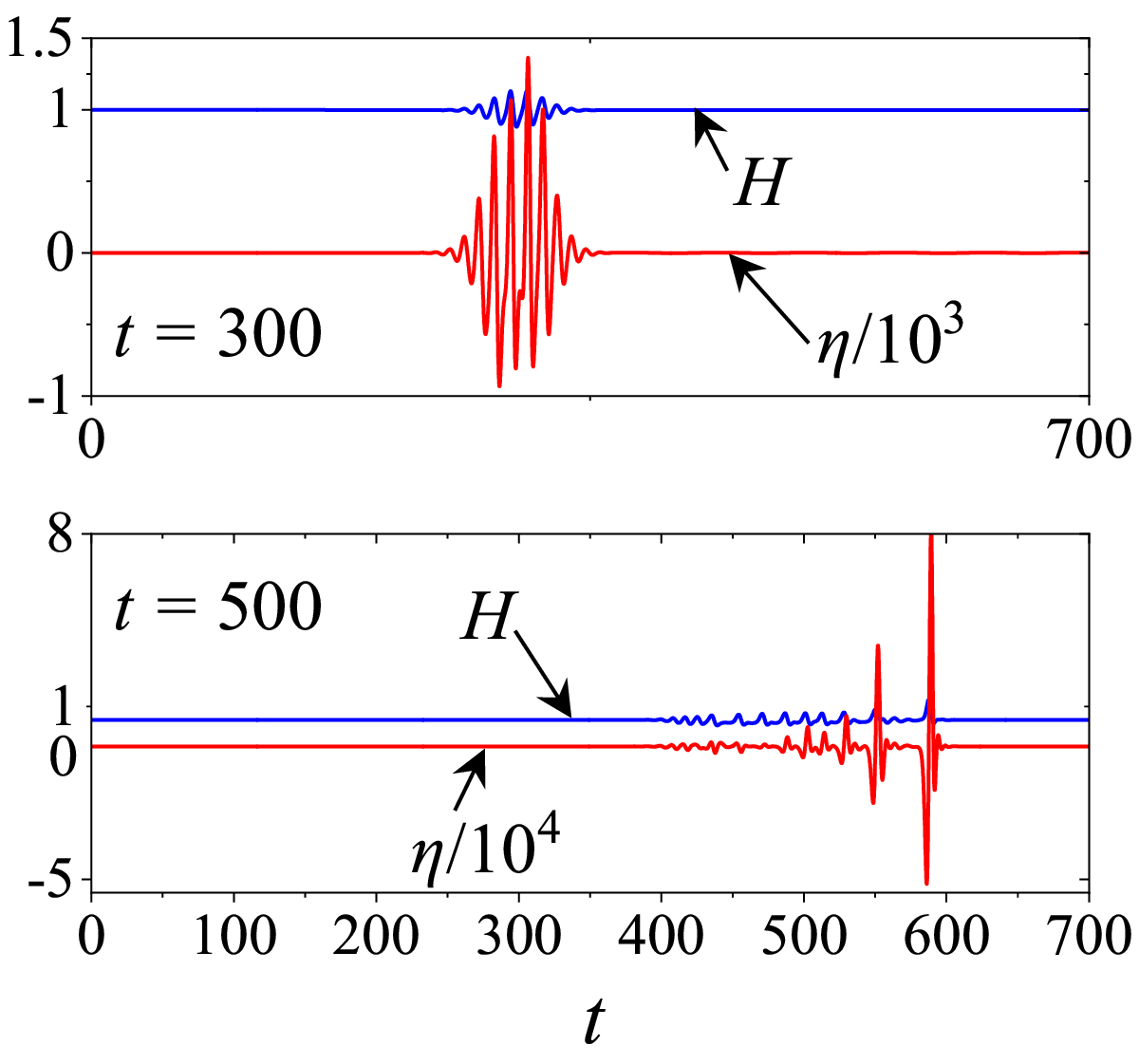}}
\qquad
\subfloat[$E=0.001$]{\includegraphics[scale=0.35]{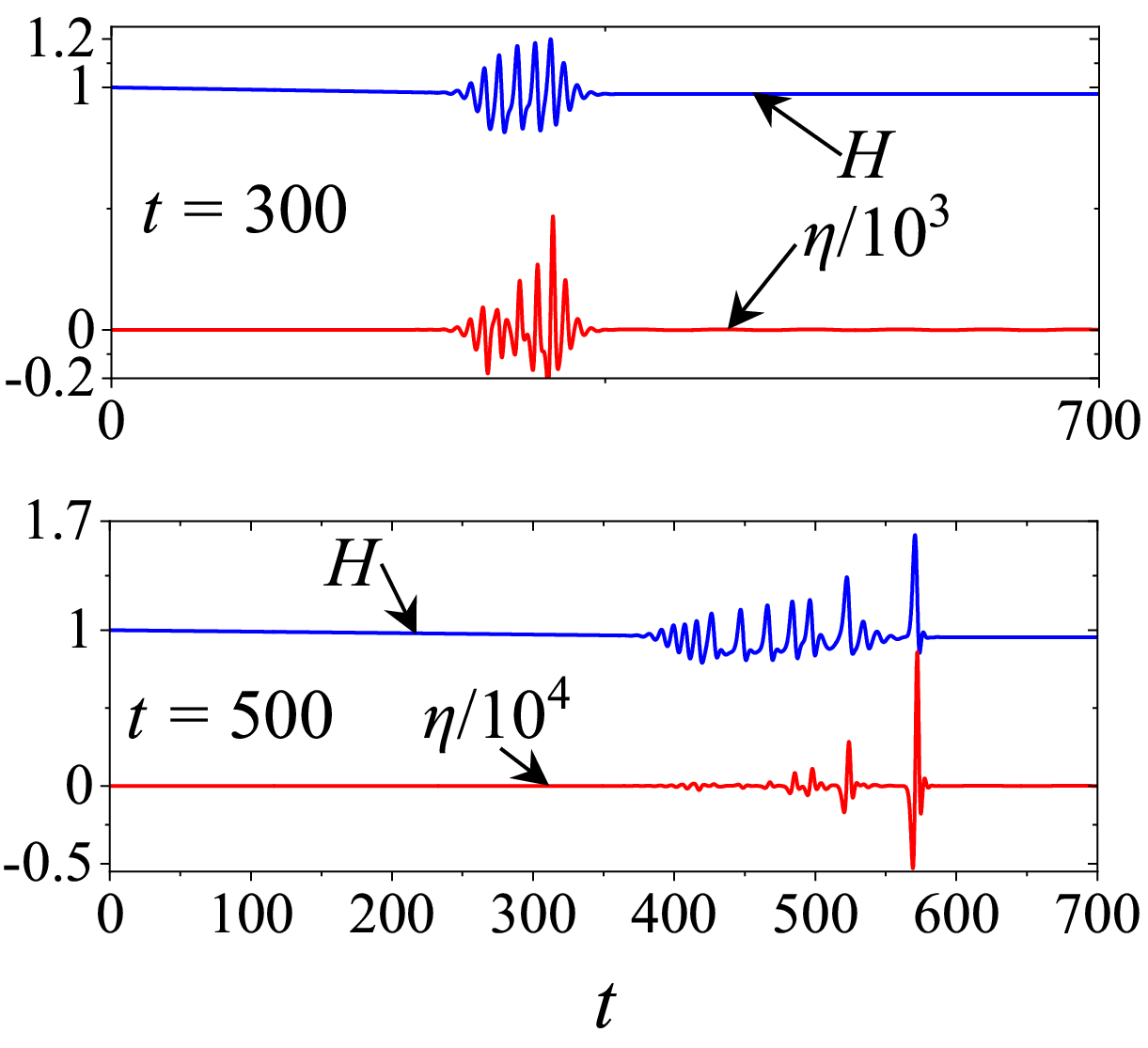}}
\caption{Typical profiles of $H$ and $\eta$ at two time instants when $k=0.1$. The other fixed parameters are  $b=0$, $\Pi=0.1$, and $K_0=10$.}
\label{fig7_nn}
\end{figure}

\par In Fig. \ref{fig7_nn}, the spatial profiles of the base film thickness $H$ and the perturbation $\eta$ are displayed at two different time instants $t=300$ and $t=500$ for the parameters $\left(b=0,\Pi=0.1, K_0=10\right)$ as Fig. \ref{fig7_n}. Fig. \ref{fig7_nn}a is presented for a very weak evaporation $(E=0.0001)$, whereas Fig. \ref{fig7_nn}b is plotted for comparatively stronger evaporation $(E=0.001)$.  It can be observed from the figure that during propagation, both $H$ and $\eta$ get amplified for any $E$, and the disturbance in $\eta$ is developed at locations where variations in the base film thickness $H$ occur.

\section{Numerical studies}\label{sec:4}
In Sec. \ref{sec:stability}, the linear stability analysis was performed on the reduced evolution equation (\ref{evolution_equation_stability}), where thermocapillary stresses were neglected. Under these assumptions, (\ref{evolution_equation_stability}) reduces to a purely hydrodynamic–phase-change model, in which interfacial dynamics are governed solely by capillarity, gravity, disjoining pressure, and evaporative mass flux. This simplification allows the fundamental instability mechanisms arising from the interplay of surface tension, gravity, and phase change to be isolated, without the additional coupling introduced by temperature gradients. However, when thermal effects are included, thermocapillary stresses arise from the interfacial temperature gradient, which, in the present study, depends on the parameters $K_0$, $\Pi$, and $b$. To capture the complete dynamics induced by phase-change effects, the full model (\ref{evolution_equation}) is considered in this section.  For simulation, the following  boundary conditions are imposed at the left boundary
\begin{subequations}\label{boundary_condition}
\begin{equation}
h\left(z=-L/2,t\right)=1, ~h_z\left(z=-L/2,t\right)=0,
\end{equation}
and the following soft conditions at the right boundary \cite{tiwari2009nonmodal,khanum2020gravity}
\begin{equation}
h_z\left(z=L/2,t\right)=0, ~h_{zzz}\left(z=L/2,t\right)=0.
\end{equation}
For all computations in this section, the domain length $L$ is considered as $50\times(2\pi/k)$.
\end{subequations}
\par The following initial condition will be used for the rest of this discussion in this section:
\begin{equation}
h(z,0)=h_0+\zeta\sin(kz),
\end{equation}
where $h_0=1$ is the base state profile.

\par
\begin{figure}[h!]
\centering
\subfloat[$E=0.001$]{
\includegraphics[scale=0.38]{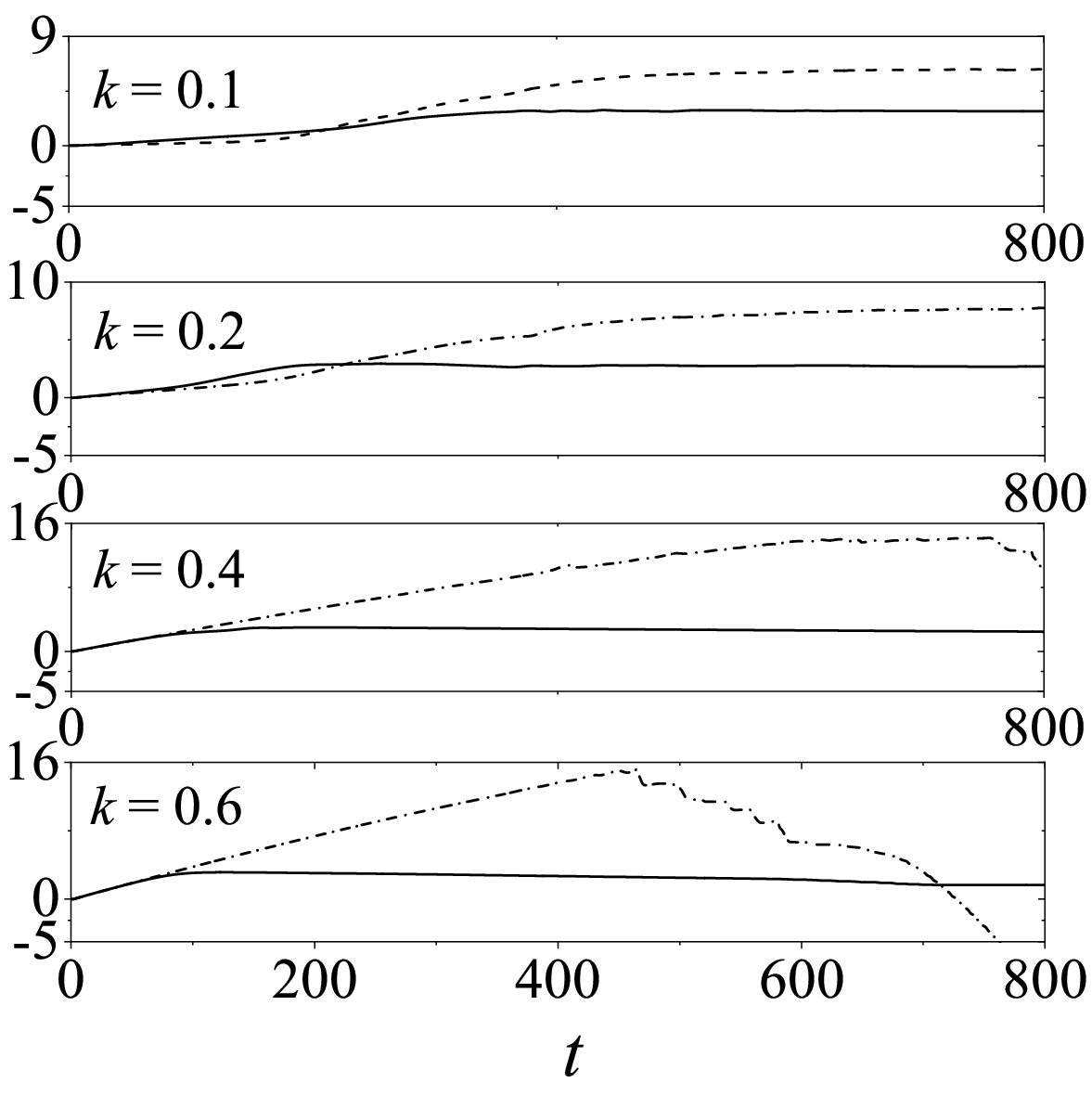}}
\quad
\subfloat[$E=0.01$]{\includegraphics[scale=0.374]{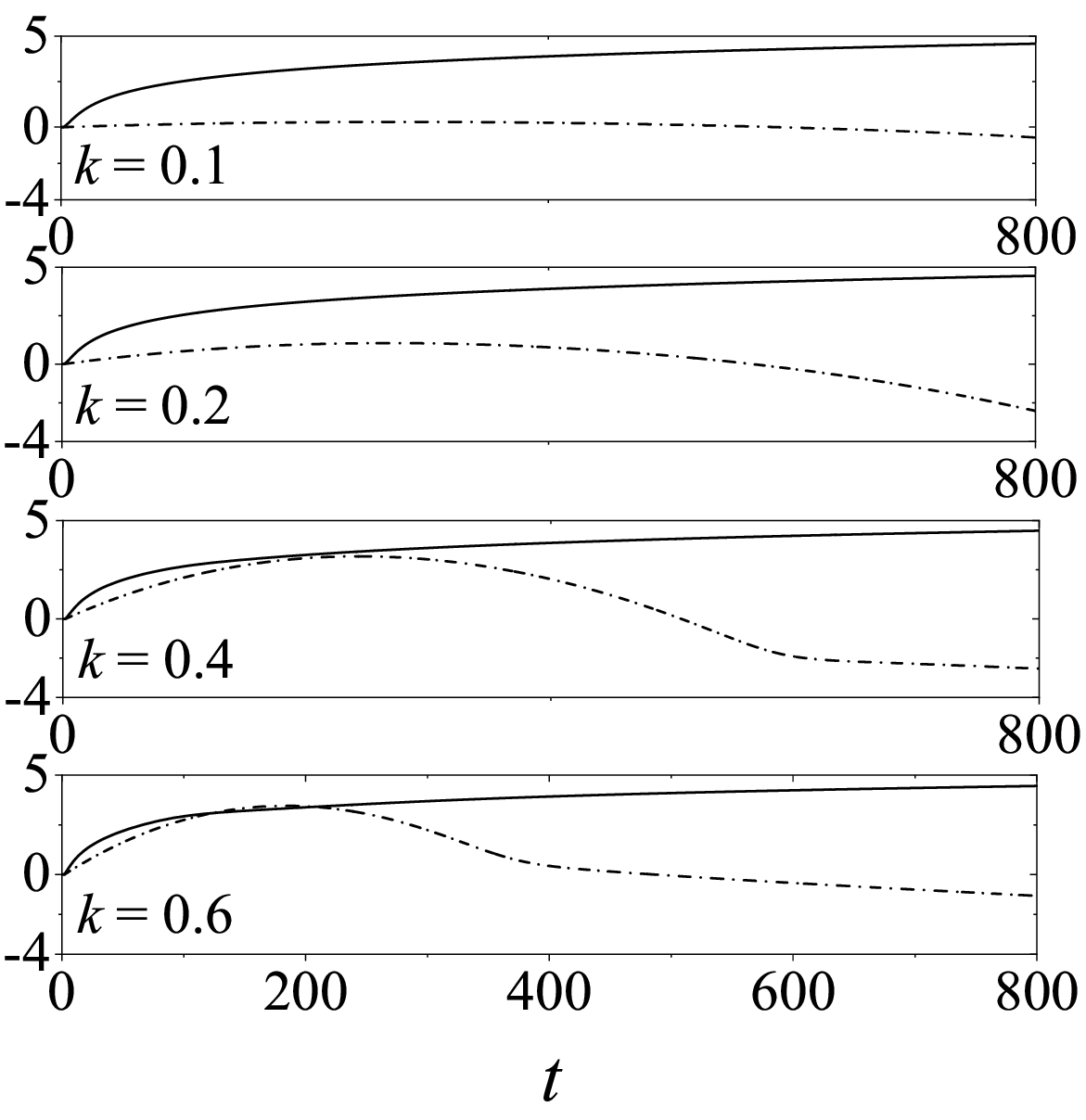}}
\caption{Comparison of growth rate of perturbation predicted by the full nonlinear model (\ref{evolution_equation}) and the linearized system (\ref{hjsc}) for an evaporating liquid. The solid line denotes $F_m$ from Eq. (\ref{nn1}), while the dashdot line denotes $N$ from Eq. (\ref{nn}). The other fixed parameters are  $b=0$, $\Pi=0.1$, $Ma=0$, $Bo=1$ and $K_0=10$.}
\label{fig10}
\end{figure}

 In Fig. \ref{fig10}, a comparison between the perturbation growth rate $F_m$ [given in Eq. (\ref{nn1})] from the full model (\ref{evolution_equation}) and the linearized system (\ref{hjsc}) is shown. It is important to note that, in the linearized system, the perturbation evolves about a space-time dependent base state $h=H(z,t)$, whereas in the full model, the growth is measured relative to the uniform base state $h=h_0=1$. Consequently, $F_m$ and $N$ do not represent identical measures of perturbation growth. Their comparison indicates the extent to which the linearized formulation, with its evolving base state, captures the temporal evolution observed in the full nonlinear simulation with uniform initial thickness.

\par To describe the nonlinear growth of the disturbance of the full model Eq. (\ref{evolution_equation}),  the amplification ratio of the disturbance is defined as
\begin{equation}\label{nn1}
F_m(t)=\ln\left[\frac{\left||h(z,t)-h_0|\right|_2}{\left||h(z,t=0)-h_0|\right|_2}\right],
\end{equation}
where $F_m$ corresponds to the full model.
\par From Fig. \ref{fig10}a, it can be observed that, under moderate evaporation, the perturbation growth predicted by the linearized system (dash-dot line) closely follows that of the full model (solid line) over a short initial time interval. For comparatively small wavenumbers ($k=0.1$ and $k=0.2$), the curve corresponding to $F_m$ is observed to lie slightly above $N$ at early times and subsequently remains below $N$ for the duration shown. As the wavenumber is increased, the two curves are observed to coincide at early times, after which the $F_m$ curves remain consistently below $N$. In contrast, from Fig. \ref{fig10}b, it can be observed that, under strong evaporation, a close correspondence between the $F_m$ and $N$ curves is retained for larger wavenumbers only during the early-time interval, whereas substantial deviations are observed for smaller wavenumbers ($k=0.1$ and $k=0.2$). Moreover, for the smaller wavenumbers ($k=0.1$ and $k=0.2$), a deviation between $F_m$ and $N$ is observed from the initial time under strong evaporation $(E=0.01)$, which was not observed for the moderate evaporation case $(E=0.001)$. Furthermore, whereas the $F_m$ curves are found to fall below $N$ at later times under moderate evaporation, the $F_m$ curves are consistently positioned above $N$ for the strong evaporation case ($E=0.01$) for all the wavenumbers considered. These observations indicate that the deviation between the two predictions becomes more pronounced as the evaporation strength is increased, particularly for smaller wavenumbers. Overall, Fig. \ref{fig10} shows that the growth rate obtained from the full model with a uniform thickness can reasonably predict the growth rate of the linearized model with an evolving base state for moderate evaporation, particularly at early times. However, this agreement is not maintained under strong evaporation or at later times under moderate evaporation.

\par
\begin{figure}[h!]
\centering
\subfloat[$E=0.001$]{
\includegraphics[scale=0.37]{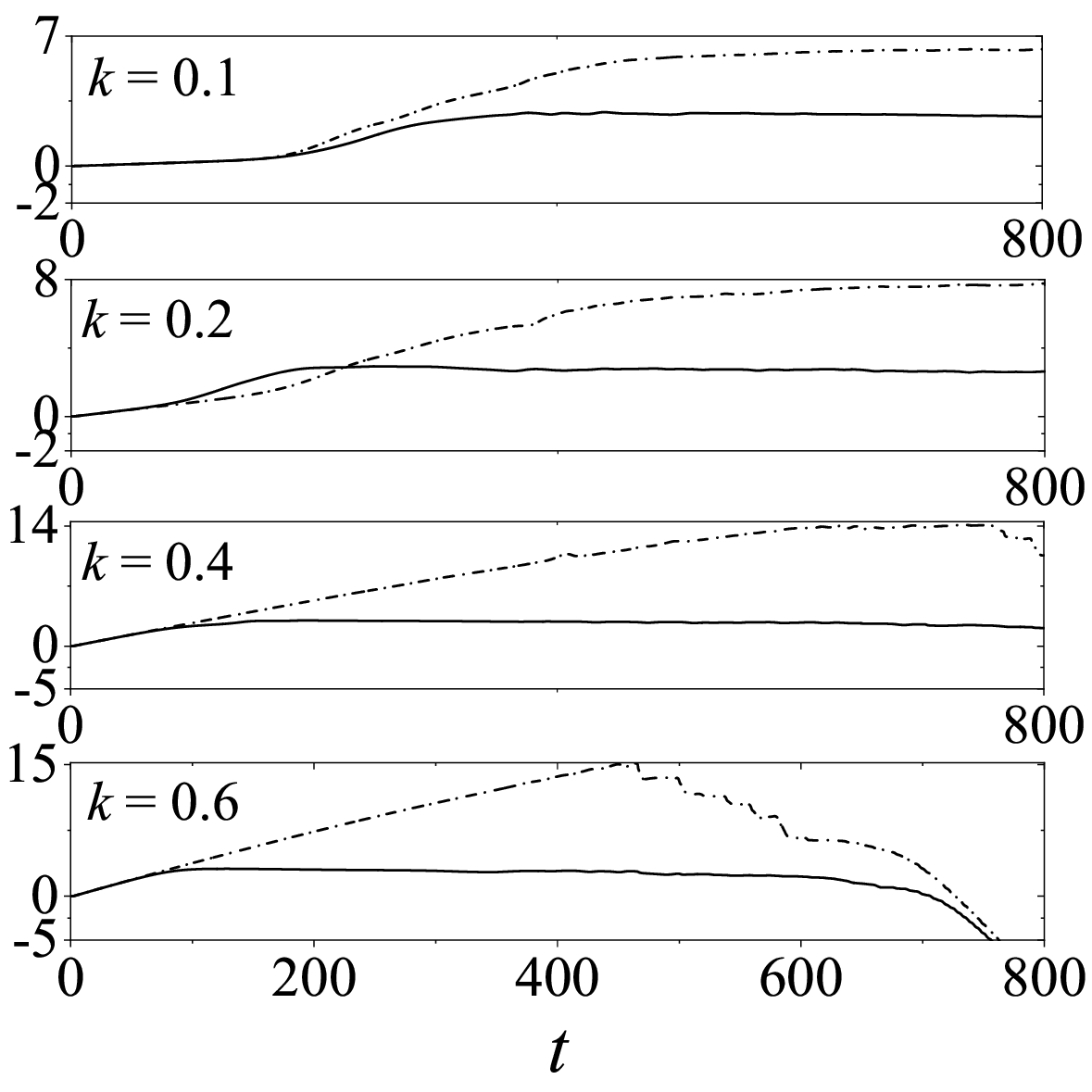}}
\qquad
\subfloat[$E=0.01$]{
\includegraphics[scale=0.37]{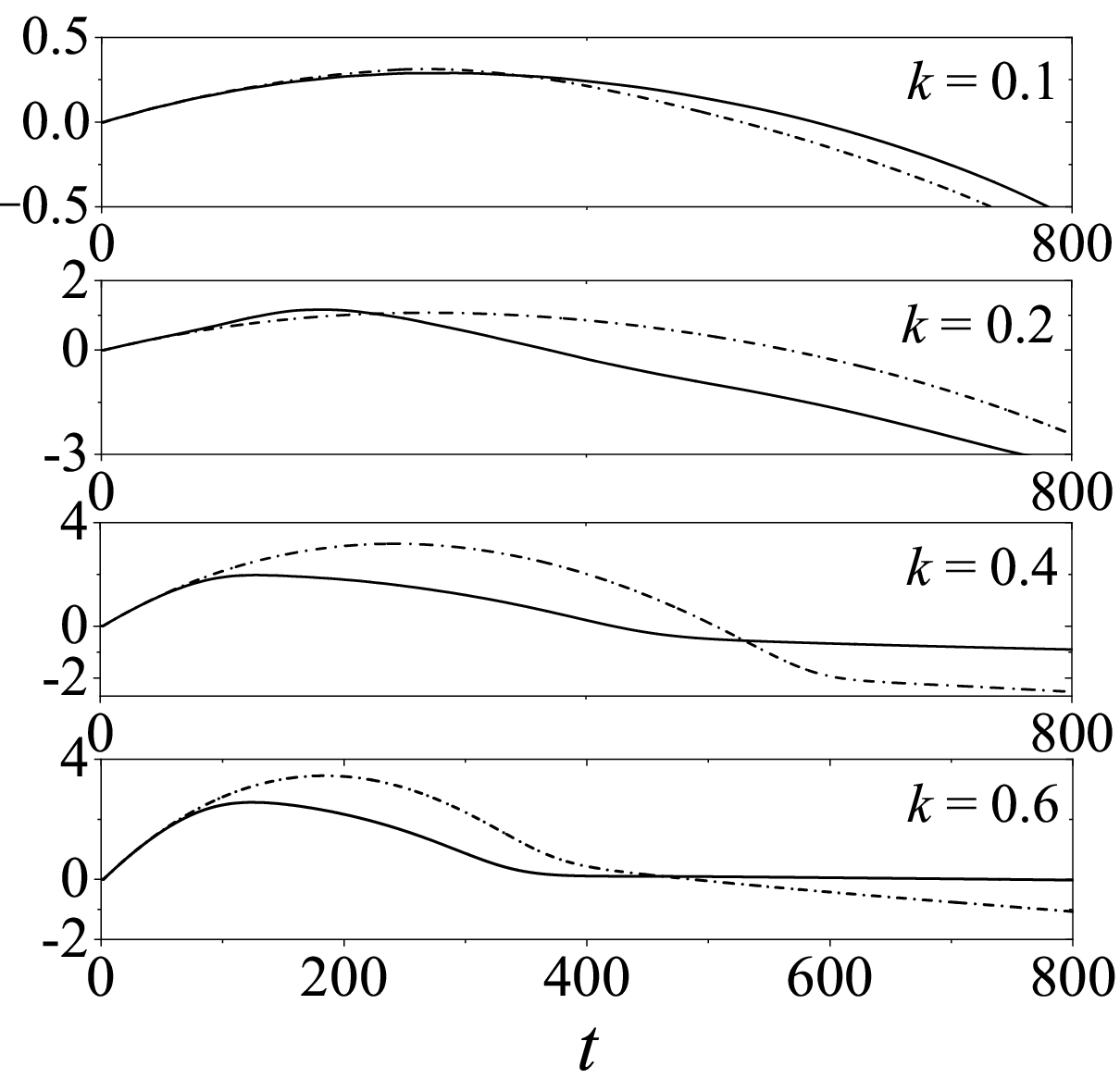}}
\caption{Comparison of the growth rate of perturbations for an evaporating liquid when $b=0$. The solid line denotes $F_{NL}$ from Eq. (\ref{nn1_sc}), while the dashdot line denotes $N$ from Eq. (\ref{nn}). The other fixed parameters are  $\Pi=0.1$, $K_0=10$, and $Ma=0$.}
\label{fig_compare_three}
\end{figure}

 In Fig. \ref{fig10}, the growth of the perturbation predicted by the coupled linearized model [Eq. (\ref{hjsc})] is compared with that obtained from the full nonlinear model [Eq. (\ref{evolution_equation})]. As discussed above, $F_m$ and $N$ do not represent identical measures of perturbation growth. In $N$, the perturbation is quantified relative to the evolving base state $H(z,t)$, whereas $F_m$ quantifies the deviation of the nonlinear solution from the initial uniform film thickness, $h_0=1$. To enable a direct comparison between the linearized and full nonlinear models, the following amplification ratio of the disturbance is defined: 
\begin{equation}\label{nn1_sc}
F_{NL}(t)=\ln\left[\frac{\left||h(z,t)-H(z,t)|\right|_2}{\left||h(z,t=0)-H(z,t=0)|\right|_2}\right].
\end{equation}
A comparison between $F_{NL}$ [Eq. (\ref{nn1_sc})], obtained from the full nonlinear model with an evolving base state, and $N$ [Eq. (\ref{nn})], obtained from the linearized model with an evolving base state, is presented in Fig. \ref{fig_compare_three} for an evaporating liquid at different wavenumbers $k$. Fig. \ref{fig_compare_three}a corresponds to $E=0.001$ (moderate evaporation), whereas Fig. \ref{fig_compare_three}b corresponds to $E=0.01$ (strong evaporation). An excellent agreement is observed between the profiles of $N$ and $F_{NL}$ during the early-time evolution, indicating that, in the small-disturbance regime, consistent perturbation growth is predicted by the linearized model and the full nonlinear model when the evolving base state is retained in both formulations. A comparison of Figs. \ref{fig10}a and \ref{fig_compare_three}a further illustrates the effect of the base-state consideration in the full nonlinear model. In Fig. \ref{fig10}a, $N$ is compared with $F_m$, where $F_m$ is obtained from the full nonlinear model with a uniform-thickness base state, whereas in Fig. \ref{fig_compare_three}a, $N$ is compared with $F_{NL}$, obtained from the full nonlinear model with an evolving base state. For smaller wavenumbers $(k=0.1,k=0.2)$, a slight discrepancy is observed between $N$ and $F_m$ during the early-time evolution in Fig. \ref{fig10}a. This discrepancy is eliminated when the evolving base state is incorporated into the full nonlinear model, as evidenced by the close agreement between $N$ and $F_{NL}$ in Fig. \ref{fig_compare_three}a. With increasing wavenumber, both $F_m$ and $F_{NL}$ are found to coincide with $N$ during the early-time evolution. Thus, the difference between the two nonlinear formulations is more evident at smaller $k$, whereas their predictions become increasingly similar as $k$ is increased. For stronger evaporation $(E=0.01)$, the distinction between the two nonlinear formulations becomes more pronounced. A comparison of Figs. \ref{fig10}b and \ref{fig_compare_three}b shows that substantially closer agreement with $N$ is obtained for $F_{NL}$ than for $F_m$. In Fig. \ref{fig10}b, a substantial deviation of $F_m$ from $N$ is observed from the beginning of the evolution at small $k$. With increasing $k$, an initial trend similar to that of $N$ is exhibited by $F_m$; however, a gradual departure from $N$ is observed after a certain time. In contrast, in Fig. \ref{fig_compare_three}b, $F_{NL}$ coincides with $N$ during the early-time evolution for both small and larger $k$. Moreover, the trends exhibited by $F_{NL}$ and $N$ remain consistent even at longer times. These comparisons indicate that the use of an evolving base state in the full nonlinear model leads to substantially better consistency with the linearized model, particularly for stronger evaporation and smaller wavenumbers.

\begin{figure}[h!]
\centering
\subfloat[$E=0.001$]{
\includegraphics[scale=0.37]{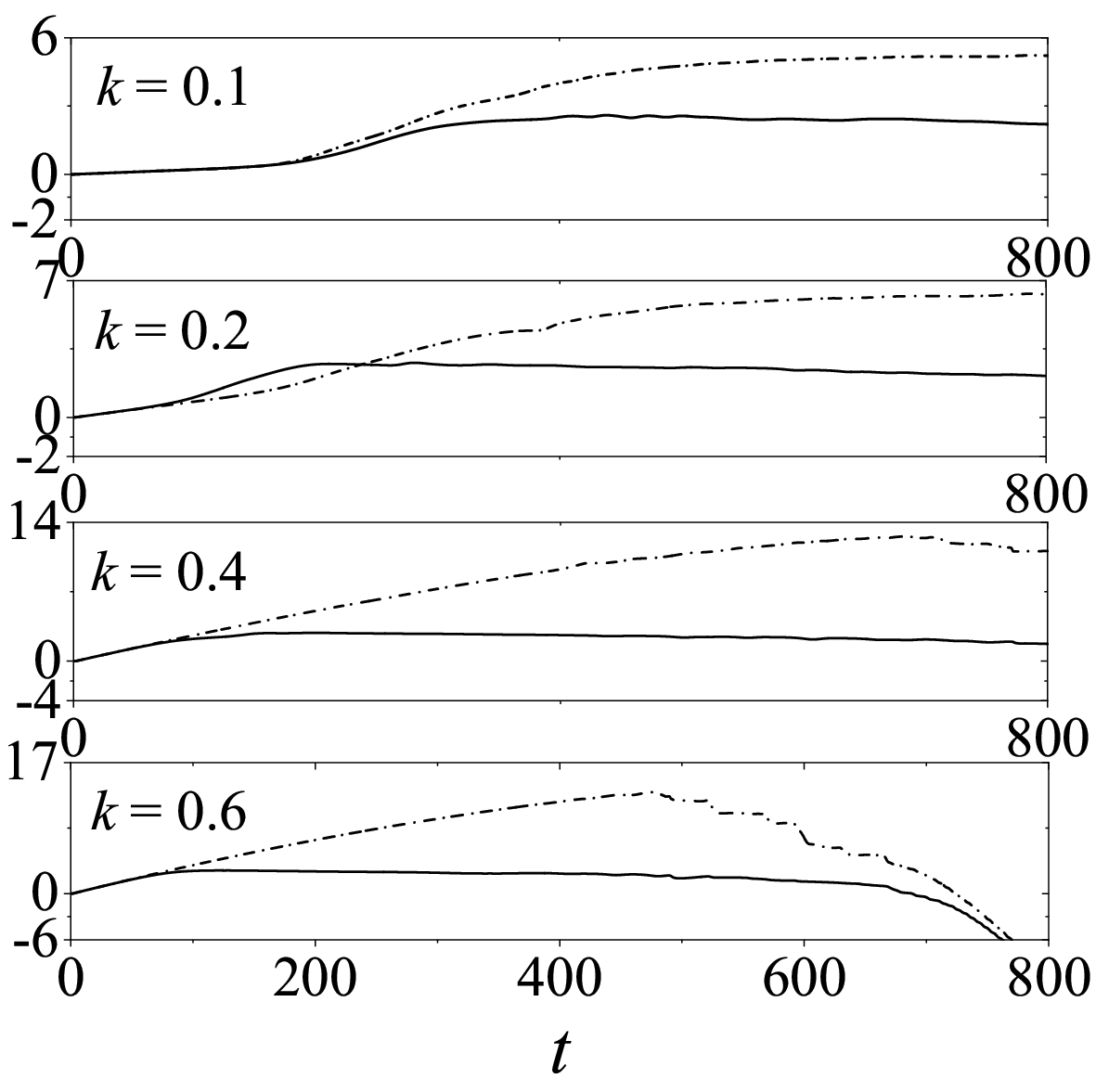}}
\qquad
\subfloat[$E=0.01$]{
\includegraphics[scale=0.37]{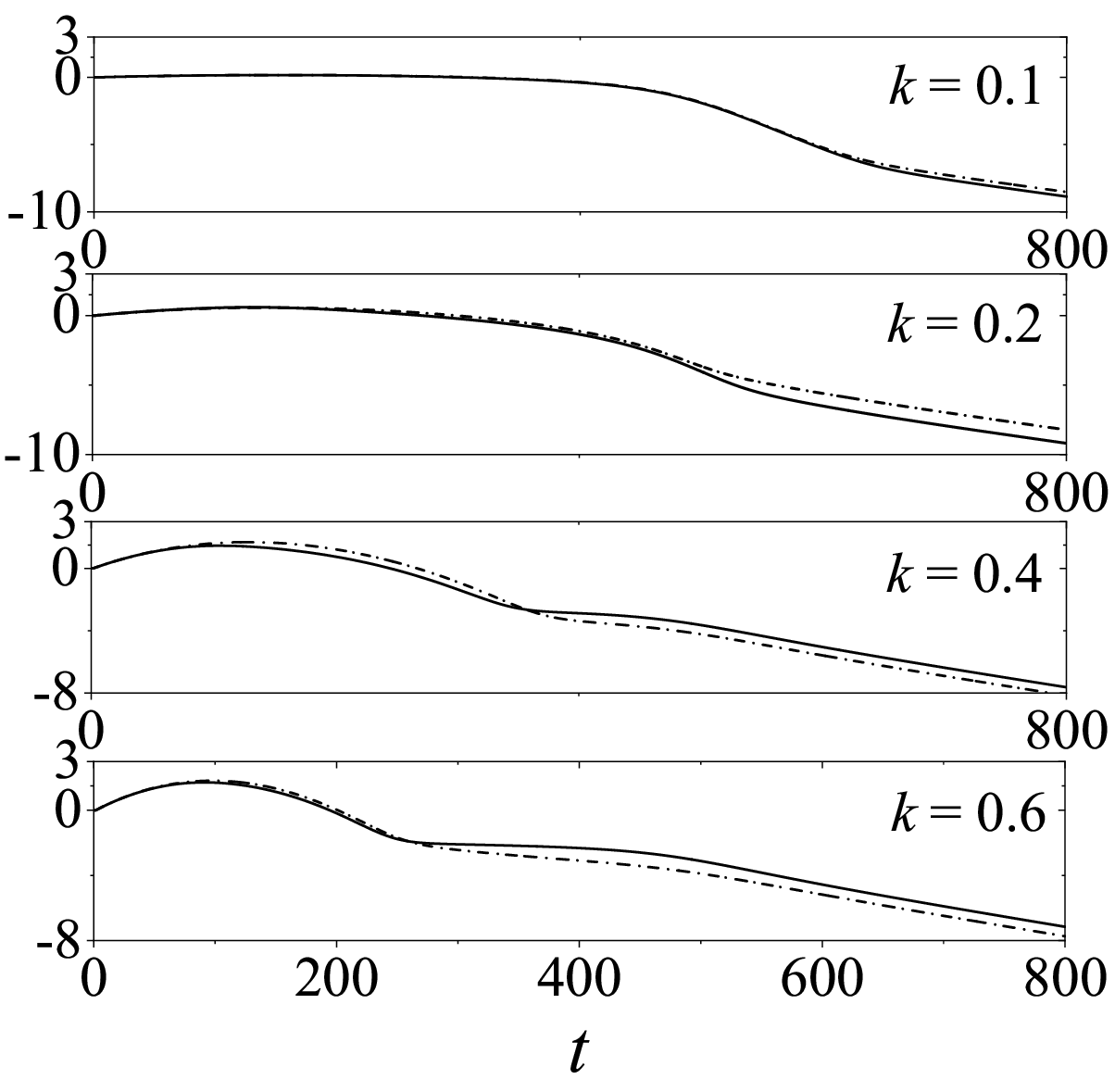}}
\caption{Comparison of the growth rate of perturbations for an evaporating liquid when $b\neq0$. The solid line denotes $F_{NL}$ from Eq. (\ref{nn1_sc}), while the dashdot line denotes $N$ from Eq. (\ref{nn}). The other fixed parameters are  $\Pi=0.1$, $b=0.1$, $\epsilon=0.1$, $K_0=10$, and $Ma=0$.}
\label{fig_compare_three_n}
\end{figure}
\par The comparison between the growth rate $N$ and $F_{NL}$ is presented in Fig. \ref{fig_compare_three} for moderate and strong evaporation scenarios with $b=0$. The corresponding comparison for $b\neq0$ is presented in Fig. \ref{fig_compare_three_n}, where $b=0.1$ is considered. It can again be observed that, for both moderate and strong evaporation cases, good agreement is obtained between the early-time behavior of the linearized and full models. Furthermore, by comparing Figs. \ref{fig_compare_three}b  and \ref{fig_compare_three_n}b, it is observed that, under strong evaporation, the trends of the two plots are in closer agreement for $b\neq0$ than for $b=0$.

\begin{figure}[h!]
\centering
\subfloat[$k=0.1$]{
\includegraphics[scale=0.28]{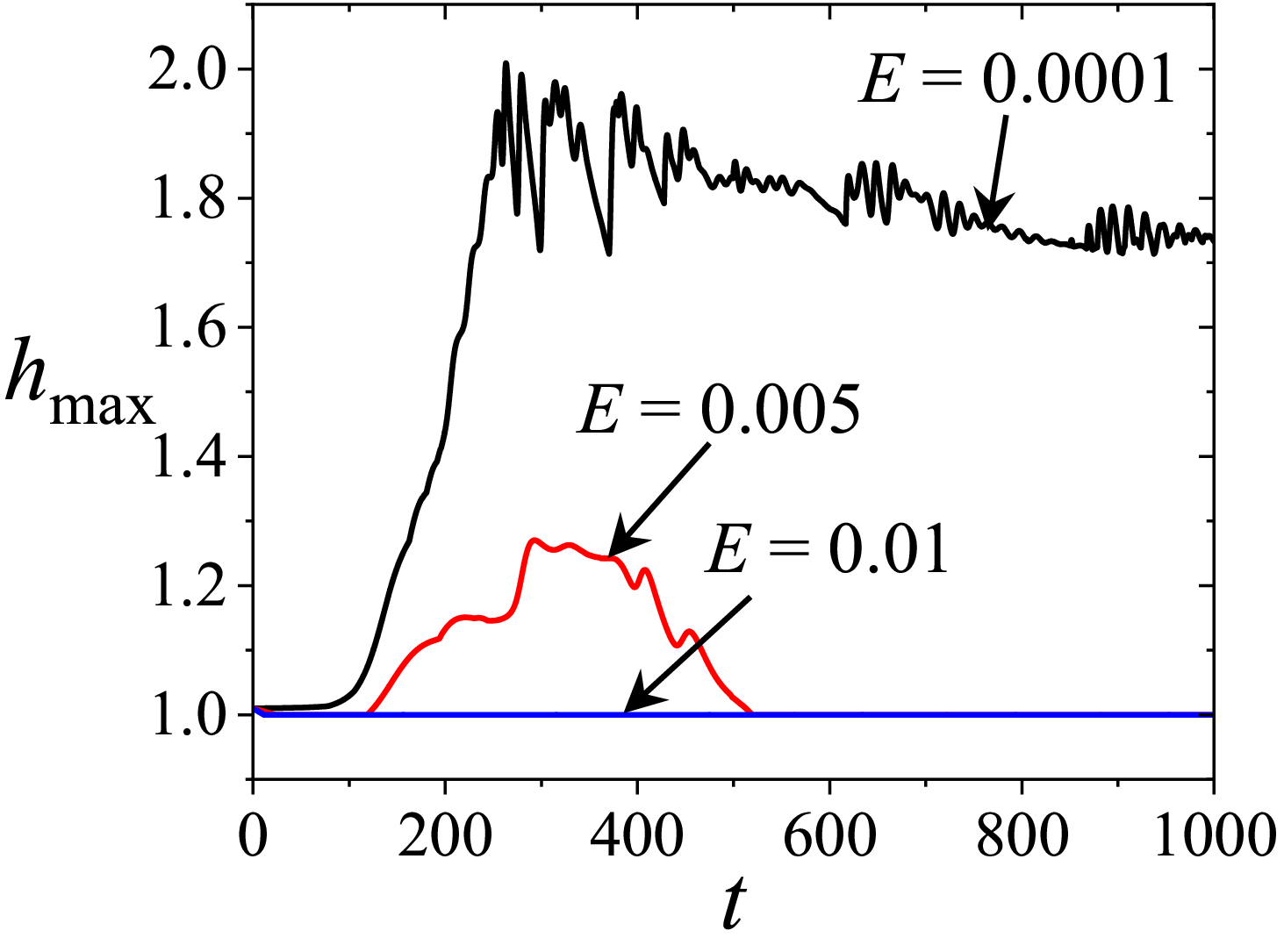}}
\subfloat[$k=0.1$]{
\includegraphics[scale=0.28]{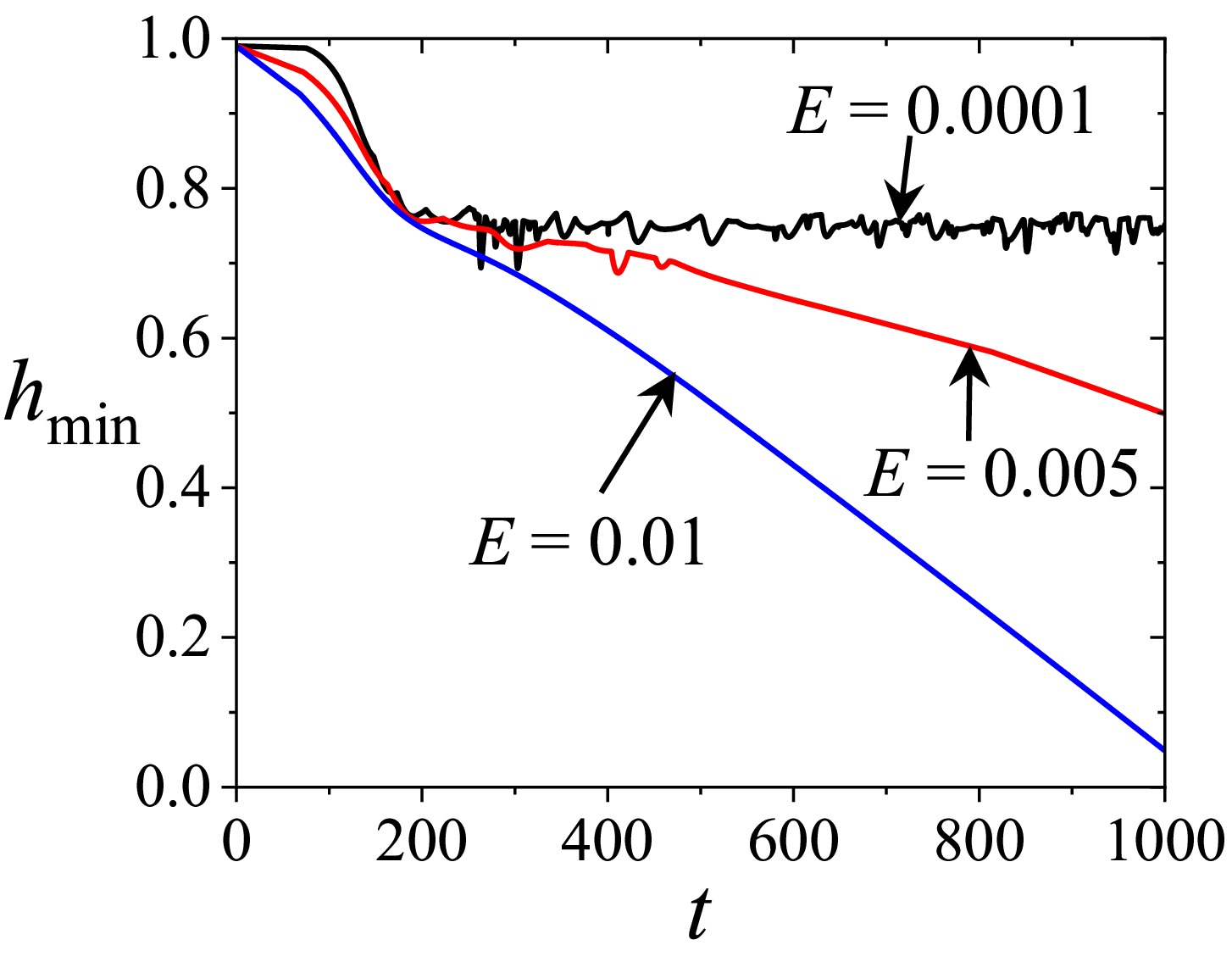}}
\quad
\subfloat[$k=0.6$]{
\includegraphics[scale=0.28]{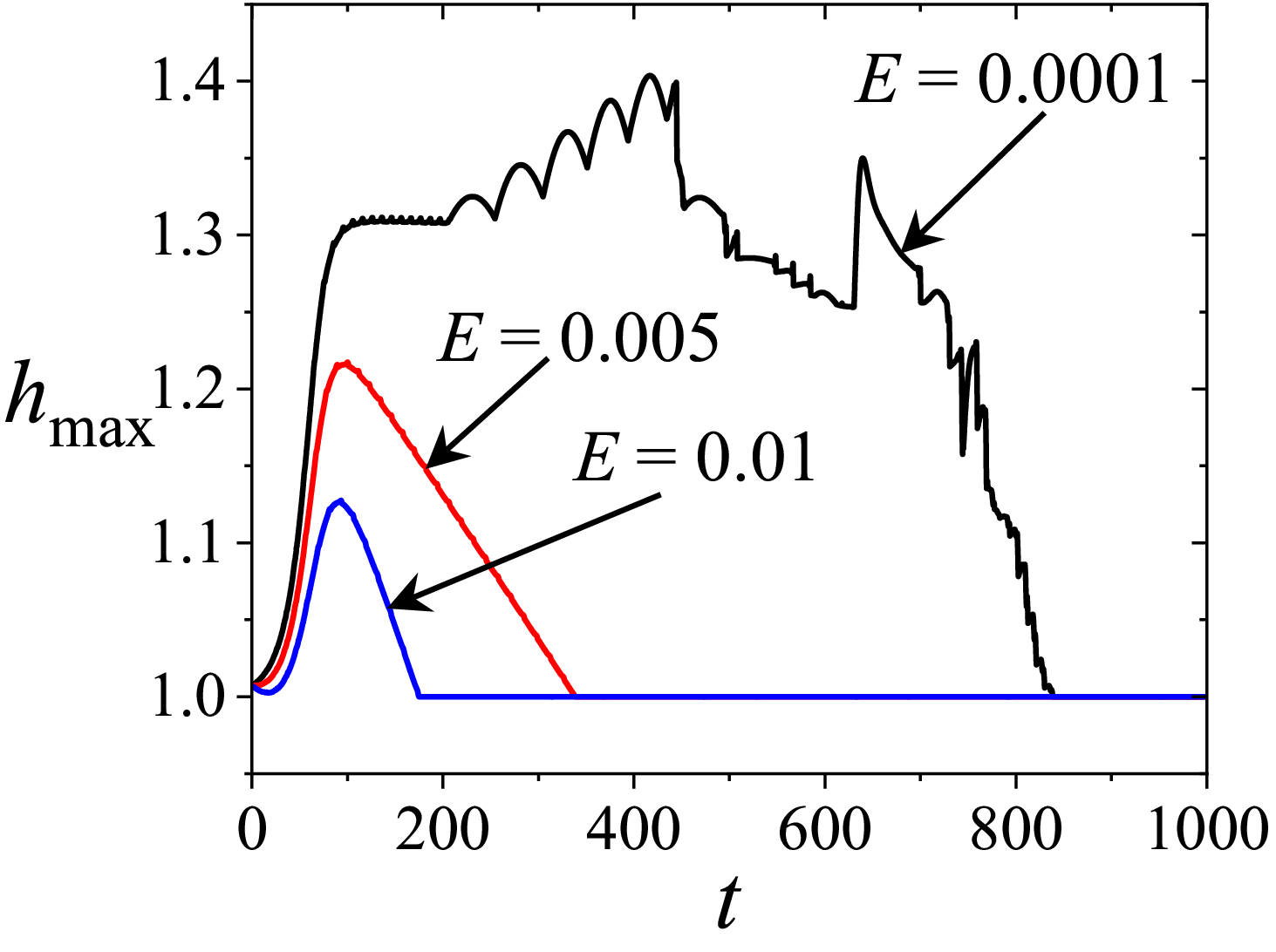}}
\subfloat[$k=0.6$]{
\includegraphics[scale=0.28]{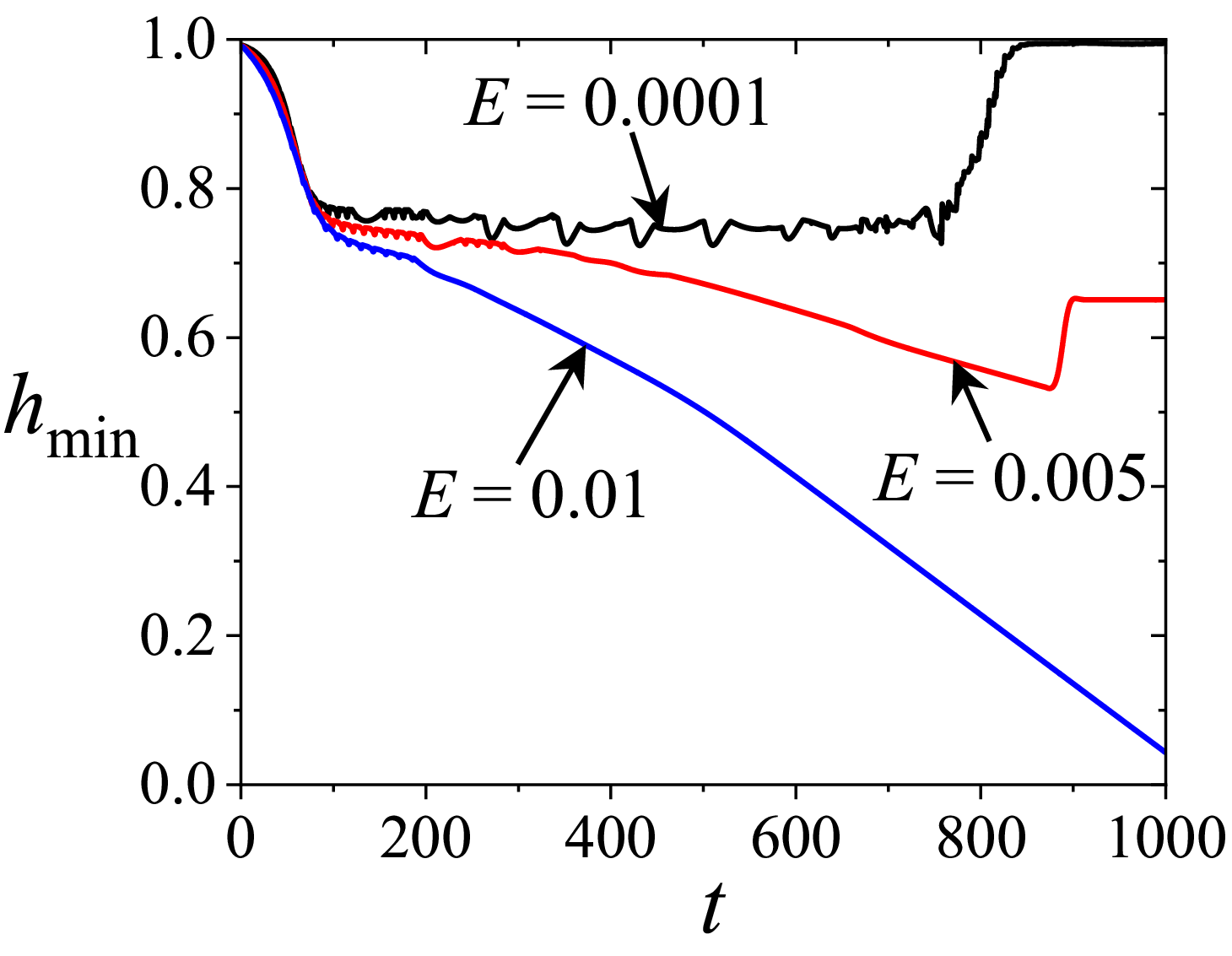}}
\caption{Evolution of the maximum $\left(h_{\max}\right)$ and minimum $\left(h_{\min}\right)$ film thickness with time for evaporating liquid $(E>0)$ when $b=0$. The other fixed parameters are $Ma=1$, $\Pi=0.1$,  $Bo=1$ and $K_0=10$.}
\label{fig11}
\end{figure}
\par The effect of the parameter $E$ on the maximum $h_{\max}(t)$ and minimum $h_{\min}(t)=\min_zh(z,t)$ amplitude of the wave is presented in Fig. \ref{fig11} for an evaporating liquid $(E>0)$ when $\Pi\neq0$ and $b=0$. The evolution of $h_{\max}$ and $h_{\min}$ is presented in Figs. \ref{fig11}a and \ref{fig11}b, respectively, when $E$ varies from weak to strong values, $E\sim O\left(10^{-4}\right)-O\left(10^{-2}\right)$. When evaporation is weak ($E=0.0001$), an increase in $h_{\max}$ is observed from the early stage until a peak value is reached. Subsequently, oscillatory evolution is observed within the computational time. When evaporation is increased to a moderate level ($E=0.005$), a similar initial growth of $h_{\max}$ is observed. However, the onset of growth is delayed compared with the weak evaporation case. For $E=0.005$, a maximum value of $h_{\max}$ is reached at approximately $t\approx217$, after which a decay is observed. When strong evaporation is considered ($E=0.01$), a similar trend in $h_{\max}$ is observed as in the moderate evaporation case. However, a much lower amplitude of the variation in $h_{\max}$ is obtained. In Fig. \ref{fig11}b, $h_{\min}$ is plotted for weak to strong evaporation numbers. For weak evaporation $(E=0.0001)$, the film initially undergoes thinning during the early-time transient. After this initial thinning, however, no further net decay of the film thickness is observed; instead, the film continues to evolve with chaotic oscillations. When the evaporation rate is increased to a moderate level $(E=0.005)$, the initial thinning remains qualitatively similar to that observed for weak evaporation. In contrast to the weak-evaporation case, the film continues to decay after the initial transient, while no chaotic oscillations are observed. Thus, increasing the evaporation rate from weak to moderate changes the long-time behavior from oscillatory evolution without further net thinning to continuous film decay. For strong evaporation, the film again exhibits an initial thinning similar to the weak and moderate cases, followed by an almost linear long-time decay of $h_{\min}$. At higher evaporation rates, the enhanced evaporative mass loss leads to more rapid depletion of the film, causing the film thickness to approach zero.

\par The top panel of Fig. \ref{fig11} is plotted for $k=0.1$ while the bottom panel is shown for $ k=0.6$. Comparing Figs. \ref{fig11}a and \ref{fig11}c, a few observations are found. Firstly, the $h_{\max}$ amplitude is larger for weak evaporation $(E=0.0001)$ when $k=0.1$, whereas amplitude is larger for strong evaporation $(E=0.01)$ when $k=0.6$. Secondly, for $k=0.6$, the initial growth of $h_{\max}$ begins at $t=0$ for all evaporation regimes (weak to strong), while for $k=0.1$, growth starts at $t>0$. Third, complete damping of $h_{\max}$ in a smaller domain requires a higher evaporation rate compared to a larger domain. Fourth, for a smaller domain $(k=0.6)$, $h_{\max}$ decays to the base state more rapidly under weak evaporation than in a larger domain $(k=0.1)$, whereas under strong evaporation, the decay to the base state is faster in the larger domain $(k=0.1)$ than in the smaller domain $(k=0.6)$. Moreover, comparing Figs. \ref{fig11}b and \ref{fig11}d, it is found that the decay is slightly faster for $k=0.6$ than $k=0.1$. However, it is noted that for most of the time, $h_{\min}$ is almost the same for both $k$.

\begin{figure}[h!]
\centering
\subfloat[$k=0.1$]{
\includegraphics[scale=0.32]{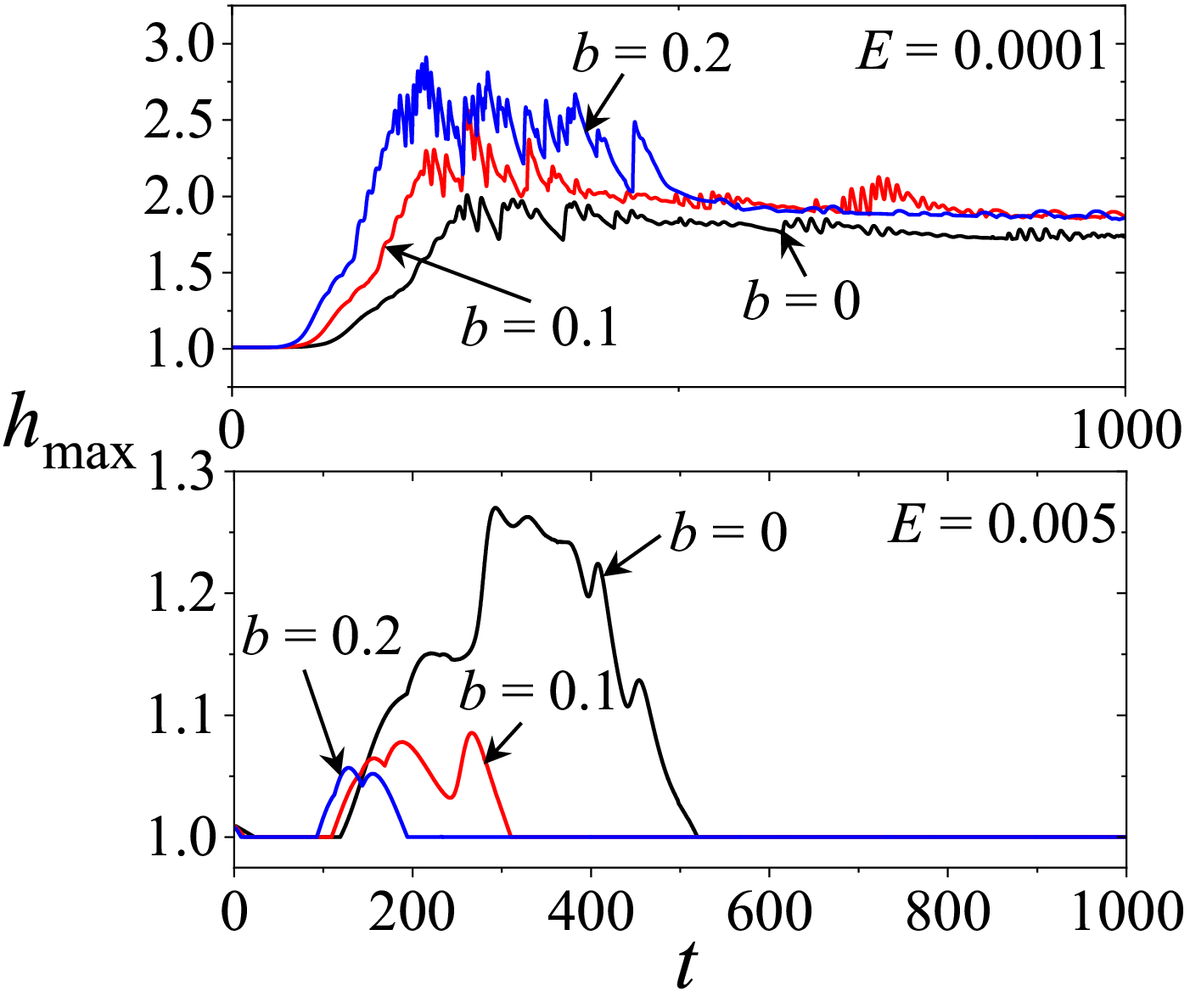}}
\subfloat[$k=0.6$]{
\includegraphics[scale=0.32]{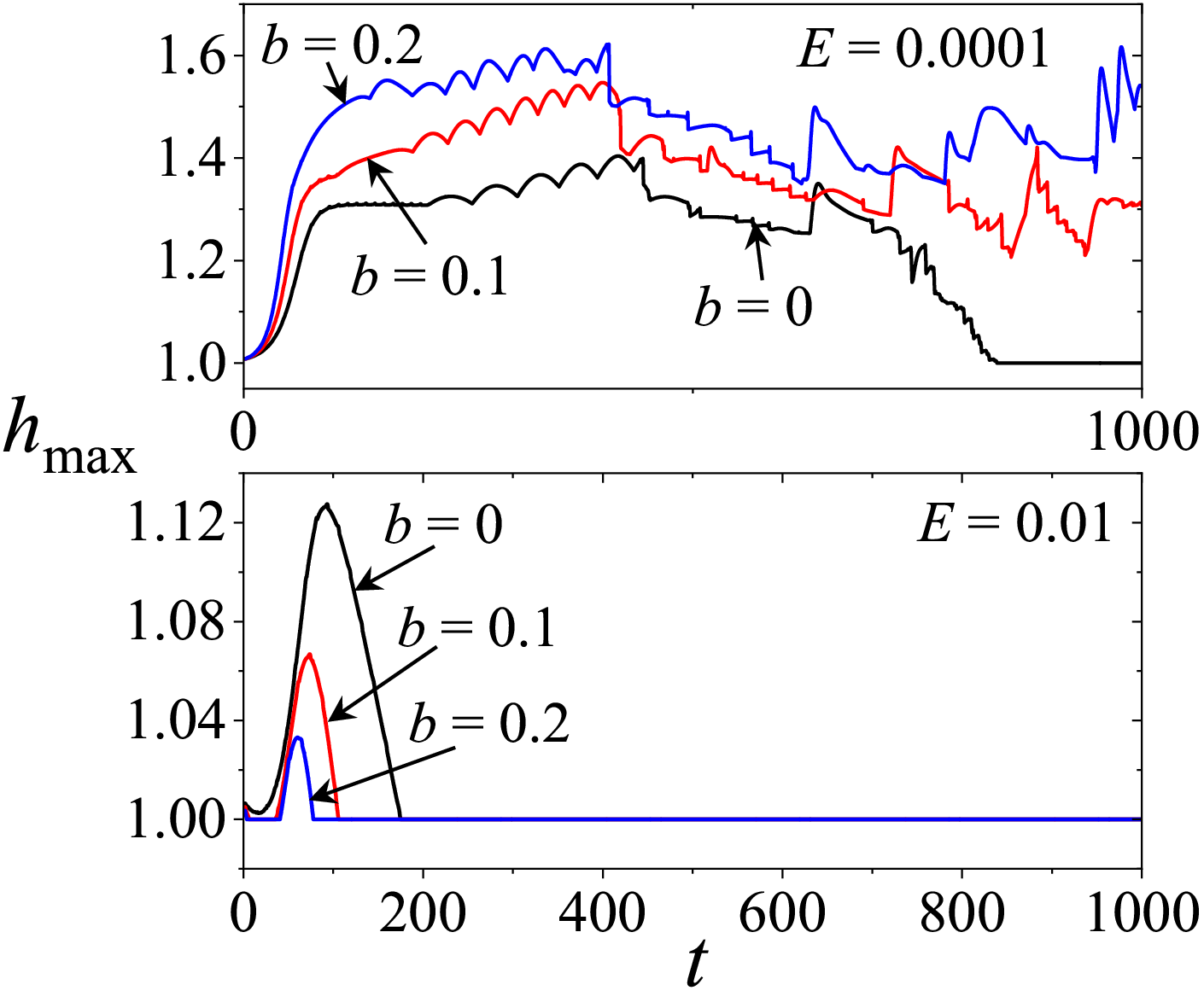}}
\quad
\subfloat[$k=0.1$]{
\includegraphics[scale=0.32]{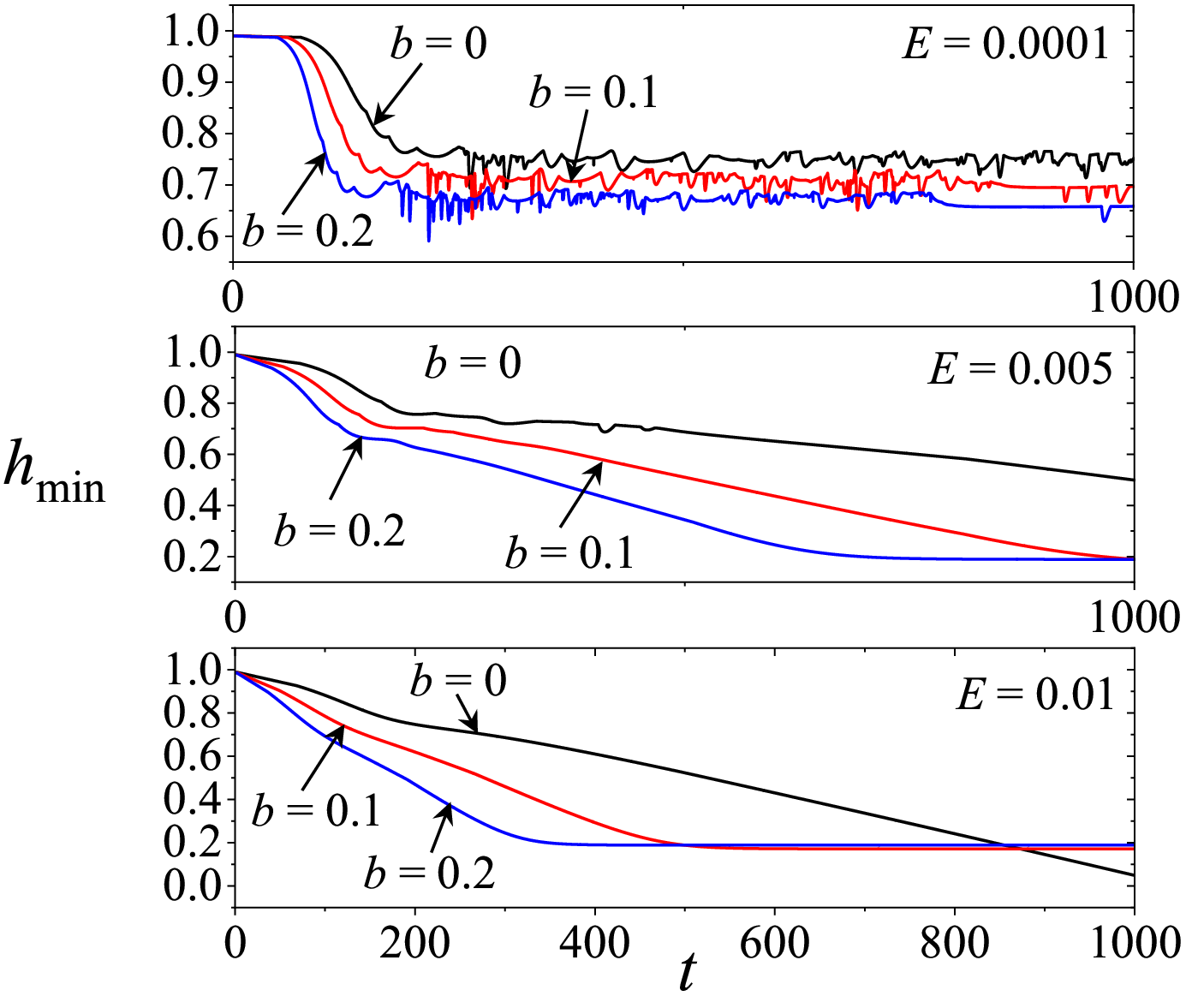}}
\subfloat[$k=0.6$]{
\includegraphics[scale=0.32]{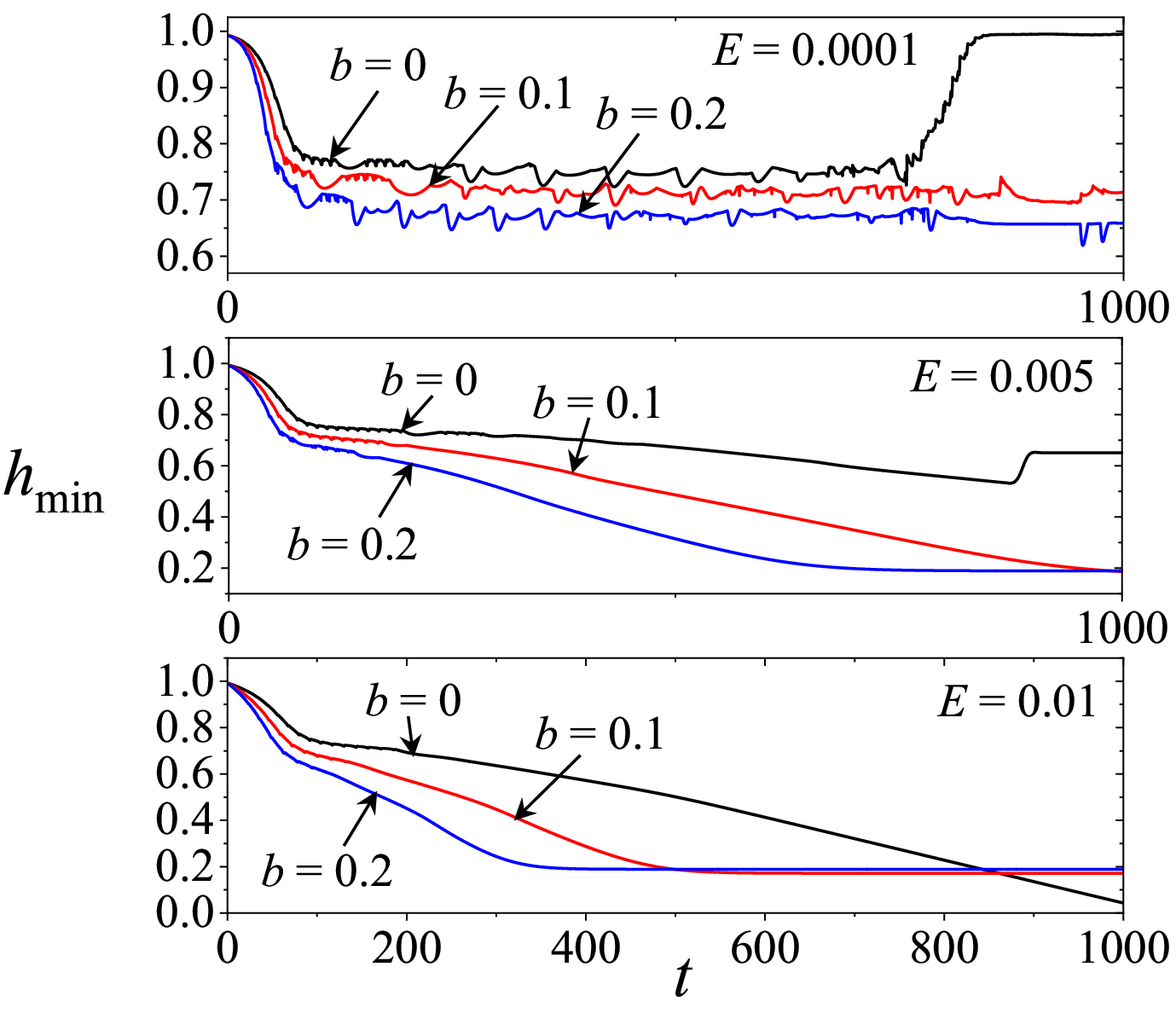}}
\caption{Evolution of the maximum $\left(h_{\max}\right)$ film thickness (top panel) and minimum $\left(h_{\min}\right)$ film thickness (bottom panel) with time for evaporating liquid $(E>0)$ when $b\neq0$. The other fixed parameters are $Ma=1$, $\Pi=0.1$, $\epsilon=0.1$, $Bo=1$ and $K_0=10$.}
\label{fig11n}
\end{figure}
\par From Eq. (\ref{mass_flux}), the evaporative mass flux is given by $J=\left(1+bp(h)\right)/\left(K_0+h\right)$. In Fig. \ref{fig11}, the temporal evolution of $h_{\max}$ and $h_{\min}$ is shown for $b=0$, i.e., when the contribution from the hydrodynamic pressure $p(h)$ vanishes. In Fig. \ref{fig11n}, the temporal evolution of $h_{\max}$ and $h_{\min}$ will be shown for $b\neq0$, while the remaining parameters will be the same as Fig. \ref{fig11}. Thus, the plots will show how hydrodynamic pressure influences evaporation. The $h_{\max}$ is shown for $k=0.1$ in Fig. \ref{fig11n}a, whereas displayed for $k=0.6$ in Fig. \ref{fig11n}b. From Figs. \ref{fig11n}a and \ref{fig11n}b, it is observed that when evaporation is weak, $h_{\max}$ is amplified with increasing $b$. On the other hand, when evaporation is moderate/strong, $h_{\max}$ is reduced with increasing $b$. These observations hold true irrespective of the domain length. In Figs. \ref{fig11n}c and \ref{fig11n}d, the evolution of $h_{\min}$ is shown with $b\neq0$ for weak to strong evaporation regimes. Fig. \ref{fig11n}c is plotted for $k=0.1$, whereas Fig. \ref{fig11n}d is shown for $k=0.6$. Both figures show that $h_{\min}$ decays faster with increasing $b$. It can be seen that when the evaporation effect is weak $(E=0.0001)$, the disturbance amplitude decays monotonically for a while before oscillating towards an equilibrium state within our computational time. When evaporation is moderate to strong, $h_{\min}$ rapidly approaches zero, and no oscillations are observed [see the $E=0.005$ and $E=0.01$ cases in Figs. \ref{fig11n}c and \ref{fig11n}d]. Although the plots were shown till $t=1000$, it was found that in the absence of $b$, $h_{\min}\to0$ at $t\approx2099.7$  for $E=0.005$ and at $t\approx1050$ for $E=0.01$ when $k=0.1$. However, when $k=0.6$, the $h_{\min}$ does not decay to zero; however, for $E=0.01$, it still reaches zero at approximately the same time as in the case $k=0.1$.

\par
\begin{figure}[h!]
\centering
\centering
\includegraphics[scale=0.33]{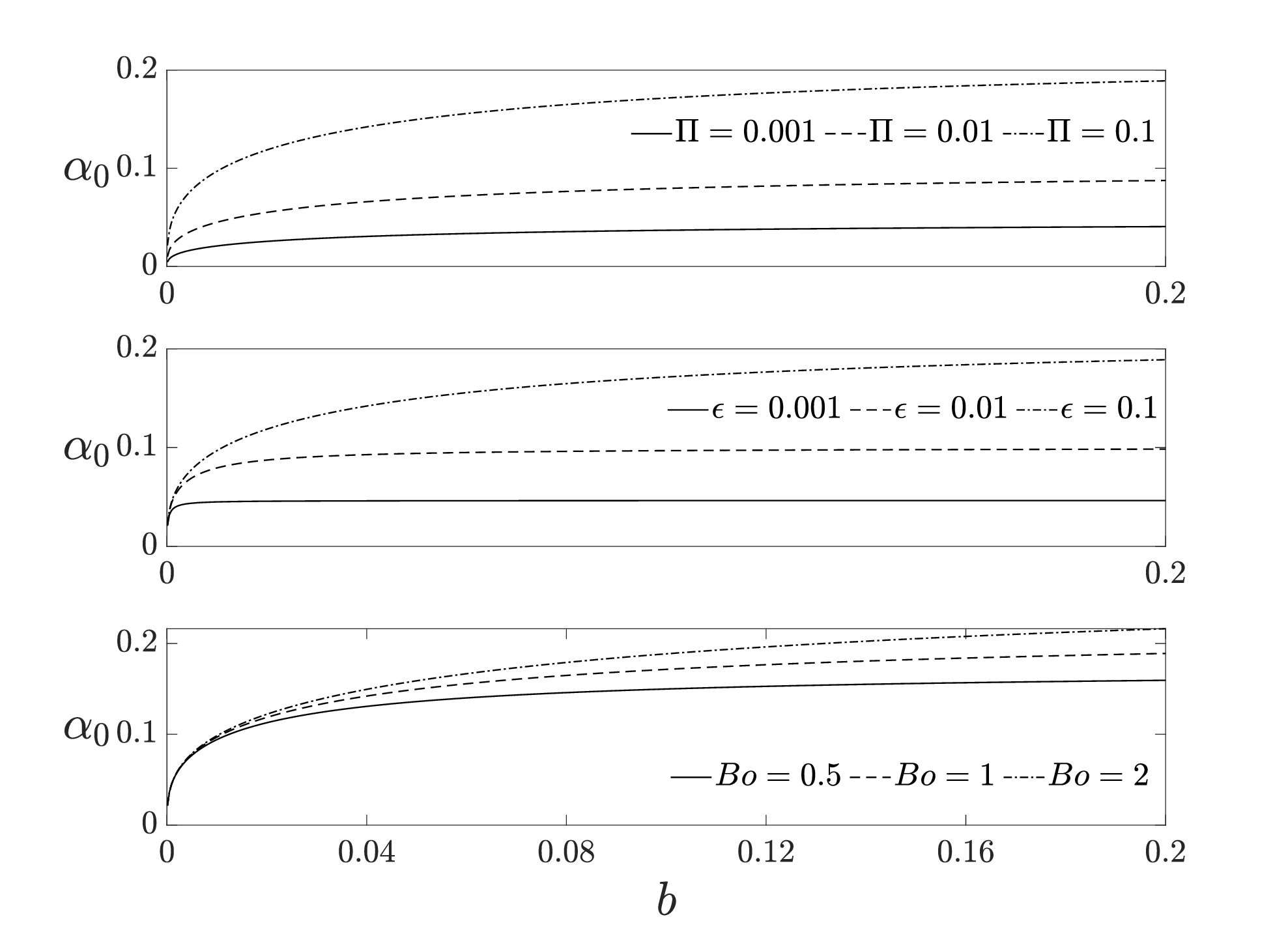}
\caption{Variation of $\alpha_0$ as $b$ varies. Top: For different $\Pi$ when $Bo=1$ and $\epsilon=0.1$; Middle: For different $\epsilon$ when $Bo=1$ and $\Pi=0.1$; Bottom: For different $Bo$ when $\Pi=0.1$ and $\epsilon=0.1$.}
\label{fig_precursor}
\end{figure}

 Moreover, from the $E=0.01$ case in these figures, it is evident that under strong evaporation, $h_{\min}$ vanishes more quickly than in the $E=0.005$ case. It can be observed from the bottom panel of Fig. \ref{fig11n} that, although $h_{\min} \to 0$ for the case $b = 0$, the film instead evolves toward a finite precursor thickness $h = \alpha_0$ for any $E>0$ when $b\neq 0$. For instance, in Figs. \ref{fig11n}c and \ref{fig11n}d, when $b = 0.1$ and $b = 0.2$, the corresponding values of $\alpha_0$ are found to be $0.1714$ and $0.189$, respectively. Physically, due to evaporation, the liquid film thins and is almost adsorbed on the solid surface. Setting $J=0$ in (\ref{mass_flux}) gives the following 
\begin{equation}\label{poly}
\alpha_0^4-\alpha_0^3\left(\frac{1}{\epsilon}+\frac{Bo}{b}\right)+Bo\Pi=0.
\end{equation}
Solving Eq. (\ref{poly}) yields $\alpha_0 = 0.1714$ and $\alpha_0 = 0.189$ for $b = 0.1$ and $b = 0.2$, respectively, when $Bo = 1$, $\Pi = 0.1$, and $\epsilon = 0.1$, consistent with the parameters used in Fig. \ref{fig11n}. The Eq. (\ref{poly}) further shows that increasing $b$ leads to a larger precursor film thickness $\alpha_0$. Moreover, $\alpha_0$ becomes thicker when $\Pi$ is increased and thinner when $Bo$ or $\epsilon$ is decreased. These observations are shown in Fig. \ref{fig_precursor}.

\begin{figure}[h!]
\centering
\includegraphics[scale=0.35]{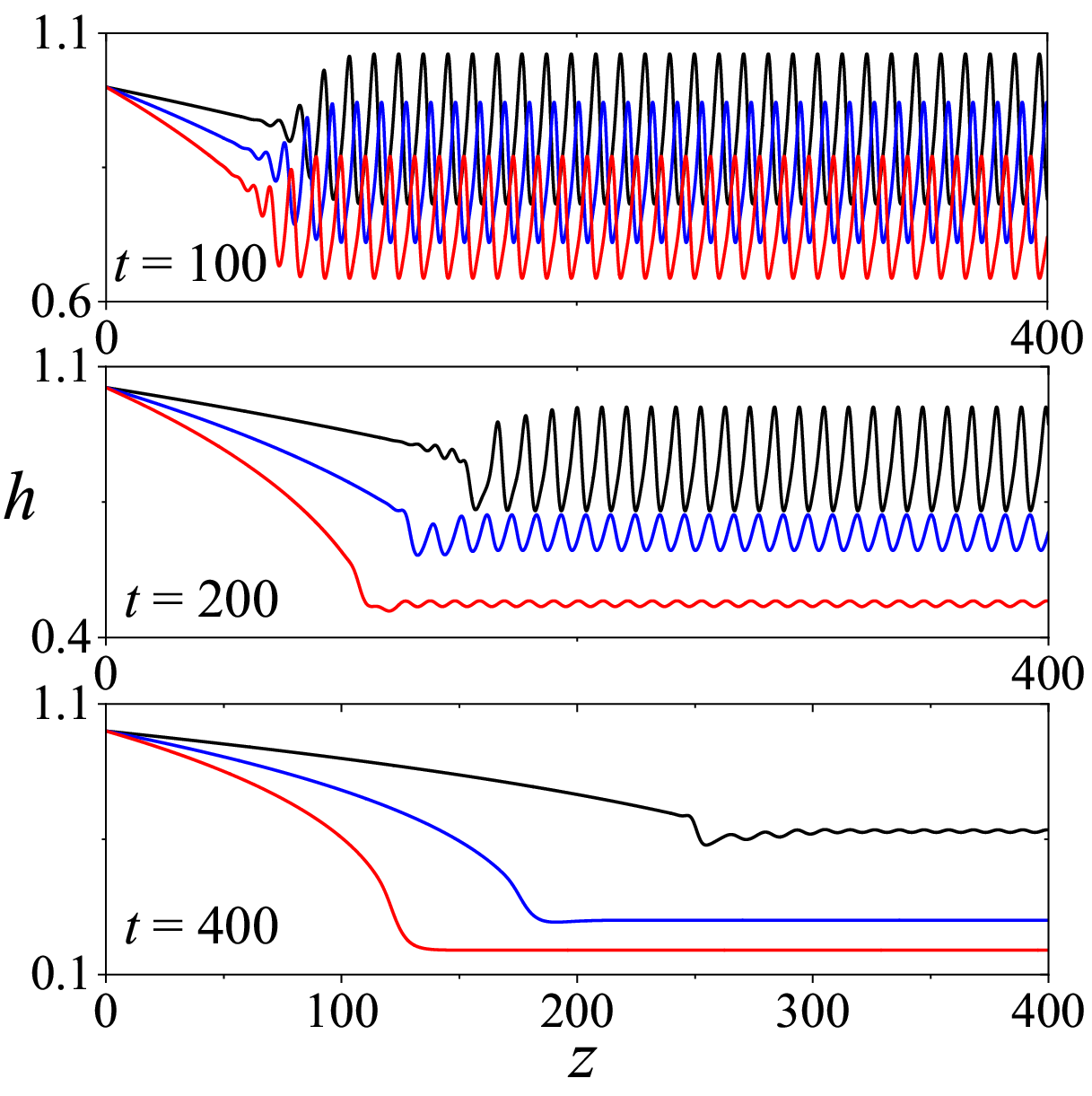}
\qquad
\includegraphics[scale=0.35]{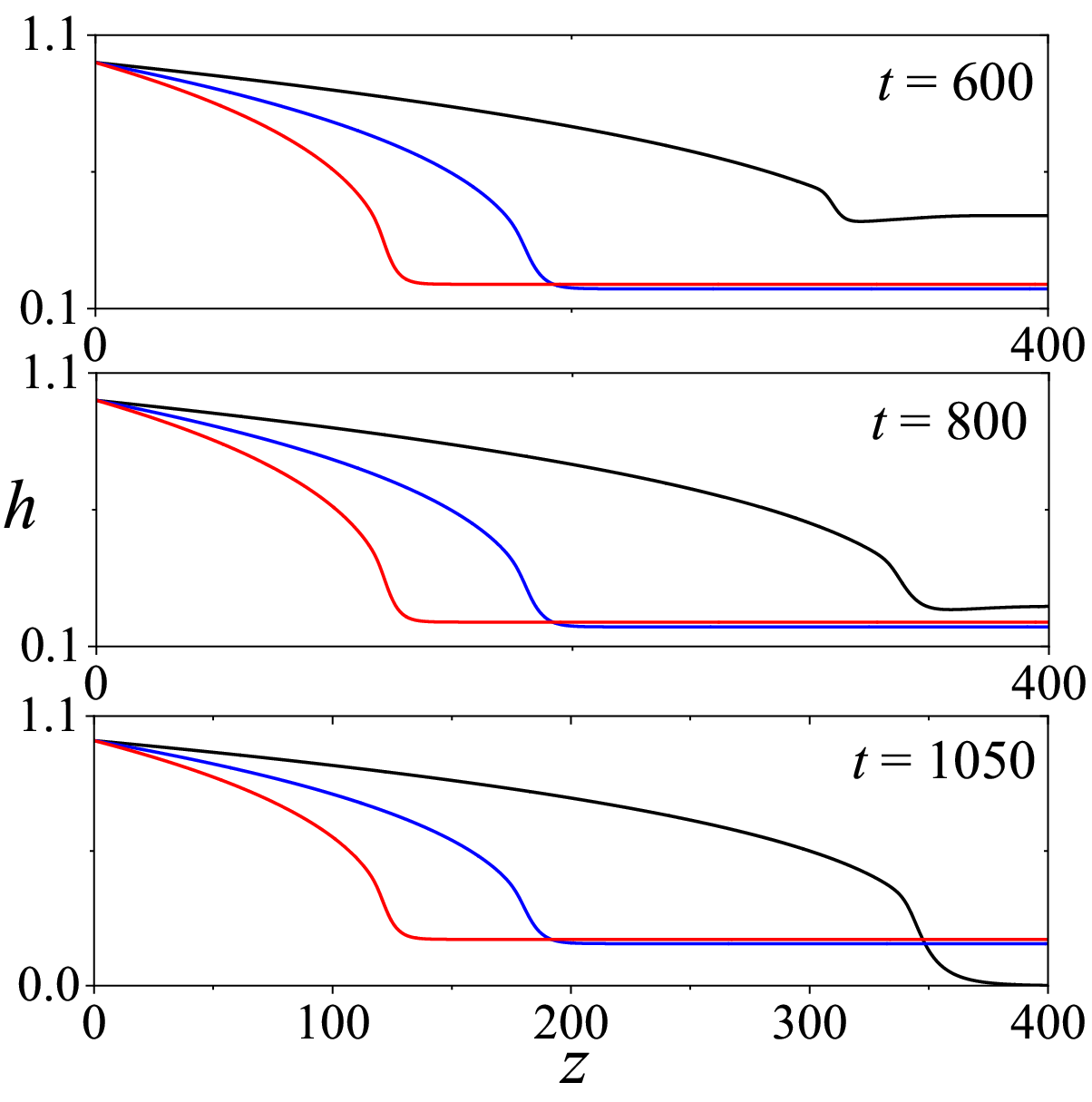}
\caption{Temporal evolution of the film profiles with or without $b$ for strong evaporation $(E=0.01)$. Here, the black line corresponds to $b=0$, the blue line corresponds to $b=0.1$, and the red line corresponds to $b=0.2$. The other fixed parameters are $k=0.6$, $Ma=1$, $\Pi=0.1$, $\epsilon=0.1$, $Bo=1$ and $K_0=10$.}
\label{fig12}
\end{figure}
\par In Fig. \ref{fig11n}, the temporal evolution of $h_{\min}$ was presented for evaporation regimes ranging from weak to strong. It was observed that $h_{\min} \to 0$ when $b = 0$, whereas $h_{\min} \to \alpha_0$ for $b \neq 0$. To better illustrate the influence of evaporation $(E > 0)$ on the film profiles, both in the presence and absence of $b$, Fig. \ref{fig12} depicts the temporal evolution of the film profiles when evaporation is strong $(E=0.01)$. It is observed that at early time $(t = 100)$, the film first exhibits a decay and then propagates across the domain while maintaining droplets of almost constant shape and size in the absence or the presence of $b$. When $t=200$, it is observed that after the initial decay, the shape and size of the droplets remain almost the same for $b=0$, while they reduce with increasing $b$. Further, as time increases, the film becomes flattened and tends towards the precursor thickness $h=\alpha_0$ ($\alpha_0=0.1714$ for $b=0.1$ and $\alpha_0=0.189$ for $b=0.2$ with $Bo=1$, $\epsilon=0.1$ and $\Pi=0.1$). At $t = 600$, only the film profile corresponding to $b = 0.2$ reaches the precursor layer thickness. By $t = 800$, both profiles with $b \neq 0$ attain the precursor thickness, whereas the $b = 0$ profile does not vanish. Finally, at $t = 1050$, $h \to 0$ for $b = 0$.

\begin{figure}[h!]
\centering
\subfloat[$Bo=0.5$]{\includegraphics[scale=0.37]{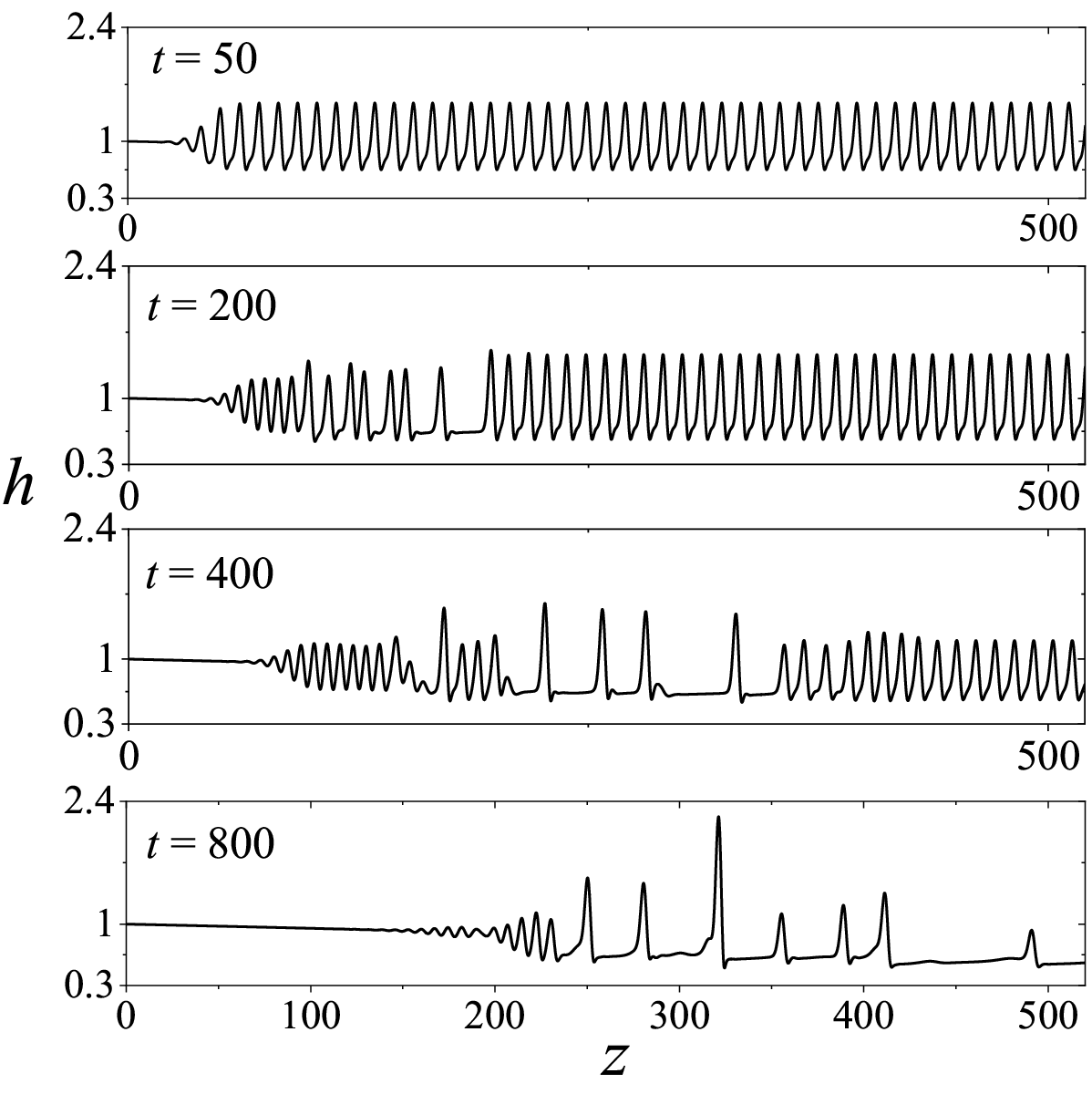}}
\qquad
\subfloat[$Bo=1$]{\includegraphics[scale=0.37]{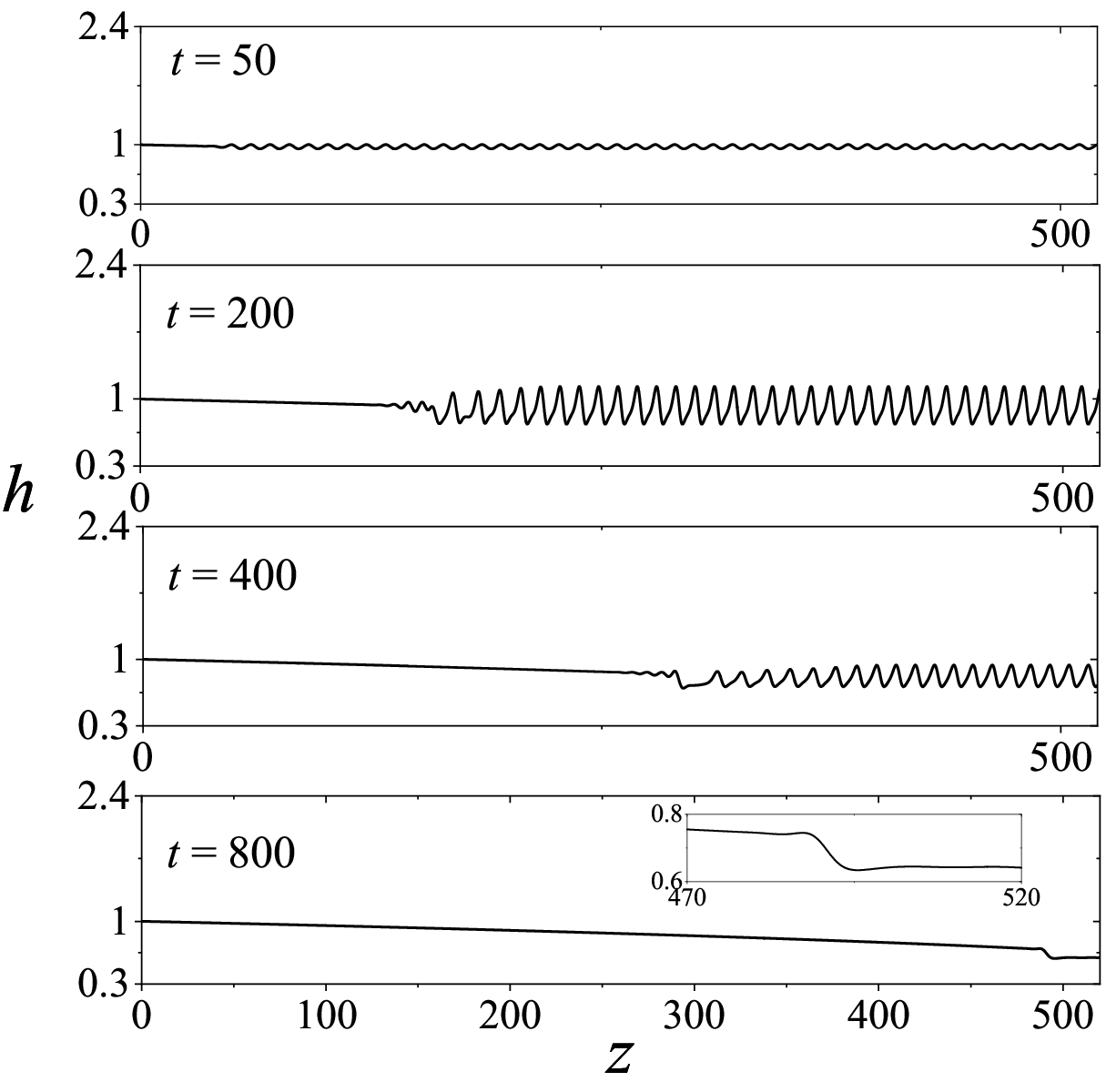}}
\caption{Temporal evolution of the film profiles for moderate evaporation. The other fixed parameters are $E=0.005$, $b=0$, $k=0.6$, $Ma=1$, $\Pi=0.1$ and $K_0=10$.}
\label{fig13}
\end{figure}
\par
\begin{figure}[h!]
\centering
\subfloat[First coalescence]{\includegraphics[scale=0.37]{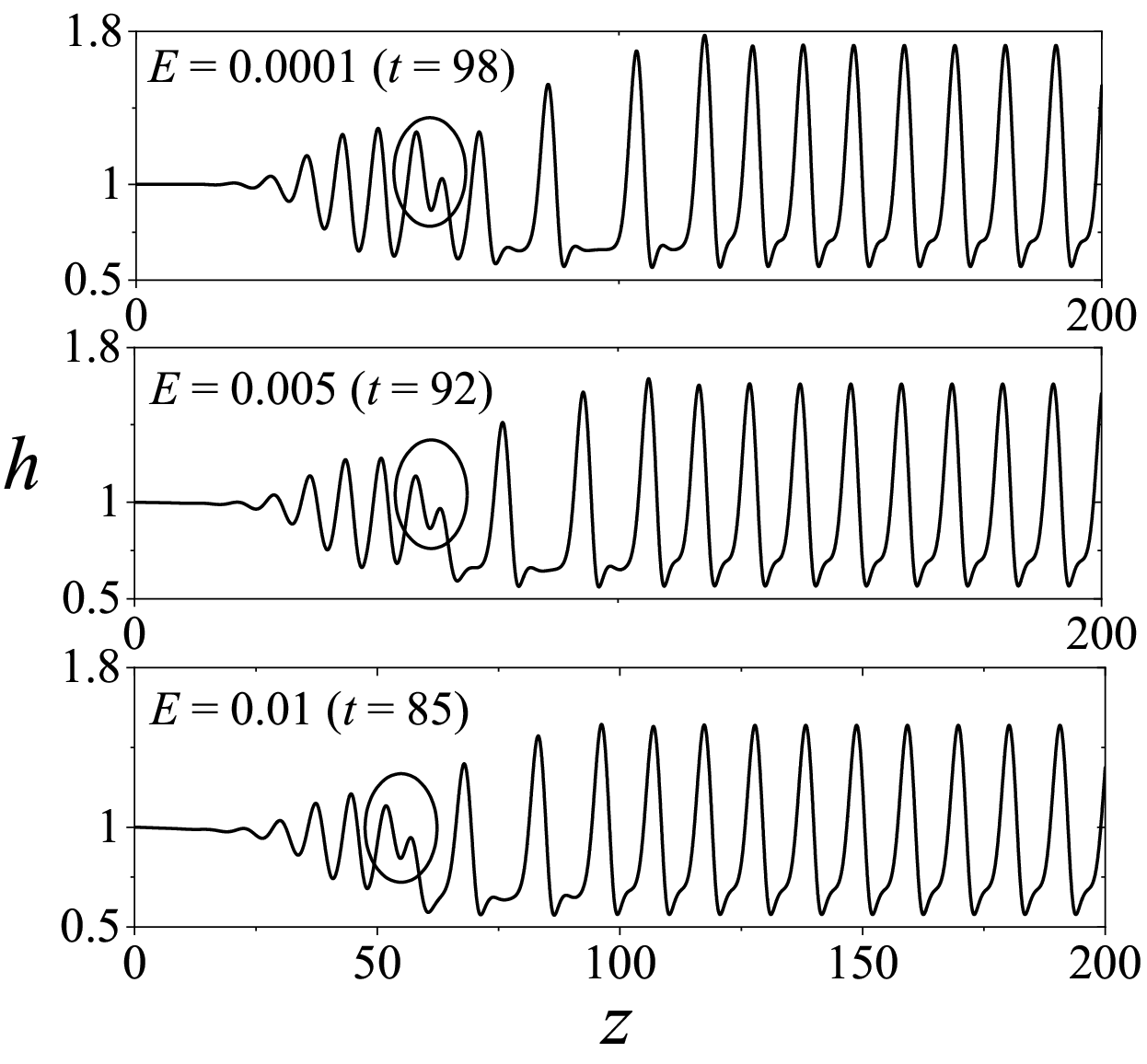}}
\quad
\subfloat[Coalescence at a later time]{\includegraphics[scale=0.37]{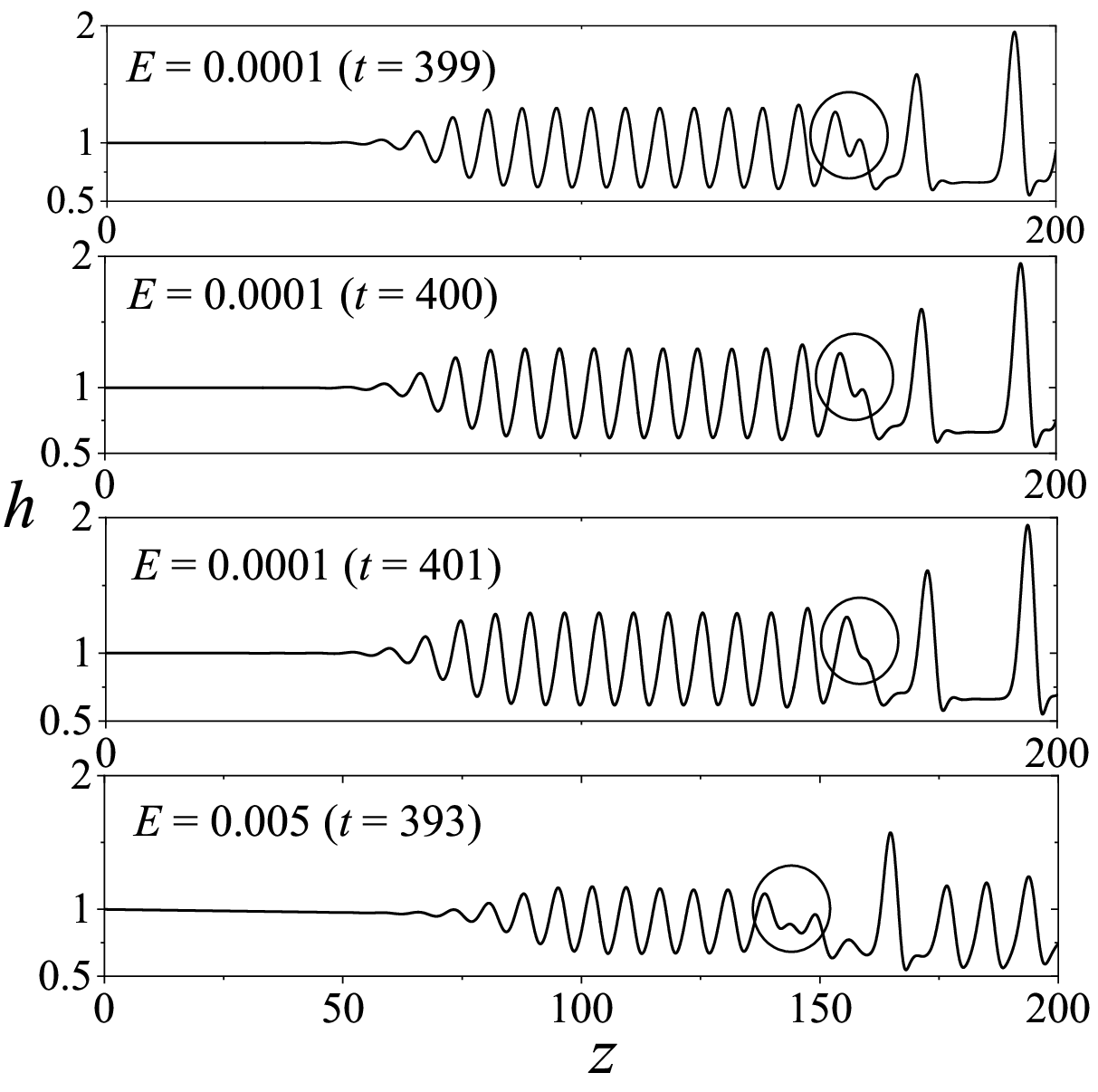}}
\caption{Droplet coalescence due to evaporation. The other fixed parameters are $b=0$, $Bo=0.5$, $k=0.6$, $Ma=1$, $\Pi=0.1$ and $K_0=10$.}
\label{fig14}
\end{figure}

 In all previous discussions, the effect due to interfacial phase change is presented for Bond number $Bo=1$, where  $Bo=\epsilon^{-1}\widehat{Bo}$, i.e. $\widehat{Bo}=\rho gR^2/\sigma_0\sim O(\epsilon)$. In experiments, the value of $\widehat{Bo}$ is found to be small ($\widehat{Bo}\lesssim 0.3$ \cite{craster2006viscous}). 
A low $Bo$ indicates the dominating effect of gravity over surface tension. The effect of a lower value, $Bo=0.5$, is shown in Figs. \ref{fig13} and \ref{fig14}. The temporal evolution of the film profiles is presented in Fig. \ref{fig13} for a moderate evaporation number $E=0.005$. The profiles corresponding to $Bo=0.5$ are shown in Fig. \ref{fig13}a, whereas profiles corresponding to $Bo=1$ are displayed in Fig. \ref{fig13}b. It is found that the interfacial waves exhibit higher humps at lower $Bo$. The droplets are more pronounced for lower $Bo$ due to the azimuthal curvature \cite{kalliadasis1994drop}. A similar trend for low $Bo$ has also been reported by Singh \& Tiwari \cite{singh2025coating} for a liquid film flow on a similar geometry. 
\par From Fig. \ref{fig13}a, it is observed that at $t=50$, the droplets exhibit uniform shape and size, with regular spacing. By $t=200$, the droplets become irregularly distributed over parts of the domain. However, the droplets continue to propagate downstream while approximately preserving their overall morphology. At $t=400$, this irregularity becomes more pronounced, and by $t=800$, the droplets are more widely spaced and developed. It can be observed that when $Bo=0.5$, the disturbance propagates downstream and leaves the domain. This behavior suggests that, for the chosen parameters, the system exhibits characteristics of convective unstable flow \cite{liu2017thermocapillary,khanum2020gravity}. However, a detailed analysis distinguishing between absolute and convective instability \cite{duprat2007absolute,ji2021thermally,khanum2020gravity} is beyond the scope of the present study. 

\par Fig. \ref{fig14} provides a clearer illustration of this collision process for an evaporating liquid at $Bo=0.5$. In Fig. \ref{fig14}a, the onset of the first droplet coalescence is illustrated as the evaporation strength increases from weak to strong. It is found that stronger evaporation leads to earlier droplet coalescence. Furthermore, Fig. \ref{fig14}b presents a representative coalescence event at a later stage. For weak evaporation, droplet interaction begins at approximately $t=399$ for the selected parameters, and complete merging occurs by $t=401$. As the evaporation rate increases to moderate levels, the collisions occur earlier and shift upstream. Therefore, Fig. \ref{fig14} demonstrates that evaporation promotes and accelerates droplet coalescence.

\section{Summary and Conclusions}\label{summary}
In this study, the dynamics of stability of a thin volatile liquid film flowing along a vertical cylinder are discussed. A thin-film model is constructed by assuming that the liquid thickness is much smaller than the cylinder's radius. To demonstrate the interfacial phase change effects, an evaporative mass flux is considered that depends on the local film thickness and the hydrodynamic pressure, including curvature, surface tension, and intermolecular forces. An equation describing the phase change effects due to liquid evaporation $(E>0)$ or vapor condensation $(E<0)$ has been derived under the lubrication approximation. In addition to the parameter $E$, this equation is influenced by thermocapillary forces (via Marangoni number $M$) and intermolecular forces (through the parameter $\Pi$). 

\par A linear stability analysis of the model has been performed by allowing the base liquid thickness to be non-constant, leading to two distinct cases. In the first case, the base film thickness is assumed to vary with time while remaining spatially uniform. The base film is found to thin with time for $E>0$ and to thicken for $E<0$. In the presence of hydrodynamic pressure, thinning during evaporation and thickening during condensation are further promoted. The corresponding linear growth rate is determined using a frozen-time approximation, under which the base film thickness is treated as fixed at each instant. An increase in the growth rate is obtained for $E>0$, whereas a decrease is obtained for $E<0$, with these effects being more pronounced when hydrodynamic pressure is included. Thus, within the frozen-time framework, a direct destabilizing contribution of evaporation to the instantaneous perturbation growth is obtained. In the second case, the base film thickness is allowed to evolve in both space and time. A coupled system of equations is then obtained, with one equation governing the evolution of the base state and the other describing the evolution of the perturbations. The perturbation growth is evaluated for different wavenumbers and compared with that of the corresponding nonvolatile film. At early times, the growth-rate profile obtained from the linear stability analysis with an evolving base state closely follows that of the nonvolatile film. However, deviations become apparent at later times. Moreover, as the evaporation rate is increased, these deviations emerge at earlier times.

\par Furthermore, the liquid–vapor interface equation is solved numerically to investigate the dynamics of the full model. The growth rate obtained from the full model is compared with that of the linearized system of equations for an evaporating liquid, and an excellent agreement is observed across all wavenumbers. The influence of evaporation on the film profiles is also demonstrated. It is shown that when evaporation is sufficiently strong, the liquid thickness decreases to zero over time when $\Pi=0$. However, when $\Pi\neq0$, a precursor film thickness is obtained, at which the liquid thickness approaches a uniform value over time under the effect of evaporation. Finally, the effect of the Bond number is also examined. It is shown that, for lower Bond numbers, the liquid interface breaks up into more droplets than at higher Bond numbers. In addition, droplet coalescence occurs during evaporation, and its timing and location depend on the evaporation rate.

\par Although this study presents a unified framework to understand the dynamics and stability of a thin, volatile liquid film flowing over a uniformly heated vertical cylinder, several limitations should be noted regarding the physical range and applicability of the model. First, all liquid properties in the present model are assumed to be constant, except for the surface tension. In reality, viscosity, density, thermal conductivity, and other thermophysical properties generally depend on temperature. Accounting for temperature-dependent fluid properties would therefore provide a more accurate description of the coupled hydrodynamic and thermal dynamics \cite{khanum2020gravity,mukhopadhyay2018long,ji2021thermally,chattopadhyay2021odd}. Second, the present formulation is developed under the assumption of negligible inertia. Thus, the present lubrication model is no longer valid when inertial effects are present. In this case, models such as weighted-residual \cite{ruyer2008modelling,ruyer2012wavy} or integral boundary-layer models \cite{sisoev2006film,ding2013viscous} would be more appropriate. Third, the vapor phase is not resolved explicitly in the present model. The phase change effects are incorporated through the interfacial mass-transfer formulation, whereas vapor-phase diffusion, convection, and heat transport are neglected. This assumption limits the model when vapor-side transport becomes comparable to the liquid-side transport or when the surrounding vapor field strongly influences the interfacial mass flux. A fully coupled liquid-vapor formulation would therefore be required to describe such regimes.

\par In addition, an important aspect concerning the estimation of the scaled Hamaker constant, $\Pi=\widetilde\Pi/(6\pi\rho g R\mathcal H^3)$, is noted. Typical values of $\widetilde\Pi$ lie between $10^{-19}$ and $10^{-21}$ N m \cite{israelachvili2011intermolecular}. This leads to very small values of $\Pi$ for both hexane and water. The order of magnitude is $O\left(10^{-9}\right)$ to $O\left(10^{-7}\right)$. A previous experimental-theoretical study reported by Ji et al. \cite{ji2019dynamics} for thin liquid films on vertical cylindrical fibres estimated $\widetilde\Pi \in \left[10^{-12},10^{-11}\right]$ N m. Thus, if a stronger $\widetilde\Pi \in \left[10^{-15},10^{-14}\right]$ N m is adopted, values of $\Pi \approx 5.15\times\left[10^{-3},10^{-2}\right]$ for hexane and $\Pi \approx 3.38\times\left[10^{-3},10^{-2}\right]$ for water are obtained. If $\widetilde\Pi=3\times10^{-14}$ N m can be chosen, $\Pi\approx0.1$ is found for both hexane and water.
Accordingly, in the present theoretical discussion, $\Pi = \left[10^{-3},10^{-1}\right]$ is considered, implying that the contribution of $\Pi$ in the model is much stronger than that associated with typical van der Waals interactions. Similarly, the physically estimated values of the parameter $b$ lie in the range $\left[10^{-7},10^{-5}\right]$, whereas larger values, such as $b\geq O\left(10^{-2}\right)$, are used to reveal appreciable effects. Thus, the results obtained for these larger $b$ and $\Pi$ values should be interpreted as demonstrating the underlying physical mechanisms rather than providing direct quantitative predictions for specific liquid systems.

\section*{Appendix}\label{appendix}
In Sec. \ref{sec:stability}, the linear stability analysis of the liquid–air interface equation was performed under the assumption that the effect of $Ma$ was negligible, so that the effect of phase change could be isolated. However, when the effect of $Ma$ is taken into account, the $O(1)$ and $O(\zeta)$ equations in (\ref{hjsc}) are given as follows:
 \begin{subequations}\label{hjsc1_Ma_Effect}
 \begin{equation}\label{leading}
O(1)~: \quad H_t+\left[\frac{H^3}{3}\left(1+H_z+H_{zzz}\right)-\frac{\Pi}{H}H_z-\frac{MaH^2}{2}\left(T_0^i\right)_z\right]_z=-EJ_0,
 \end{equation}
\begin{equation}\label{zeta1_Ma_Effect}
O(\zeta)~: \quad\eta_t+\left[H^2\left(1+H_z+H_{zzz}\right)\eta+\frac{H^3}{3}\left(\eta_z+\eta_{zzz}\right)-\Pi\left(\frac{\eta_z}{H}-\frac{H_z}{H^2}\eta\right)-Ma\left(H\eta\left(T_0^i\right)_z+\frac{H^2}{2}\left(T_1^i\right)_z\right)\right]_z=EJ_1,
 \end{equation}
 where
\begin{equation}
T_0^i=\frac{K_0}{K_0+H}\left[1-\frac{bH}{K_0}\left(\frac{1}{\epsilon}-H-H_{zz}\right)+\frac{b\Pi}{K_0H^2}\right],
\end{equation}
\begin{equation}
T_1^i=-\frac{b}{K_0+H}\left[\left(\frac{1}{\epsilon}-2H-H_{zz}\right)\eta-H\eta_{zz}+\frac{2\Pi}{H^3}\eta\right]-\frac{\eta}{K_0+H} T_0^i,
\end{equation}
\end{subequations}
and the expressions of $J_0$ and $J_1$ are provided in Eqs. (\ref{hjsc}c) and (\ref{hjsc}d), respectively.

\par For spatially uniform films as discussed in Sec. \ref{sec:stability_a}, the system of equations (\ref{hjsc1_Ma_Effect}) reduces to the following form
\begin{subequations}\label{sc111_Ma_Effect}
 \begin{equation}
O(1)~: \quad H_{0,t}=-EJ_0\left(H_0\right),
 \end{equation}
\begin{equation}
O(\zeta)~: \quad\eta_t+H_0^2\eta_z+\gamma_1\eta_{zz}+\gamma_2\eta_{zzzz}=EJ_1\left(H_0\right),
 \end{equation}
where 
\begin{equation}\label{sscc}
\gamma_1=\frac{H_0^3}{3}-\frac{\Pi}{H_0}+\frac{MaH_0^2}{2\left(K_0+H_0\right)}\left[b\left(\frac{1}{\epsilon}-2H_0+\frac{2\Pi}{H_0^3}\right)+T_u^i\right],
\end{equation}
\begin{equation}
\gamma_2=\frac{H_0^3}{3}\left(1-\frac{3Ma b}{2\left(K_0+H_0\right)}\right),
\end{equation}
\begin{equation}
T_u^i=\frac{K_0}{K_0+H_0}\left[1-\frac{bH_0}{K_0}\left(\frac{1}{\epsilon}-H_0\right)+\frac{b\Pi}{K_0H_0^2}\right]
\end{equation}
\end{subequations}
and $J_0,J_1$ are provided in Eqs. (\ref{sc11}c) and (\ref{sc11}d), respectively. 

\begin{figure}[h!]
\centering
\subfloat[$b=0$]{
\includegraphics[scale=0.45]{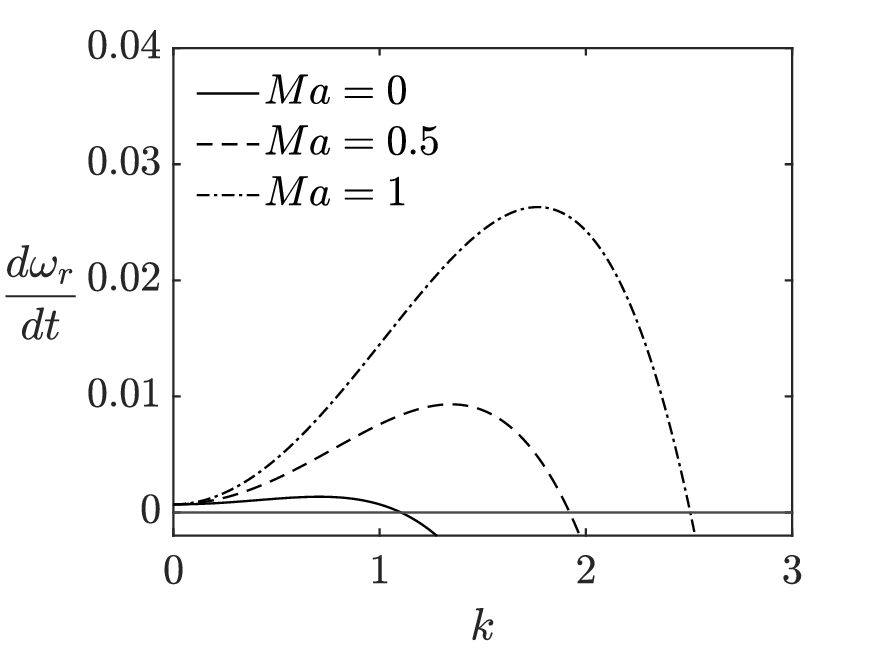}}
\subfloat[$b=0.01$]{
\includegraphics[scale=0.45]{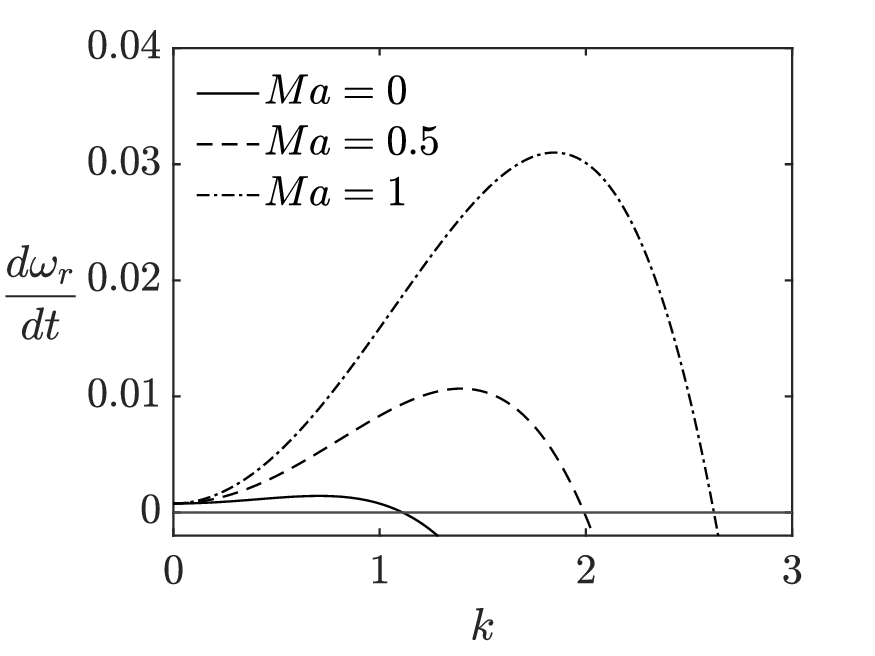}}
\caption{Linear growth rate as a function of wavenumber $k$ for different $Ma$ when $H_0$ is evaluated from Eq. (\ref{sc0}) at $t=1$. The other fixed parameters are $E=0.001$, $\Pi=0$ and $K_0=1$.}
\label{fig5nn_Ma_new}
\end{figure}

\par Consequently, for a spatially uniform film, the effective growth rate $d\omega_r/dt$ is as follows:
\begin{equation}\label{growth1_Ma_new}
\frac{d\omega_r}{dt}=\left(\gamma_1-\frac{Eb}{K_0+H_0}\right)k^2-\gamma_2k^4+\frac{bE}{K_0+H_0}\left(1-\frac{3\Pi}{H_0^4}\right)+\frac{EJ_0\left(H_0\right)}{K_0+H_0}.
\end{equation}
\par In the absence of $E,b,\Pi$ and considering $H_0=1,K_0=1/Bi$, Eq. (\ref{growth1_Ma_new}) reduces to $\frac{d\omega_r}{dt}=k^2\left[\frac{1}{3}+\frac{MaBi}{2\left(1+Bi\right)^2}-\frac{k^2}{3}\right]$, which agrees well with the growth rate obtained in the previous studies on similar geometries under different setups \cite{chao2018dynamics,chattopadhyay2026stability,chattopadhyay2024modeling,khanum2020gravity}.

\par In Fig. \ref{fig5nn_Ma_new}, the linear growth rate is plotted as a function of the wavenumber for an evaporating liquid $(E>0)$ to investigate the effect of the Marangoni number $Ma$. It was previously observed from Fig. \ref{fig4}a that the flow instability is amplified by evaporation; it can be seen from Fig. \ref{fig5nn_Ma_new}a that the evaporation-induced instability is further enhanced by the presence of $Ma$. A comparison of the two cases presented in Fig. \ref{fig5nn_Ma_new} reveals that, for $b\neq0$, the instability is further amplified, as also observed in Fig. \ref{fig5}a. When condensation is considered instead of evaporation, the same role of $Ma$ is observed (the corresponding figure is not presented here). Thus, it can be concluded from Fig. \ref{fig5nn_Ma_new} that a destabilizing effect is exerted by $Ma$ which is consistent with the findings of previous studies on similar geometries \cite{chao2018dynamics,chattopadhyay2026stability}.

\begin{figure}[h!]
\centering
\includegraphics[scale=0.42]{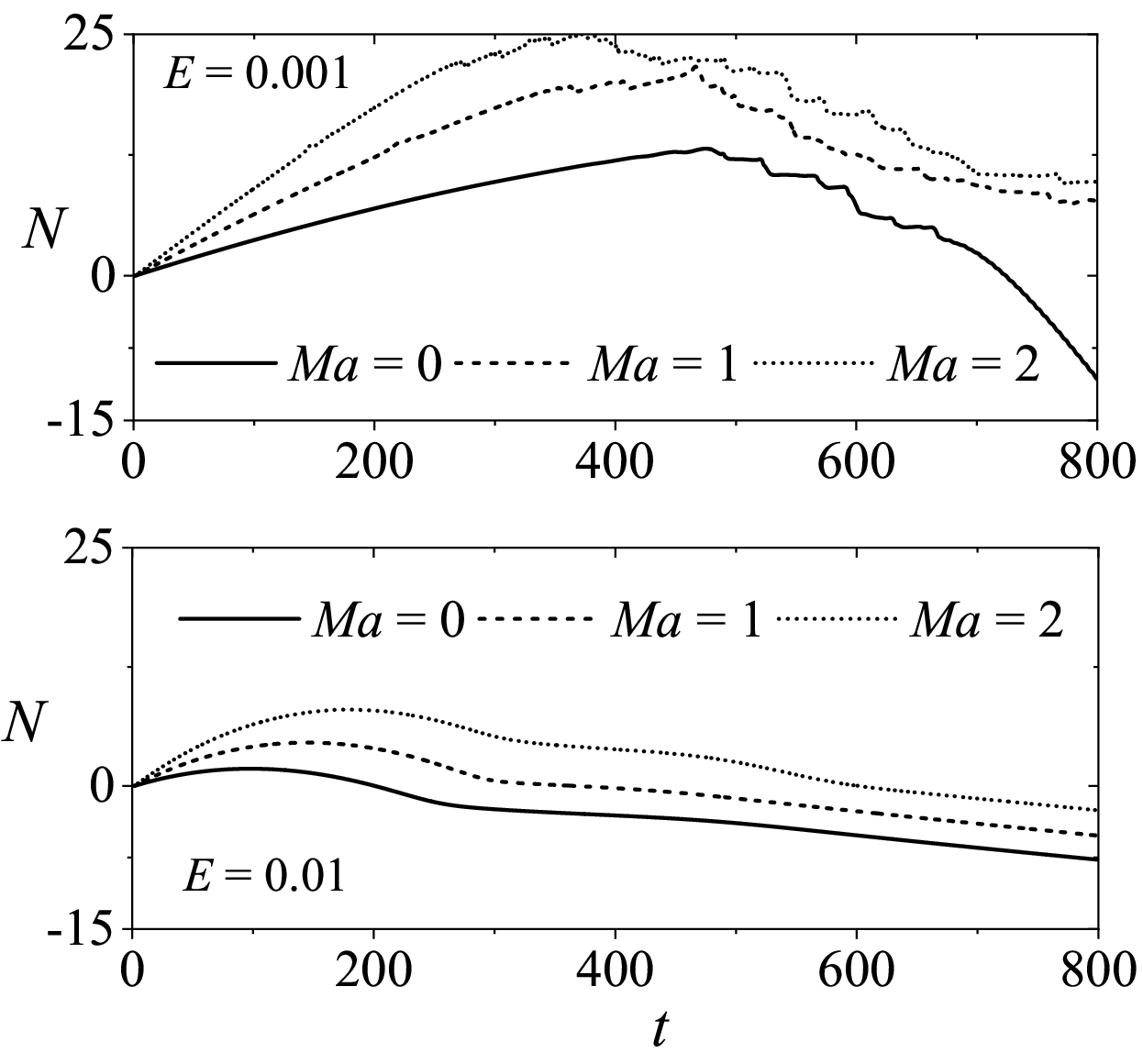}
\caption{Temporal evolution of $N$ from Eq. (\ref{nn}) corresponding to Eq. (\ref{hjsc1_Ma_Effect}) for an evaporating liquid at different $Ma$. The other fixed parameters are $k=0.6$, $b=0.1$, $\epsilon=0.1$, $\Pi=0.1$, and $K_0=10$.}
\label{fig7_n_sc_abcd}
\end{figure}
\begin{figure}[h!]
\centering
\subfloat[$E=0.001$]{
\includegraphics[scale=0.34]{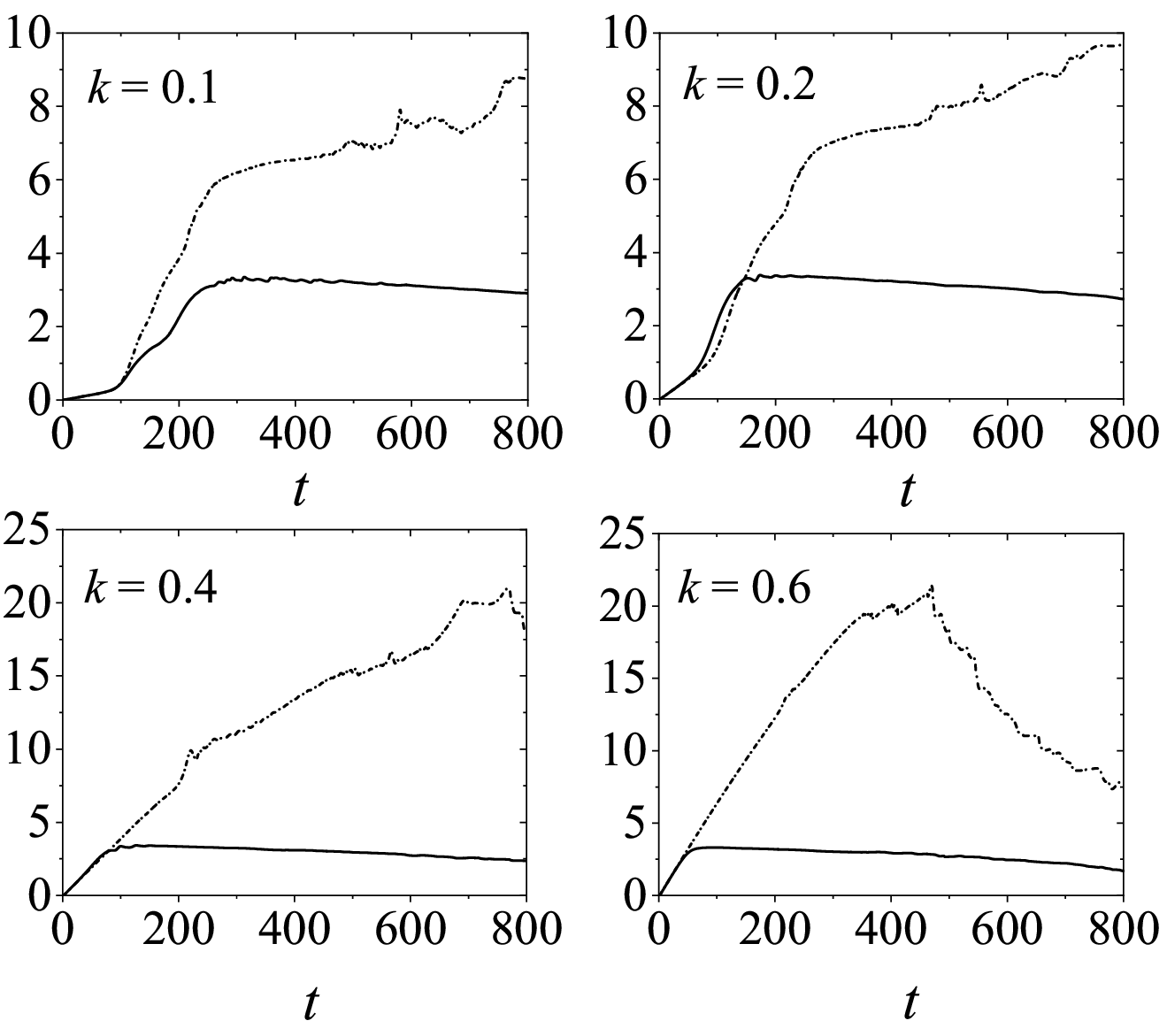}}
\quad
\subfloat[$E=0.01$]{\includegraphics[scale=0.34]{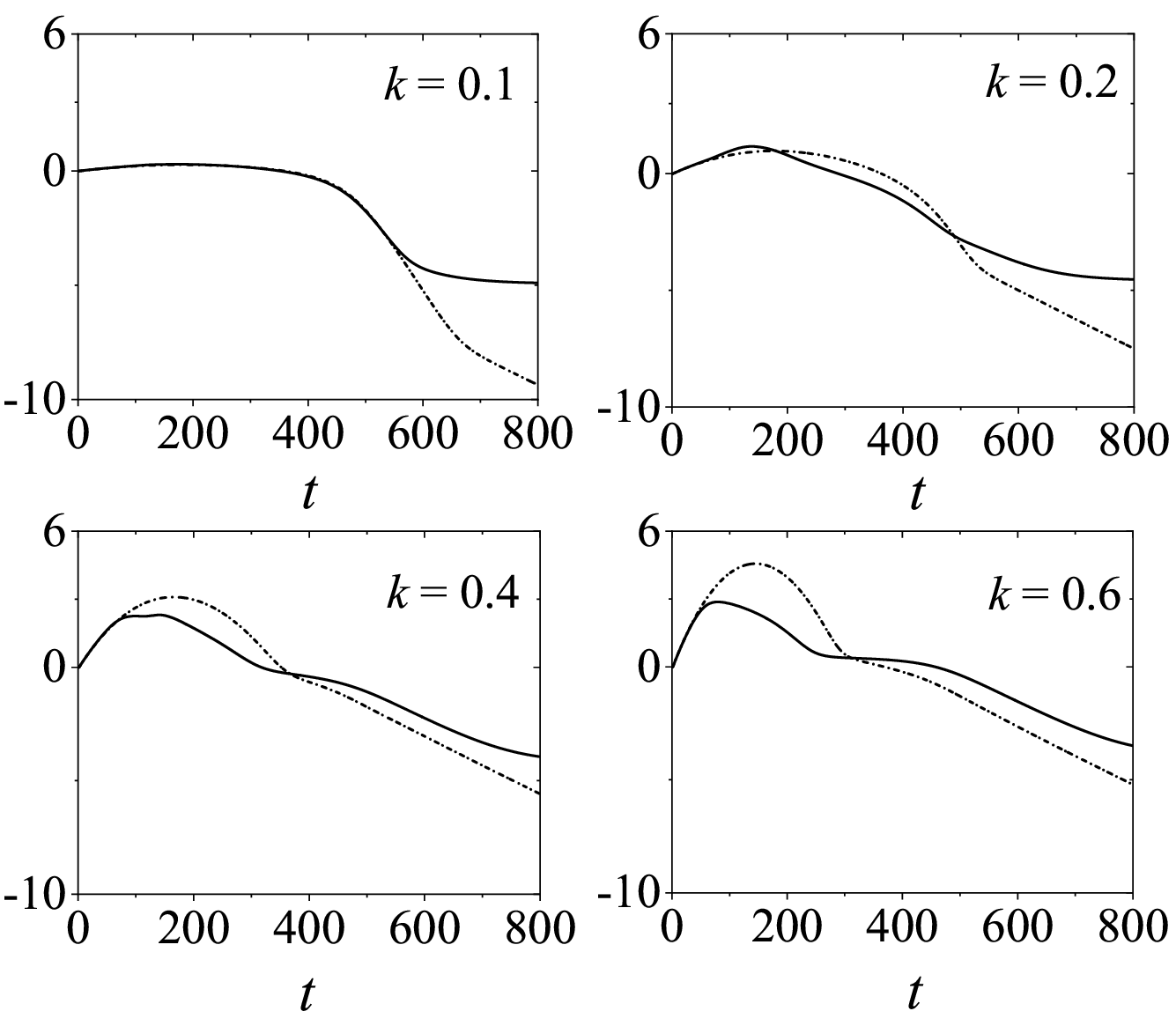}}
\caption{Comparison of the growth rate of perturbations for an evaporating liquid when $Ma\neq0$. The solid line denotes $F_{NL}$ from Eq. (\ref{nn1_sc}), while the dashdot line denotes $N$ from Eq. (\ref{nn}) corresponding to Eq. (\ref{hjsc1_Ma_Effect}). The other fixed parameters are  $\Pi=0.1$, $b=0.1$, $\epsilon=0.1$, $K_0=10$ and $Ma=1$.}
\label{fig7_nnn_scc_abcd}
\end{figure}
\par In Fig. \ref{fig7_n_sc_abcd}, the temporal evolution of the growth rate $N$ is presented for $Ma\neq0$ under two different evaporation conditions: moderate evaporation $(E=0.001)$ and strong evaporation $(E=0.01$). It is observed that higher growth rates are obtained with increasing $Ma$ for both evaporation conditions. Furthermore, in Fig. \ref{fig7_nnn_scc_abcd}, the growth rates obtained from the linearized system (\ref{hjsc1_Ma_Effect}) are compared with those predicted by the full model (\ref{evolution_equation}) for different wavenumbers when $Ma\neq0$. Similar to the good agreement observed in Fig. \ref{fig_compare_three_n} for $Ma=0$, a similarly good match between the two models is also observed for $Ma\neq0$.

\section*{Declaration of Interests}
The authors report no conflicts of interest.



\bibliographystyle{elsarticle-num}
\bibliography{cylinder}

\end{document}